\documentclass[aps,pre,onecolumn,superscriptaddress,longbibliography,hidelinks]{revtex4-2}

\usepackage{xcolor}
\usepackage{graphicx}
\usepackage{siunitx}
\usepackage{amsmath}
\usepackage{hyperref}

\begin{document}

\title{Pattern selection and the route to turbulence in incompressible polar active fluids}

\author{Henning Reinken}
\email{henning.reinken@ovgu.de}
\affiliation{Institut für Physik, Otto-von-Guericke-Universit\"at Magdeburg, Universit\"atsplatz 2, 39106 Magdeburg, Germany}
\affiliation{Institut f\"ur Theoretische Physik, Technische Universit\"at Berlin, Stra{\ss}e des 17. Juni 135, 10623, Berlin, Germany} 

\author{Sebastian Heidenreich}
\affiliation{Department of Mathematical Modelling and Data Analysis,
Physikalisch-Technische Bundesanstalt Braunschweig und Berlin, Abbestr. 2-12,
10587 Berlin, Germany}

\author{Markus B\"ar}
\affiliation{Department of Mathematical Modelling and Data Analysis,
Physikalisch-Technische Bundesanstalt Braunschweig und Berlin, Abbestr. 2-12,
10587 Berlin, Germany}

\author{Sabine H. L.  Klapp}
\affiliation{Institut f\"ur Theoretische Physik, Technische Universit\"at Berlin, Stra{\ss}e des 17. Juni 135, 10623, Berlin, Germany} 

\date{\today}

\begin{abstract}
Active fluids, such as suspensions of microswimmers, are well known to self-organize into complex spatio-temporal flow patterns. An intriguing example is mesoscale turbulence, a state of dynamic vortex structures exhibiting a characteristic length scale.
Here, we employ a minimal model for the effective microswimmer velocity field to explore how the turbulent state develops from regular, stationary vortex patterns when the strength of 
activity \textit{resp.} related parameters such as nonlinear advection or polar alignment strength - is increased.
First, we demonstrate analytically that the system, without any spatial constraints, develops a stationary square vortex lattice in the absence of nonlinear advection.
Subsequently, we perform an extended stability analysis of this nonuniform ``ground state'' and uncover a linear instability, which follows from the mutual excitement and simultaneous growth
of multiple perturbative modes.
This extended analysis is based on linearization around an approximation of the analytical vortex lattice solution and allows us to calculate a critical advection or alignment strength, above which the square vortex lattice becomes unstable.
Above these critical values, the vortex lattice develops into mesoscale turbulence in numerical simulations.
Utilizing the numerical approach, we uncover an extended region of hysteresis where both patterns are possible depending on the initial condition.
Here, we find that turbulence persists below the instability of the vortex lattice.
We further determine the stability of square vortex patterns as a function of their wavenumber and represent the results analogous to the well-known Busse balloons known from classical pattern-forming systems such as Rayleigh--B\'enard convection experiments and corresponding models such as the Swift--Hohenberg equation.  Here, the region of stable periodic patterns shrinks and eventually disappears with increasing activity parameters. 
Our results show that the strength of activity plays a similar role for active turbulence as the Reynolds number does in driven flow exhibiting inertial turbulence.
\end{abstract}


\maketitle

\section{Introduction}
\label{sec: introduction}

One of the recurring questions in fluid dynamics research is how turbulence arises from a laminar base flow upon increase of the Reynolds number~\cite{eckert2010troublesome,barkley2015rise}.
In driven fluid systems described by the Navier--Stokes equation, the geometry of the problem and the specific setup have a crucial influence on the relevant mechanisms.
For example, in Taylor--Couette flow dominated by inner-cylinder rotation, the development of turbulence follows from a series of subsequent linear instabilities~\cite{taylor1923viii,barkley2016theoretical,feldmann2023routes}.
This is in contrast to dominant outer-cylinder rotation, where linear instabilities are absent and, instead, finite perturbations are necessary to trigger turbulent flow~\cite{feldmann2023routes}.
The situation is similar to channel~\cite{sano2016universal}, pipe~\cite{barkley2015rise,barkley2016theoretical}, and plane Couette flow~\cite{grossmann2000onset,lemoult2016directed,manneville2015transition} where mechanisms beyond linear stability are at work~\cite{drazin2004hydrodynamic}.
Recent large-scale experimental and numerical studies of such systems~\cite{nishi2008laminar,avila2011onset,sano2016universal,lemoult2016directed,avila2023transition} have revealed that the mechanism is better understood in terms of a ``statistical phase transition''. For pipe flow,  the dynamics of turbulent puffs, e.g., appears analogous  with the much-studied  nonequilibrium phase transitions in the directed percolation universality class ~\cite{hinrichsen2000non}.

Turbulent states are also a central topic in active matter, a research field that focuses on living or artifical systems consisting of moving constituents~\cite{marchetti2013hydrodynamics,cates2015motility,bechinger2016active,bar2020self,gompper20202020,chate2020dry}.
Thus, active matter systems are intrinsically out of equilibrium. 
As a consequence, they display diverse pattern formation, including turbulent flows commonly labelled as ``active turbulence''~\cite{alert2021active} that were found experimentally in suspensions of microswimmers~\cite{dombrowski2004self,wensink2012meso} and in active nematics~\cite{doostmohammadi2018active}.
Contrary to the above-mentioned turbulence in ``passive'' fluids, which arises at high Reynolds numbers, active turbulence occurs in the overdamped regime.
Still, similarly to some of the passive systems, a recent study on the transition to active turbulence 
in a model for active nematic liquid crystals (such as microtubule protein mixtures) has revealed analogies to directed percolation as well~\cite{doostmohammadi2017onset}.
Recent studies of detailed ``microscopic'' models for the collective dynamics of interacting self-propelled microswimmers \cite{qi2022emergence,zantop2022emergent} as well as a qualitative model with competing alignment interactions \cite{grossmann2014vortex, grossmann2015pattern} have also shown turbulent states with qualitatively similar energy spectra and correlation functions as continuum models for active polar fluids \cite{wensink2012meso, dunkel2013minimal,dunkel2013fluid}.

Besides active turbulence, stable vortex lattices with different symmetries have been reported in earlier experiments with swimming sperm cells~\cite{riedel2005self} and an active microtubule-myosin motor mixture~\cite{sumino2012large} and attributed to chiral motion of the active agents. Subsequently, many simulation studies of continuum models for active nematics~\cite{doostmohammadi2016stabilization,thijssen2020role, caballero2023vorticity} as well as of agent-based simulations~\cite{grossmann2014vortex,grossmann2015pattern} and continuum models for active polar fluids~\cite{james2021emergence} exhibited vortex patterns of different symmetries. While most of continuum model studies assume constant density, recent work also has shown that vortex lattices precede turbulent clustering also in a model for compressible active fluids that allows for density fluctuations~\cite{worlitzer2021turbulence}. A stable vortex lattice could also be realized experimentally by introducing a periodic array of small obstacles into a bacterial suspensions that exhibits active turbulence~\cite{nishiguchi2018engineering}. This phenomenon could be reproduced with simulations in a model of an active polar fluid~\cite{reinken2020organizing,reinken2022ising}. More recently, several numerical studies have suggested control schemes to tame and control turbulence in active nematics~\cite{partovifard2024controlling,schimming2024vortex}.

Most experimental studies of transitions between regular patterns and turbulence in active systems have been carried out exploiting confinement effects, i.e., gradually changing dimensions and aspect ratio of the experimental system.  
In two-dimensional systems, a single vortex was shown to be stable in experiments of small systems~\cite{wioland2013confinement} often connected with edge currents~\cite{beppu2021edge} and transitions to more complex multiple-vortex states when increasing the system size~\cite{opathalage2019self,beppu2022exploring, beppu2023geometric, shiratani2023route}. 
In three-dimensional channels several experimental and theoretical studies with active nematics have shown that widening the smallest extension of the channel can lead to a transition between coherent patterns and apparently turbulent flows~\cite{wu2017transition,chandragiri2020flow, chandrakar2020confinement}.
For the polar active fluid, simulations of the transition between turbulent states and regular vortex patterns, which are externally stabilized by small obstacles in the flow, have been shown to exhibit features of a non-equilibrium phase transition in the Ising universality class~\cite{reinken2022ising}.
The latter study has utilized a model for suspensions of microswimmers 
(such as \textit{Bacillus Subtilis}) that supports regular structures even in the absence of external stabilization, as well as turbulent states~\cite{reinken2020organizing}.
To summarize, while some knowledge has been
gained on the transition to turbulence in active fluids, we are far from a general understanding, including effects of the type of order parameter and the role of confinement and geometrical constraints. In particular, the situation for (polar) active fluids in the absence of spatial constraints is essentially unclear. This problem is addressed in the present paper.
In particular, we tackle the question of how the turbulent state arises from an underlying regular pattern.

The utilized model is formulated in terms of an effective microswimmer velocity field $\mathbf{v}(\mathbf{x},t)$ and was first introduced on phenomenological grounds in Ref.~\cite{wensink2012meso}.
It captures the main features of mesoscale turbulence, a state of highly dynamic vortex patterns that has been observed in suspensions of bacteria~\cite{wensink2012meso,sokolov2012physical} and artifical swimmers~\cite{nishiguchi2015mesoscopic}.
The emergent flow patterns are characterized by the presence of a characteristic length scale much larger than the size of single swimmers but smaller than the system size~\cite{wensink2012meso,sokolov2012physical}.
Recently, the model was substantiated by a derivation from a generic microscopic theory~\cite{heidenreich2016hydrodynamic,reinken2018derivation}.
Although there are a large number of follow-up studies using the model~\cite{dunkel2013fluid,bratanov2015new,james2018turbulence,james2018vortex,reinken2019anisotropic,reinken2020organizing,mukherjee2021anomalous,james2021emergence,reinken2022optimal}, it is still unclear how mesoscale turbulence emerges from the underlying regular structures that the model supports.
To answer this question, we here focus on the stability of possible structures, which include stripe patterns, as well as square, triangular and hexagonal lattices of vortices.

As in the Navier--Stokes equation, the emergence of turbulent states is facilitated by nonlinear advection.
In analogy to the Reynolds number in classical turbulence, we identify the strength of self-advection (advection of the velocity field via the field itself) as the central parameter that determines the stability of regular flow patterns and, thus, the emergence of the active turbulent state.
Analytical investigations of this connection, however, are severely complicated by the fact that the regular base state is already nonuniform.
By means of an extended stability analysis, we are able to show that the mesoscale-turbulent state arises as a consequence of a linear instability of the underlying base pattern.
The key finding here is that the route to turbulence includes multiple modes that only grow simultaneously via mutual excitement.
Analogously to a critical Reynolds number, we then determine the critical advection strength, which signifies the onset of the instability.
Our results are in excellent agreement with numerical solution of the full nonlinear equation (see Fig.~\ref{fig: bifurcation and snapshots}(b) for examples).
In parallel, we unravel the role of active polar alignment.
As it turns out, the alignment strength can serve as an alternative control parameter to drive the onset of turbulence.
To give a first impression,
we present in Fig.~\ref{fig: bifurcation and snapshots}(a) a schematic overview of the behavior at fixed advection strength. Upon increasing the alignment strength from small values, there is first a transition between the  disordered (isotropic) state and the regular vortex lattice. The latter becomes unstable at a critical alignment strength, beyond which a turbulent state emerges. Going back, we find a hysteresis loop that, as our results show, represents a characteristic feature related to the onset and disappearance of mesoscale turbulence.
Combining the results from the extendend linear stability analysis with a numerical investigation of the hysteretic behavior, we construct a state diagram of the system.
We also discuss the stability of vortex lattices with arbitrary wavenumbers on the basis of stability balloons.

\section{Model}
\label{sec: model}

\begin{figure}
\includegraphics[width=0.95\linewidth]{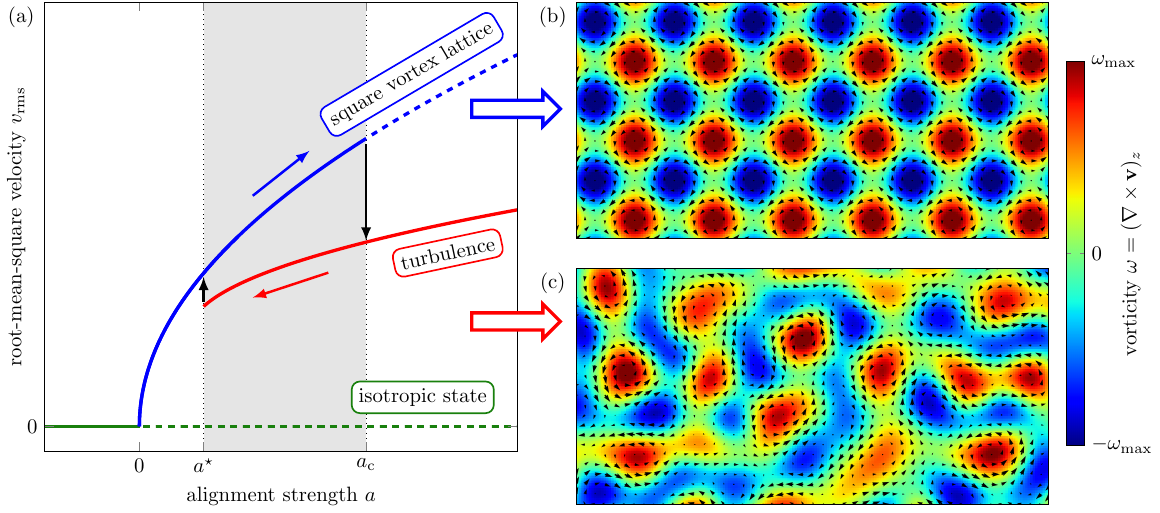}
\caption{\label{fig: bifurcation and snapshots}Schematic overview of spatiotemporal states and their occurrence in the present model [Eq.~(\ref{eq: dynamic equation})]. (a) We here employ $a$ as the control parameter and distinguish between different spatiotemporal states using the root-mean-square velocity observed in the system.
Solid lines represent stable branches, dashed lines indicate unstable branches.
Increasing $a$, the isotropic state becomes unstable against the emergence of a regular square vortex lattice at $a=0$.
This regular pattern, in turn becomes unstable at $a_\mathrm{c}$ and the system develops turbulence.
Following the turbulence branch while decreasing $a$, we observe a hysteresis loop as indicated by the arrows.
(b)/(c) Representative snapshots of the vorticity field $\omega(\mathbf{x},t)$ in the square vortex lattice and in the mesoscale-turbulent state at $a = 0.5$ and $\lambda = 3.0$ obtained via numerical solution of Eq.~(\ref{eq: dynamic equation}).
The superimposed arrows show the velocity field $\mathbf{v}(\mathbf{x},t)$ and the size of the snapshots is $12\pi \times 6\pi$.}
\end{figure}

We utilize an established model for dense  suspensions of microswimmers~\cite{wensink2012meso,dunkel2013fluid,dunkel2013minimal,bratanov2015new,heidenreich2016hydrodynamic,reinken2018derivation,reinken2019anisotropic,james2018vortex,james2018turbulence} where the dynamics is described via the effective microswimmer velocity field $\mathbf{v}(\mathbf{x},t)$~\cite{reinken2018derivation}.
The model can be derived from microscopic Langevin dynamics including coupling to the solvent flow~\cite{reinken2018derivation} and has been shown to capture experiments on bacteria in unconfined bulk flow~\cite{wensink2012meso} as well as in the presence of geometric confinement, e.g., lattices of obstacles~\cite{reinken2020organizing}.
The dynamics of $\mathbf{v}(\mathbf{x},t)$ can be conveniently written as
\begin{equation}
\label{eq: dynamic equation}
\partial_t \mathbf{v} + \lambda \mathbf{v}\cdot\nabla\mathbf{v} = -\frac{\delta \mathcal{F}}{\delta \mathbf{v}}\, ,
\end{equation}
where the functional $\mathcal{F}$ is given by
\begin{equation}
\label{eq: functional}
\begin{aligned}
\mathcal{F} = \int f(\mathbf{v},\nabla^2 \mathbf{v}) \, d \mathbf{x} \, , \qquad
f = q \nabla \cdot \mathbf{v} - \frac{a}{2} |\mathbf{v}|^2 + \frac{1}{4} |\mathbf{v}|^4 + \frac{1}{2} |(1 + \nabla^2) \mathbf{v}|^2 \, .
\end{aligned}
\end{equation}
Additionally, we assume that $\mathbf{v}(\mathbf{x},t)$ is divergence-free,
$\nabla \cdot \mathbf{v} = 0$.
Thus, the pressure-like quantity $q$ acts as a Lagrange multiplier ensuring the incompressibility.
Note that $\mathcal{F}$ plays a role similar to a free energy density, however, as the activity of the swimmers makes this a fundamentally non-equilibrium system, it is in fact not a free energy, but rather a more general non-equilibrium potential.
This model is sometimes referred to as the Toner--Tu--Swift--Hohenberg model~\cite{alert2021active}, because characteristic terms from both the  Toner--Tu and the Swift--Hohenberg equations appear in the above equations.
As Eq.~(\ref{eq: dynamic equation}) conveniently shows, the dynamics of $\mathbf{v}(\mathbf{x},t)$ is governed by a competition between gradient dynamics determined by the functional $\mathcal{F}$ and nonlinear advection $\lambda \mathbf{v}\cdot\nabla\mathbf{v}$.
We denote the coefficient $\lambda$ as advection strength, which can be related to mesoscopic parameters such as the self-propulsion speed~\cite{reinken2018derivation,reinken2022ising}.
The coefficient $a$ can be positive or negative and encodes the strength of polar alignment of the microswimmers.
Generally, it increases with the activity (self-propulsion speed and active stresses)~\cite{reinken2018derivation,reinken2022ising}, which leads to pattern formation for $a>0$, as we will see in the next section.
The quartic term in the functional $\mathcal{F}$, see Eq.~(\ref{eq: functional}), yields a cubic term in the evolution equation~(\ref{eq: dynamic equation}).
This term has a negative sign and, thus, leads to the saturation of the emerging patterns.
Note that in most studies~\cite{wensink2012meso,dunkel2013fluid,bratanov2015new,james2018turbulence,james2018vortex,reinken2019anisotropic,reinken2020organizing,mukherjee2021anomalous,james2021emergence,reinken2022optimal}, there is an additional positive coefficient in front of the cubic term.
However, this coefficient can be scaled out by an appropriate rescaling of $\mathbf{v}$ and $\lambda$. 

Before starting with an in-depth analysis of the model, let us briefly discuss the mesoscale-turbulent state obtained via numerical solution of Eq.~(\ref{eq: dynamic equation}), see Appendix~\ref{app: numerical methods} for information on the numerical methods.
Figure~\ref{fig: bifurcation and snapshots}(c) shows a typical snapshot of the vorticity field $\omega = (\nabla \times \mathbf{v})_z$ in the turbulent state at $a = 0.5$ and $\lambda = 3.16$.
The disordered vortex patterns display a characteristic length scale or vortex size, as observed in experiments on bacterial suspensions~\cite{wensink2012meso,sokolov2012physical}.

\section{Stationary pattern selection}
\label{sec: stationary patterns}

Before discussing the onset of the turbulent state from an analytical point of view, we start by establishing the kinds of stationary states that can be expected in our model.
First, let us note that Eq.~(\ref{eq: dynamic equation}) always has an isotropic, stationary homogeneous solution, $\mathbf{v} = \mathbf{0}$.
Depending on the value of $a$, one further finds a polar stationary homogeneous solution, i.e., $|\mathbf{v}| = \sqrt{a - 1}$.
The transition to the polar state at $a  = 1$ is analogous to the transition to a state of collective motion in the Toner--Tu equations~\cite{toner1998flocks,toner2005hydrodynamics}.
The emergence of such a polar state breaks the rotational symmetry of the system.
In this work, we restrict ourselves to the case $a < 1$, hence we can assume rotational invariance.

Performing a linear stability analysis of the isotropic solution of Eq.~(\ref{eq: dynamic equation}) is straightforward, see Refs.~\cite{dunkel2013minimal,reinken2019anisotropic} for more details.
We start by adding a small perturbation to the isotropic solution, i.e, 
\begin{equation}
\label{eq: add perturbations isotropic solution}
\mathbf{v} = \delta \hat{\mathbf{v}} e^{\sigma t} e^{i \mathbf{k} \cdot \mathbf{x}}\, ,\qquad q = \delta \hat{q} e^{\sigma t} e^{i \mathbf{k} \cdot \mathbf{x}} \, ,
\end{equation}
where $\delta \hat{\mathbf{v}}$ and $\delta \hat{q}$ are the perturbation amplitudes, $\mathbf{k}$ denotes the wavevector and $\sigma$ is the growth rate to be determined.
We here take into account a perturbation of $q$ because the incompressibility condition is part of the dynamical system.
After linearization around the isotropic state, we obtain the growth rate as a function of the wavevector, i.e.,
\begin{equation}
\label{eq: growth rate isotropic solution}
\sigma(\mathbf{k}) = a - 1 + 2 |\mathbf{k}|^2 - |\mathbf{k}|^4\, .
\end{equation}
As a first observation, note that the growth rate is real-valued.
This indicates that the instability is not oscillatory and instead stationary patterns start to grow.
Due to the chosen scaling of Eq.~(\ref{eq: dynamic equation}), the maximum of $\sigma(\mathbf{k})$ is located at a critical wavenumber $k_\mathrm{c} = 1$.
The sign of the coefficient $a$ determines whether the isotropic state is linearly stable.
For $a < 0$, i.e., low activity, the growth rate remains negative for all $\mathbf{k}$.
Increasing $a$ above the critical value $a = 0$, the growth rate becomes positive at $k_\mathrm{c}$, signifying the onset of a finite-wavelength instability.
Subsequently, for high activity, i.e., $a > 0$, a band of unstable modes develops, limited by the wavenumbers
\begin{equation}
\label{eq: band of unstable modes}
k_\pm = \sqrt{1 \pm \sqrt{a}}\, .
\end{equation}
Note that, for the here considered case $a < 1$, the growth rate only depends on $|\mathbf{k}|$ and not on the direction of the wavevector due to rotational invariance.

Building on these results, we continue the analysis by determining what kind of patterns might actually emerge.
As we are dealing with a stationary instability, we will restrict ourselves to stationary patterns.
Further, we will set $\lambda = 0$ for now, thus  eliminating the nonlinear advection term.
As a result, the instabilities that lead to the turbulent state are disregarded and the emerging states are completely determined as minima of $\mathcal{F}$, see Eq.~(\ref{eq: dynamic equation}).
In the following, we will investigate different patterns with the help of derived amplitude equations to determine if they are indeed minima.

We start by constructing possible emergent states in the system on the basis of the result from the linear stability analysis.
As we have seen, for $a > 0$ the isotropic solution $\mathbf{v} = \mathbf{0}$ is unstable to perturbations characterized by a band of wavenumbers, with $k_\mathrm{c}$ denoting the critical as well as the fastest-growing mode.
Therefore, close to the instability, it is reasonable to assume that the emerging patterns will be dominated by a characteristic length scale given by the wavelength $\Lambda_\mathrm{c} = 2\pi/k_\mathrm{c}$.
As the system is translationally invariant, we can arbitrarilly set the phase and, thus, a growing mode can be represented in two dimensions as
\begin{equation}
\label{eq: general mode cos}
\begin{aligned}
v_x(\mathbf{x}) &= A_x e^{i\mathbf{k}\cdot \mathbf{x}} +  \mathrm{c.c.}\, ,\\
v_y(\mathbf{x}) &= A_ye^{i\mathbf{k}\cdot \mathbf{x}} +  \mathrm{c.c.}\, ,
\end{aligned}
\end{equation}
where $A_x$ and $A_y$ are the amplitudes in $x$- and $y$-direction, $\mathrm{c.c.}$ denotes the complex conjugate and $\mathbf{k}\cdot\mathbf{k} = k_\mathrm{c}^2$.
The incompressibility condition, $\nabla\cdot\mathbf{v} = 0$, imposes a constraint on the relation of the amplitudes $A_x$ and $A_y$.
To meet the condition we set 
\begin{equation}
\label{eq: incompressibility condition new amplitude}
A_x = A \frac{k_y}{k_\mathrm{c}}\, , \quad A_y = - A \frac{k_x}{k_\mathrm{c}} \, .
\end{equation}
Thus, the amplitude can be expressed by a single quantity $A$.

In the following we will analyze the stability of patterns that can be characterized by the superposition of $N$ modes of the form given in Eqs.~(\ref{eq: general mode cos}) and (\ref{eq: incompressibility condition new amplitude}). 
Each mode $i = 1, \dots N$ is characterized by its amplitude $A_i$ and a polar angle $\varphi_i$ relative to the $x$-axis.
Thus, the wavevector corresponding to that mode can be written as
\begin{equation}
\label{eq: wavevector angle }
\mathbf{k}_i = k_\mathrm{c} \begin{pmatrix} \cos(\varphi_i) \\ \sin(\varphi_i) \end{pmatrix}\, ,
\end{equation}
where $k_\mathrm{c} = 1$.
The patterns are then represented via
\begin{equation}
\label{eq: superposition modes}
\begin{aligned}
v_x(\mathbf{x}) &= \sum_{i=1}^N A_i \sin(\varphi_i) e^{i[\cos(\varphi_i)x + \sin(\varphi_i)y]} + \mathrm{c.c.}\, ,\\
v_y(\mathbf{x}) &= - \sum_{i=1}^N A_i \cos(\varphi_i) e^{i[\cos(\varphi_i)x + \sin(\varphi_i)y]} + \mathrm{c.c.}\, .
\end{aligned}
\end{equation}
In the following, we will restrict the considerations to three modes, i.e., $N = 3$, representing different directions of $\mathbf{k}$.
As a consequence, we only search for the minimum of the functional $\mathcal{F}$ in this predefined space.
However, as we will see, three modes are sufficient to represent a wide variety of patterns.

\begin{figure}
\includegraphics[width=0.999\linewidth]{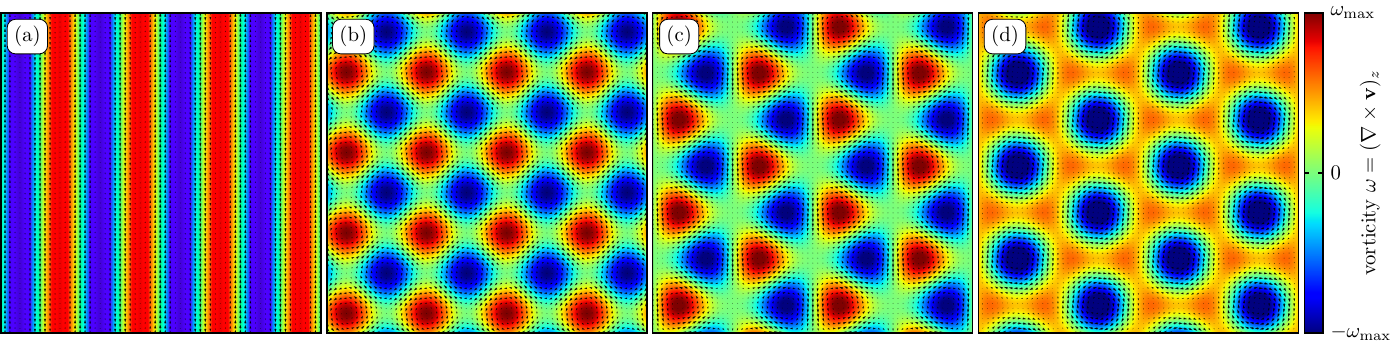}
\caption{\label{fig: patterns}Possible stationary patterns. The vorticity field $\omega(\mathbf{x})$ is visualized as color map. The superimposed arrows show the velocity field $\mathbf{v}(\mathbf{x})$. (a) Stripe pattern composed of a single mode in $x$-direction. (b) Square vortex lattice according to Eq.~(\ref{eq: square lattice modes}), which consists of clockwise- and counterclockwise-rotating vortices in an antiferromagnetic arrangement. (c) Triangular vortex pattern consisting of three modes. (d) Hexagonal vortex lattice with Kagome-like symmetry. Both vorticity and velocity are rescaled for visualization purposes. The box size for all patterns is $8\pi\times 8\pi$.}
\end{figure}

Due to rotational invariance, we can set one of the angles to an arbitrary value without loss of generality.
Here, we use $\varphi_1 = 0$.
The remaining five parameters $A_1$, $A_2$, $A_3$, $\varphi_2$ and $\varphi_3$ now completely describe the pattern under investigation.
For example, we can depict a stripe-like pattern via 
\begin{equation}
\label{eq: stripes modes}
\begin{aligned}
v_x(\mathbf{x}) &= 0 \, ,\\
v_y(\mathbf{x}) &= - 2 A_1 \cos(x)\ .
\end{aligned}
\end{equation}
This is shown in Fig.~\ref{fig: patterns}(a), where we plot the vorticity field $\omega = (\nabla \times \mathbf{v})_z$ as color map and superimpose the velocity field $(v_x,v_y)$ using arrows.
Fig.~\ref{fig: patterns}(b) visualizes a ``square lattice'', obtained by setting $\varphi_2 = \pi/2$ and $A_3 = 0$, i.e.,
\begin{equation}
\label{eq: square lattice modes}
\begin{aligned}
v_x(\mathbf{x}) &= 2 A_2 \cos(y)\, ,\\
v_y(\mathbf{x}) &= - 2 A_1 \cos(x)\ .
\end{aligned}
\end{equation}
The so-described square pattern consists of clockwise- and counterclockwise-rotating vortices that are arranged in an antiferromagnetic manner. 
Thinking of cogwheels, such an antiferromagnetic arrangement makes intuitive sense due to the mutual friction of rotating vortices. 
Further, a ``triangular lattice'' is obtained, e.g., by setting $\varphi_2 = 2\pi/3$ and $\varphi_2 = 4\pi/3$.
Again, as is visualized in Fig.~\ref{fig: patterns}(c), neighboring vortices always display opposing senses of rotation.
Note that there are different possibilities to represent such square and triangular patterns.
The common property of these representations is that the angle between the two modes in the square pattern is a multiple of $\pi/2$, whereas the angles between the  modes in the triangular pattern is a multiple of $\pi/3$.
Further, note that by phase-shifting one of the modes in the triangular lattice by $\pi/2$, we obtain a pattern that exhibits the symmetries of a Kagome lattice, a structure that has been explored in the context of spatially constrained microswimmer suspensions in an earlier work~\cite{reinken2020organizing}.

We will now exploit the fact that the dynamics without nonlinear advection ($\lambda = 0$) is completely determined by the relaxation to the minimum of the functional $\mathcal{F}$, compare Eq.~(\ref{eq: dynamic equation}).
This feature can be used to derive a system of evolution equations for the amplitudes $A_i$ and angles $\varphi_i$.
To this end, we insert the representation of three superimposed modes given by Eq.~(\ref{eq: superposition modes}) into Eq.~(\ref{eq: functional}).
We will refrain from writing down the full form resulting for $f$ as it is quite involved.
However, let us state a few points to motivate the final result.
First, we already ensured incompressibility, i.e., $\nabla\cdot\mathbf{v} = 0$, see Eq.~(\ref{eq: incompressibility condition new amplitude}). 
Thus, the first term in $f$ [given by Eq.~(\ref{eq: functional})] vanishes.
Second, the term $|(1 + \nabla^2) \mathbf{v}|^2$ vanishes for modes characterized by $k_\mathrm{c}$, as is also demonstrated by the fact that $\sigma(a=0,|\mathbf{k}|=k_\mathrm{c}) = 0$ [see Eq.~(\ref{eq: growth rate isotropic solution})].
This also holds for a linear superposition of multiple modes, provided all have wavenumber $k_\mathrm{c}$.
The remaining terms appearing in $f$, i.e., $-a|\mathbf{v}|^2/2 + |\mathbf{v}|^4/4$, are of even order. 
Thus, inserting Eq.~(\ref{eq: superposition modes}) into Eq.~(\ref{eq: functional}) yields only even modes of the form $e^{2il\mathbf{k}\cdot\mathbf{x}}$, where $l$ is an integer number, i.e.,
\begin{equation}
\label{eq: modes in f}
f = f_0(A_1,A_2,A_3,\varphi_2,\varphi_3) + \mathcal{O}\big(e^{2i\mathbf{k}\cdot\mathbf{x}}\big)\, .
\end{equation}
In the following, we neglect higher order modes with $|l| \geq 1$.
This simplification is justified by the wavenumber dependence of the growth rate [see Eq.~(\ref{eq: growth rate isotropic solution})], which exhibits strong damping for larger wavenumbers due to the quartic term, i.e., $-|\mathbf{k}|^4$.
As a result of this simplification, $f$ loses its dependence on the spatial variable $\mathbf{x}$.
Instead of the functional $\mathcal{F}[\mathbf{v},\nabla^2\mathbf{v}]$, we now have in lowest order the function $f_0(A_1,A_2,A_3,\varphi_2,\varphi_3)$ determining the dynamics, see Appendix~\ref{app: amplitude equations} for the explicit form.
This approach has thus reduced the phase space to five dimensions.
Recall that the five variables $A_1$, $A_2$, $A_3$, $\varphi_2$ and $\varphi_2$ were introduced in the ansatz for the emerging patterns as a superposition of modes characterized by $k_\mathrm{c}$ [see Eq.~(\ref{eq: superposition modes})].
The precise form of $f_0$ thus depends on the type of patterns considered.

Building on this reduction of phase space, we are now able to derive evolution equations for the amplitudes and angles.
To this end, we first express the dynamic equation for $\mathbf{v}(\mathbf{x},t)$ [Eq.~(\ref{eq: dynamic equation})] in terms of $f_0$ instead of $\mathcal{F}$.
In this context, we again set $\lambda = 0$.
As the explicit function $f_0 = f_0(A_1,A_2,A_3,\varphi_2,\varphi_3)$ is known [Eq.~(\ref{eq: f_0})], this equation can be utilized to obtain evolution equations for $A_i$ and $\varphi_i$, see Appendix~\ref{app: amplitude equations}.
The final equations are:
\begin{equation}
\label{eq: amplitude and angle equations three modes}
\begin{aligned}
\frac{\partial A_1}{\partial t} &= a A_1 - A_1 \Big(3 A_1^2 + 2\big[1 + 2\cos^2(\varphi_2)\big] A_2^2 + 2\big[1 + 2\cos^2(\varphi_3)\big] A_3^2\Big)\, ,\\
\frac{\partial A_2}{\partial t} &= a A_2 - A_2\Big(3 A_2^2 + 2\big[1 + 2\cos^2(\varphi_2)\big] A_1^2  + 2\big[1 + 2\cos^2(\varphi_3 - \varphi_2)\big] A_3^2\Big)\, ,\\
\frac{\partial A_3}{\partial t} &= a A_3 - A_3\Big(3 A_3^2 + 2\big[1 + 2\cos^2(\varphi_3)\big] A_1^2 + 2\big[1 + 2\cos^2(\varphi_3 - \varphi_2)\big] A_2^2\Big)\, ,\\
\frac{\partial \varphi_2}{\partial t} &= 2 A_1^2 \sin(2\varphi_2) + 2 A_3^2 \sin(2\varphi_2 - 2\varphi_3)\, ,\\
\frac{\partial \varphi_3}{\partial t} &= 2 A_1^2 \sin(2\varphi_3) - 2 A_2^2 \sin(2\varphi_2 - 2\varphi_3)\, .
\end{aligned}
\end{equation}
Possible stationary patterns emerging in our original system described by the evolution equation for the effective microswimmer velocity field are now given as stationary solutions of Eqs.~(\ref{eq: amplitude and angle equations three modes}), provided $\lambda = 0$.
Recall that Eq.~(\ref{eq: dynamic equation}) is governed by gradient dynamics, i.e., a relaxation to a minimum of the potential $\mathcal{F}$, which, in the framework of the amplitude equations, corresponds to a linearly stable solution.
Thus, the next step is to investigate the stability of the stationary solutions of Eqs.~(\ref{eq: amplitude and angle equations three modes}), which represent the isotropic state and the three possible patterns already introduced:
The stripe pattern consisting of one mode, the square vortex lattice consisting of two and the triangular lattice consisting of three modes.

To this end, we determine the linear stability of each of these patterns with respect to small perturbations of the amplitudes and angles.
As Eqs.~(\ref{eq: amplitude and angle equations three modes}) constitute a set of ordinary differential equations, we restrict ourselves to purely time-dependant perturbations that can be expressed in the reduced five-dimensional phase space consisting of $A_1$, $A_2$, $A_3$, $\varphi_1$ and $\varphi_2$. (More complex inhomogeneous perturbations will be later considered in Section~\ref{sec: extended linear stability analysis}.)
The detailed calculations are shown in Appendix~\ref{app: amplitude equations}.
First, we find that the stripe pattern is unstable with respect to the growth of an additional mode perpendicular to the first.
This already hints at the next result, which is that the square lattice in fact constitutes a stable solution of Eq.~(\ref{eq: amplitude and angle equations three modes}).
In contrast, the hexagonal pattern is unstable against perturbations that change both amplitudes and angles.

Thus, in our predefined set of states represented by the superposition of modes characterized by the critical wavenumber $k_\mathrm{c}$, only the square lattice pattern is actually linearly stable for $a > 0$.
Therefore, it represents the only minimum of $\mathcal{F}$. 
This result of course relies on the validity of the representation of the velocity field $\mathbf{v}(\mathbf{x},t)$ in terms of up to three modes according to Eq.~(\ref{eq: superposition modes}). 
It is not guaranteed that a higher number
of modes does not also yield a linearly stable state and thus represents a minimum of $\mathcal{F}$ as well. However, the fact that already the three-mode, triangular state is linearly unstable indicates otherwise. 
A final answer would require an analysis going beyond the scope of this work.
Meanwhile, numerically solving the full dynamic equation~(\ref{eq: dynamic equation}), we not only observe a square vortex lattice for $\lambda = 0$~\cite{reinken2022optimal}, but also for small values of $\lambda$ below the threshold to turbulence.
In fact, the numerical calculations show that $\lambda$ has no impact on the observed patterns (including the lattice constant) in this stationary regime whatsoever, see Appendix~\ref{app: weakly nonlinear analysis}.

To analytically explore the relevance of the square vortex lattice as given in Eq.~(\ref{eq: square lattice modes}) for the full nonlinear dynamics governed by Eq.~(\ref{eq: dynamic equation}), we perform a weakly nonlinear analysis, the details of which are given in Appendix~\ref{app: weakly nonlinear analysis}.
To this end, we employ the standard procedure utilizing multiple scales~\cite{cross1993pattern,newell1993order} and define the small expansion parameter $\varepsilon$ characterizing the distance to the linear instability of the uniform state using the growth rate of perturbations, i.e, $\varepsilon^2 = \sigma(|\mathbf{k}| = k_\mathrm{c}) = a$, see Eq.~(\ref{eq: growth rate isotropic solution}).
The basis of the analysis is an expansion of the velocity field with respect to $\varepsilon$, i.e., close to the instability.
The full expansion is given as
\begin{equation}
\label{eq: expansion velocity WNLA}
\begin{aligned}
v_x(\mathbf{x},t,\mathbf{X},T) = \sum_{l=1}^{\infty}\sum_{m=0}^{l}\sum_{n=0}^{l} \varepsilon^l A_{x,l,m,n}(\mathbf{X},T) e^{i m x + i n y} + \mathrm{c.c.}\, ,\\
v_y(\mathbf{x},t,\mathbf{X},T) = \sum_{l=1}^{\infty}\sum_{m=0}^{l}\sum_{n=0}^{l} \varepsilon^l A_{y,l,m,n}(\mathbf{X},T) e^{i m x + i n y} + \mathrm{c.c.}\, ,\\
\end{aligned}
\end{equation}
where $\mathrm{c.c.}$ denotes the complex conjugate.
Here, the indices $m$ and $n$ are used to introduce higher harmonics that may become increasingly relevant when moving further from the instability.
For the sake of a simplified notation, we use $A_1 = A_{y,1,1,0}$ and $A_2 = A_{x,1,0,1}$ for the dominant modes of the square vortex lattice in the following.
This mirrors the notation used above and in Appendix~\ref{app: amplitude equations}.
We have further introduced the slow time, $T$, and long spatial scale, $\mathbf{X}$, of amplitude modulations.
These are given as $T = \varepsilon^2 t$ and $\mathbf{X} = \varepsilon \mathbf{x}$ (motivated by the fact that the growth rate scales in lowest order quadratic in $\mathbf{k}$).
Inserting the expansion~(\ref{eq: expansion velocity WNLA}) into Eq.~(\ref{eq: dynamic equation}) and matching terms with the same order of $\varepsilon$ (see Appendix~\ref{app: weakly nonlinear analysis} for details), we obtain evolution equations for the two dominating amplitudes $A_1$ and $A_2$ in $\mathcal{O}(\varepsilon^3)$,
\begin{equation}
\label{eq: amplitude equations WNLA}
\begin{aligned}
\partial_T A_1 = a A_1 - 3 |A_1|^2 A_1 - 2 |A_2|^2 A_1 + 4 \partial_X^2 A_1\, , \\
\partial_T A_2 = a A_2 - 3 |A_2|^2 A_2 - 2 |A_1|^2 A_2 + 4 \partial_Y^2 A_2\, , 
\end{aligned}
\end{equation}
where we already scaled back to the fast scales.
Compared to the amplitude equations that we derived above to investigate the stationary patterns selected by the functional $\mathcal{F}$, Eqs.~(\ref{eq: amplitude equations WNLA}) additionally contain gradient terms.
This is because we here consider also modulations of these amplitudes in space.

Eqs.~(\ref{eq: amplitude equations WNLA}) are real-valued Ginzburg--Landau equations~\cite{aranson2002world} and thus describe simple relaxation towards the solution given by $|A_1| = |A_2| = \sqrt{a/5}$.
Remarkably, $\lambda$ has no impact on the amplitude equations.
This result is in line with the full numerical solution of Eq.~(\ref{eq: dynamic equation}), where we also observe no impact of $\lambda$ below the threshold to turbulence.
Representing the square vortex lattice solely by two perpendicular modes [see Eq.~(\ref{eq: square lattice modes})] is of course an approximation.
In this context, the weakly nonlinear analysis enables us to calculate corrections to this approximation in the form of higher harmonics, which result from the nonlinearities in Eq.~(\ref{eq: dynamic equation}).
However, as we discuss in detail in Appendix~\ref{app: weakly nonlinear analysis}, the amplitudes of these higher harmonics are orders of magnitude smaller than the amplitudes of the dominating modes.
We conclude that we can safely continue our analysis using the two-mode approximation of the square vortex lattice.
In the following, we will denote this lowest-order square pattern as 
$\mathbf{v}_\mathrm{s}(\mathbf{x})$, i.e., 
\begin{equation}
\label{eq: lowest-order square lattice}
\mathbf{v}_\mathrm{s}(\mathbf{x}) = \sqrt{\frac{a}{5}} \begin{pmatrix}
e^{iy} \\ e^{ix} 
\end{pmatrix} + \mathrm{c.c.}\, ,
\end{equation}
and employ it as an approximative solution of Eq.~(\ref{eq: dynamic equation})

\section{Extended stability analysis}
\label{sec: extended linear stability analysis}

Having established the presence of the square vortex lattice for small values of $\lambda$, we now turn to the stability of this pattern with respect to an increase in $\lambda$.
As the discussion above has shown, amplitude equations derived via standard procedures are only of limited use in this context.
Instead, we will again employ a linear stability analysis of the full nonlinear equation (\ref{eq: dynamic equation}).
Taking into account the effect of the advection term is, however, not straightforward.
This can be shown by naively considering space- and time-dependant perturbations denoted as $\delta \mathbf{v}$ and $\delta q$ given by a single mode [analogously to Eq.~(\ref{eq: add perturbations isotropic solution})], i.e.,
\begin{equation}
\label{eq: perturbation to square lattice}
\begin{aligned}
\mathbf{v} = \mathbf{v}_\mathrm{s} + \delta \mathbf{v}\, , \qquad  \delta \mathbf{v} = \delta \hat{\mathbf{v}} e^{\sigma t} e^{i \mathbf{k} \cdot \mathbf{x}}\, , \qquad 
q = \delta q\, , \qquad  \delta q = \delta \hat{q} e^{\sigma t} e^{i \mathbf{k} \cdot \mathbf{x}}\, ,
\end{aligned}
\end{equation}
where $\delta \hat{\mathbf{v}}$ and $\delta \hat{q}$ are the perturbation amplitudes, $\sigma$ is the growth rate, $\mathbf{k}$ the wavevector of the perturbative mode and $\mathbf{v}_\mathbf{s}$ is given by Eq.~(\ref{eq: lowest-order square lattice}).

Linearizing the evolution equation for $\mathbf{v}(\mathbf{x},t)$ [Eq.~(\ref{eq: dynamic equation})] around $\mathbf{v}_\mathrm{s}(\mathbf{x})$ yields
\begin{equation}
\label{eq: LSA square lattice linearized}
\partial_t \delta \mathbf{v} = -\nabla \delta q + \mathbf{J}\vert_{\mathbf{v}_\mathrm{s}} \cdot \delta \mathbf{v} \, ,
\end{equation}
where the Jacobian $\mathbf{J}$ excludes the term in Eq.~(\ref{eq: dynamic equation}) involving $q$.
As we have specified the spatial and temporal functional form of the perturbation [see Eq.~(\ref{eq: perturbation to square lattice})], we are able to evaluate the spatial and temporal derivatives.
Thus, after canceling out the factor $e^{i\mathbf{k}\cdot\mathbf{x}}$, Eq.~(\ref{eq: LSA square lattice linearized}) can be written in terms of the perturbation amplitudes $\delta \hat{\mathbf{v}}$ and $\delta \hat{q}$ as
\begin{equation}
\label{eq: LSA square lattice one mode step 1}
\sigma \delta \hat{\mathbf{v}} = -i \mathbf{k}\delta \hat{q} + \mathbf{B}(\mathbf{x}) \cdot \delta \hat{\mathbf{v}}\, .
\end{equation}
We introduce the matrix $\mathbf{B}(\mathbf{x})$ as
\begin{equation}
\label{eq: LSA square lattice one mode B lowest order}
\mathbf{B}(\mathbf{x}) = \mathbf{B}_0 + \mathrm{h.o.t.}\, ,
\end{equation}
where $\mathbf{B}_0$ contains all terms independent of the spatial variable $\mathbf{x}$ and $\mathrm{h.o.t.}$ denotes higher-order terms, which still involve $\mathbf{x}$, see Appendix~\ref{app: details linear stability} for the explicit components of $\mathbf{B}(\mathbf{x})$.
To continue, we neglect these higher-order terms and obtain an equation independent of $\mathbf{x}$,
\begin{equation}
\label{eq: LSA square lattice one mode step 2}
\sigma \delta \hat{\mathbf{v}} = -i \mathbf{k}\delta \hat{q} + \mathbf{B}_0 \cdot \delta \hat{\mathbf{v}}\, .
\end{equation}

As a final step, we deal with the perturbation of the quantity $q$.
To this end, we utilize the scalar product of $\mathbf{k}$ with Eq.~(\ref{eq: LSA square lattice one mode step 2}).
Using the incompressibility condition, in Fourier space given as $\mathbf{k} \cdot \delta \hat{\mathbf{v}} = 0$, we obtain
\begin{equation}
\label{eq: LSA square lattice one mode delta q}
\delta \hat{q} = - i \frac{\mathbf{k}}{|\mathbf{k}^2|} \cdot \mathbf{B}_0 \cdot \delta \hat{\mathbf{v}}\, .
\end{equation}
Inserting Eq.~(\ref{eq: LSA square lattice one mode delta q}) into Eq.~(\ref{eq: LSA square lattice one mode step 2}), we finally arrive at the familiar eigenvalue problem
\begin{equation}
\label{eq: LSA square lattice one mode final}
\sigma \delta \hat{\mathbf{v}} = \mathbf{M} \cdot \delta \hat{\mathbf{v}}\, ,
\end{equation}
where the matrix $\textbf{M}$ is calculated as
\begin{equation}
\label{eq: LSA square lattice one mode matrix M}
\mathbf{M} =  \big( \mathbf{I} - \mathbf{k}\mathbf{k}/|\mathbf{k}^2|\big) \cdot \mathbf{B}_0 \, .
\end{equation}
Determining the eigenvalues of $\mathbf{M}$, we obtain 
\begin{equation}
\label{eq: growth rate square lattice one mode}
\sigma(\mathbf{k}) = a - 1 + 2 |\mathbf{k}|^2 - |\mathbf{k}|^4 - 8 A^2\, .
\end{equation}
Compared to the growth rate of perturbations to the isotropic state [Eq.~(\ref{eq: growth rate isotropic solution})], the growth rate in Eq.~(\ref{eq: growth rate square lattice one mode}) is again real-valued but contains the additional term $-8A^2$.
This term is a result of the nonlinear cubic term in Eq.~(\ref{eq: dynamic equation}) and describes a strong damping of perturbations dependent on the amplitude $A$ of the square vortex lattice pattern that is already present.
Importantly, Eq.~(\ref{eq: growth rate square lattice one mode}) does not contain any contribution from the nonlinear advection term, i.e., is independent of $\lambda$.
This is because all contributions to $\mathbf{B}$ that stem from nonlinear advection represent higher-order terms that are neglected in the analysis above, see Eqs.~(\ref{eq: LSA square lattice one mode Bxx}) to (\ref{eq: LSA square lattice one mode Byy}) in Appendix~\ref{app: details linear stability}.
These terms result from the coupling between the perturbative mode and the modes constituting the vortex lattice.
In order to consistently take these contributions into account we have to extend the linear stability analysis.
In the following, we will outline our approach to achieve this.

\begin{figure}
\includegraphics[width=0.999\linewidth]{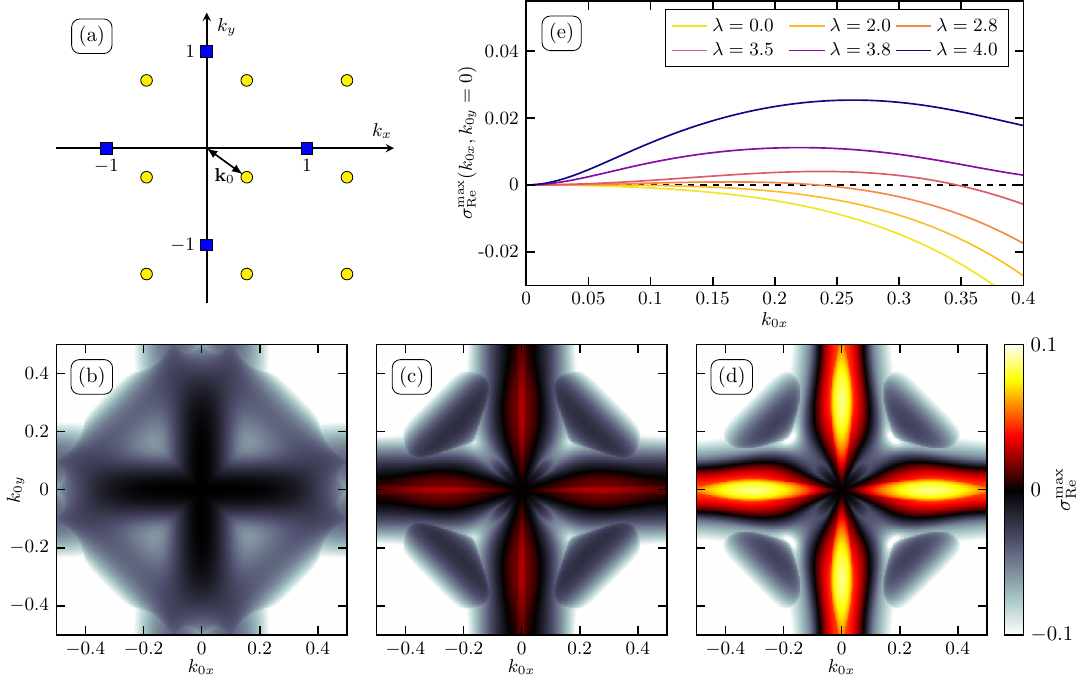}
\caption{\label{fig: largest eigenvalue}(a) Visualization of a perturbation consisting of multiple modes in Fourier space.
The blue squares show the base vectors of the reciprocal lattice of the square lattice pattern, whereas the yellow circles show the wavevectors of perturbative modes that are located on a grid of size $3\times 3$.
The center of this grid is shifted by $\mathbf{k}_0$ from the origin.
(b)/(c)/(d) Maximum growth rate of perturbations to the square lattice pattern at $a = 0.1$ and (b) $\lambda = 2$, (c) $\lambda = 4$ and (d) $\lambda = 4.8$.
The growth rate is given by the largest real part of the eigenvalues $\sigma_\mathrm{Re}^\mathrm{max}$ of the matrix $\mathbf{M}^\ast$ in Eq.~(\ref{eq: eigenvalue equation square lattice multiple modes}) and is plotted here as a function of $\mathbf{k}_0$.
When increasing $\lambda$, the square lattice becomes unstable to perturbations with offset $\mathbf{k}_0$ on the axes $\mathbf{k}_0 = (k_{0x},0)$ and $\mathbf{k}_0 = (0,k_{0y})$, as indicated by the emergence of red and yellow colors.
To illustrate, the maximum growth rate is shown for $k_{0y} = 0$ in (e) at different values of $\lambda$.}
\end{figure}

At this point, it is important to understand how modes in the system are coupled as a result of the mathematical form of the dynamic equation and the already present vortex lattice state.
As this state is not uniform, a perturbative mode can excite or dampen other modes via the nonlinear terms even in the linearized case.
This is clearly demonstrated by the components of the matrix $\mathbf{B}$, which is introduced above and explicitely given in Appendix~\ref{app: details linear stability}.
As was previously pointed out~\cite{james2020phdthesis}, a single-wavevector perturbation is thus not sufficient to capture this effect.
We here use an ansatz that can not be represented by a single wavevector but rather consists of a more complex subset of Fourier space.
To find an appropriate subset, it makes sense to visualize the wavevectors corresponding to the square vortex pattern in Fourier space, see Fig.~\ref{fig: largest eigenvalue}(a).
The primary modes that constitute this lattice, i.e., $2 \cos(x) = e^{ix} + e^{-ix}$ and $2 \cos(y) = e^{iy} + e^{-iy}$ are shown as blue squares in the figure.
By examining the combination of the nonuniform base pattern and the kinds of nonlinear terms that are present, we determine the subset of wavevectors that captures the coupling between modes relevant for the linear stability analysis.
In our case, due to the form of the square vortex lattice and the presence of the cubic and the advection term, we have to take into account wavevectors that are arranged on a grid in Fourier space with a spacing of $k_\mathrm{c} = 1$, oriented the same way as the vortex lattice modes.
To illustrate the suitability of this grid of modes, let us consider one of the emerging terms as an example.
After inserting the sum of the square vortex lattice and the perturbation given via Eq.~(\ref{eq: perturbation to square lattice}), we linearize the nonlinear advection term $\mathbf{v}\cdot \nabla \mathbf{v}$, which yields, among other contributions, a term $\propto v_{sx} \partial_x \delta v_x$.
Here, $v_{sx}$ and $\delta v_x$ are the $x$-components of the square vortex lattice and the perturbation, respectively.
Evaluating this term gives a contribution $ \propto i k_x  e^{i k_x x + i (k_y + 1)}$, i.e., a mode that is shifted by $\Delta \mathbf{k} = (0,k_\mathrm{c}) = (0,1)$ compared to the original mode of the perturbation $\mathbf{k} = (k_x,k_y)$.
When combining the non-uniform base state with a perturbative mode, all nonlinear terms lead to similar contributions characterized by such a shift.
As a result, modes that are additionally excited or damped lie on a grid in Fourier space.
Taking a closer look at the terms in the matrix $\mathbf{B}$ in Appendix~\ref{app: details linear stability} further demonstrates the relevance of this particular subset.

The ansatz for the perturbation is visualized in Fig.~\ref{fig: largest eigenvalue}(a) via the yellow dots and can be written as
\begin{equation}
\label{eq: ansatz perturbation multiple modes}
\delta \mathbf{v} = e^{\sigma t} \sum\limits_{m = -M}^{M} \, \sum\limits_{n = -M}^{M} \, \delta \hat{\mathbf{v}}_{mn} e^{i\mathbf{k}_{mn}\cdot \mathbf{x}}\, .
\end{equation}
The wavevectors $\mathbf{k}_{mn}$ consituting the grid are expressed relative to the wavevector of the center mode $\mathbf{k}_0 = (k_{0x},k_{0y})$ at $m=0$ and $n=0$ via
\begin{equation}
\label{eq: ansatz wavevectors multiple modes}
\mathbf{k}_{mn} = \begin{pmatrix}
k_{0x} + k_\mathrm{c} m \\
k_{0y} + k_\mathrm{c} n
\end{pmatrix}\, .
\end{equation}
Note that, for $\mathbf{k}_0 = \mathbf{0}$, the ansatz (\ref{eq: ansatz perturbation multiple modes}) just describes vectors of the reciprocal lattice of the original square lattice in real space.
In order to describe a perturbation from this pattern, $\mathbf{k}_0$ must be non-zero.
The total number of perturbative modes described in Eq.~(\ref{eq: ansatz perturbation multiple modes}) is given by $N_\mathrm{p} = (2M+1)^2$.
For the remainder of this paper, we choose $M = 1$, thus considering a perturbation that can be represented by $9$ wavevectors, see Fig.~\ref{fig: largest eigenvalue}(a).

Now, inserting the perturbation [Eq.~(\ref{eq: ansatz perturbation multiple modes})] into the dynamic equation~(\ref{eq: dynamic equation}), the nonlinear coupling between modes can be taken into account even after linearization.
To continue with the stability analysis, we write down a system of $2 N_\mathrm{p}$ linearized equations by collecting terms of $\mathcal{O}(e^{i \mathbf{k}_{mn} \cdot \mathbf{x}})$ from the equations for $v_x$ and $v_y$, respectively.
Note that, again, the incompressibility condition, $\nabla \cdot \mathbf{v} = 0$, has to be satisfied by employing a projection operator as in Eq.~(\ref{eq: LSA square lattice one mode matrix M}).
Finally, we obtain the eigenvalue problem
\begin{equation}
\label{eq: eigenvalue equation square lattice multiple modes}
\sigma \delta \hat{\mathbf{V}} = \mathbf{M}^\ast \cdot \delta \hat{\mathbf{V}}\, ,
\end{equation}
where the vector $\delta \hat{\mathbf{V}}$ now contains the perturbation amplitudes of all modes in both $x$- and $y$-direction, i.e.,  
\begin{equation}
\label{eq: perturbation amplitude vector}
\hat{\delta\mathbf{V}} = \begin{pmatrix}
\delta \hat{v}_{-M,-M,x}\\
\delta \hat{v}_{-M,-M,y}\\
\vdots\\
\delta \hat{v}_{M,M,x}\\
\delta \hat{v}_{M,M,y}
\end{pmatrix}\, .
\end{equation}
By calculating the eigenvalues of the matrix $\mathbf{M}^\ast$, we now determine the linear stability of the square vortex lattice with respect to the simultaneous growth of multiple modes.
The calculation outlined above is quite involved as the matrix $\mathbf{M}^\ast$ has dimensions $2N_\mathrm{p}\times 2N_\mathrm{p}$.
This necessitates the use of symbolic computation software such as the Python library Sympy~\cite{sympy}.
Writing down a closed form for the eigenvalues is also not feasible due to the high dimensionality.
Rather, we evaluate the matrix $\mathbf{M}^\ast$ for a specific set of parameters $a$, $\lambda$ and $\mathbf{k}_0$ and calculate the eigenvalues numerically.

The stability of the square lattice base pattern is then determined by the largest real part of any of these eigenvalues, which we denote as $\sigma^\mathrm{max}$ in the following.
Fig.~\ref{fig: largest eigenvalue}(b) to (d) shows the largest real part $\sigma_\mathrm{Re}^\mathrm{max}$ as a function of the grid offset $\mathbf{k}_0$ at $a = 0.5$ and different values of $\lambda$.
Note that the four-fold rotational symmetry is expected due to the symmetry of the square lattice pattern.
Taking a closer look at Fig.~\ref{fig: largest eigenvalue}(b), we observe that $\sigma_\mathrm{Re}^\mathrm{max}$ stays negative for all $\mathbf{k}_0$ at smaller advection strength $\lambda$.
However, increasing the advection strength, $\sigma_\mathrm{Re}^\mathrm{max}$ becomes positive for certain $\mathbf{k}_0$, see Fig.~\ref{fig: largest eigenvalue}(c) and (d).
In particular, the perturbations that start to grow first are characterized by $\mathbf{k}_0$ located on the axes $\mathbf{k}_0 = (k_{0x},0)$ or $\mathbf{k}_0 = (0,k_{0y})$.
Picking one of these axes, we plot $\sigma_\mathrm{Re}^\mathrm{max}$ as a function of $k_{0x}$ for $k_{0y} = 0$ in Fig.~\ref{fig: largest eigenvalue}(e) to be able to visualize the dependency on $\lambda$ in more detail.
Here, we see that the growth rate becomes positive at some critical value $\lambda_\mathrm{c}$ ($2.0 < \lambda_\mathrm{c} < 2.8$) and develops a clear maximum at $k_{0x} \approx 0.25$ for larger values of $\lambda$.
This picture stays qualitatively the same even for larger values of $a$.
We always observe the development of clear maxima on the positive and negative axes [$\mathbf{k}_0 = (k_{0x},0)$ and $\mathbf{k}_0 = (0,k_{0y})$], respectively.

At this point, let us briefly discuss how the number of perturbative modes might affect the stability analysis.
A subset of Fourier space consisting of a grid of $3\times 3$ modes represents the lowest-order choice that allows for an investigation of instabilities caused by the nonlinear advection term, as we show in this work.
In principle, the accuracy of the results might be increased by allowing for a larger grid of perturbative modes.
However, these additional modes are characterized by larger and larger wavevectors.
As a result, these will be increasingly damped by the biharmonic term in Eq.~(\ref{eq: dynamic equation}) and thus only have a diminishing influence.
Further, when aiming for a consistent stability analysis including such higher-order perturbations, the representation of the square lattice as given in Eq.~(\ref{eq: lowest-order square lattice}) should also be amended by the inclusion of higher-order modes.
Note that an infinite grid of perturbative modes implies a translation symmetry in Fourier space for wavevector shifts of $\Delta k_{0x} = 1$ or $\Delta k_{0y} = 1$.
Thus, it is sufficient to explore the region given by $-k_\mathrm{c}/2 < k_{0x} < k_\mathrm{c}/2$ and $-k_\mathrm{c}/2 < k_{0y} < k_\mathrm{c}/2$, as we have done in Fig.~\ref{fig: largest eigenvalue}.
As a perturbation consisting of $3\times 3$ modes already leads to quite accurate results concerning the stability as compared to the numerical solution of the full Eq.~(\ref{eq: dynamic equation}), we restrict our study to this case.
In order to corroborate our results, we have additionally performed a limited analysis using $5 \times 5$ perturbative modes, where we find qualitatively the same results.
The quantitative differences are small, in particular close to the onset of the instability.

So far we have concentrated on the real part of the eigenvalues, as this gives access to the growth rate.
Appendix~\ref{app: imaginary} further discusses the imaginary part of the eigenvalue with largest real part $\sigma^\mathrm{max}$.
We find that $\sigma_\mathrm{Im}^\mathrm{max}$ vanishes for perturbations with $\mathbf{k}_0$ located on the axes $\mathbf{k}_0 = (k_{0x},0)$ or $\mathbf{k}_0 = (0,k_{0y})$, but is nonzero in large regions of $\mathbf{k}_0$-space.
In particular, $\sigma^\mathrm{max}_\mathrm{Im} \neq 0$ for most unstable perturbations close to the fastest-growing perturbation, indicating the emergence of oscillatory behavior or even more complex dynamic states.

\section{Onset of turbulence}

\begin{figure}
\includegraphics[width=0.45\linewidth]{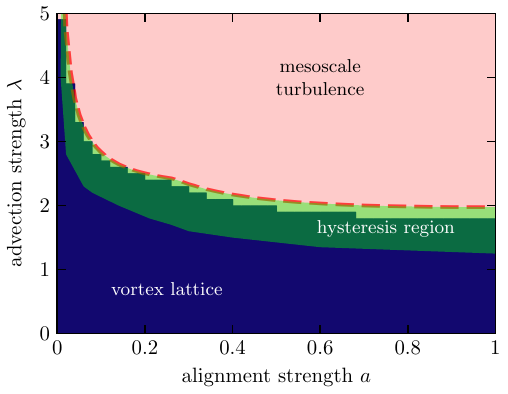}
\caption{\label{fig: state diagram}State diagram in the space spanned by the coefficient $a$ and advection strength $\lambda$.
The background color is used to show numerical results: dark blue for a stationary vortex lattice and light red for mesoscale turbulence.
These results are obtained by numerically solving the evolution equation for $\mathbf{v}(\mathbf{x},t)$ directly, starting from a square vortex lattice with small perturbations and observing if the lattice becomes unstable and turbulence develops.
The dashed red line shows the results from the linear stability analysis, i.e., the point where the real part of any of the eigenvalues of the matrix $\mathbf{M}^\ast$ in Eq.~(\ref{eq: eigenvalue equation square lattice multiple modes}) becomes positive.
The area shaded in green shows the region where both stable vortex lattice and turbulent state can be encountered depending on the initial state of the system. We thus observe hysteretic behavior.
This region is determined via additional numerical simulations discussed in Section~\ref{sec: hysteresis}.}
\end{figure}

The instability of the square vortex lattice with respect to the simultaneous growth of multiple modes signifies the start of the development of mesoscale turbulence in our model.
Having established the procedure of our extended stability analysis, we now turn to the determination of the critical point where perturbative modes start to grow.
To this end, we calculate all eigenvalues of the matrix $\textbf{M}^\ast$ as outlined above for a set of parameters $a$, $\lambda$ and $\mathbf{k}_0$.
If all of the eigenvalues for all $\mathbf{k}_0$ have a negative real part, the square vortex lattice is stable.
However, if any of the eigenvalues for any $\mathbf{k}_0$ has a positive real part, it is unstable and mesoscale turbulence is expected to develop via the mutual excitement of multiple modes.
On this basis we can determine the critical advection strength $\lambda_\mathrm{c}$ as a function of the coefficient $a$.

\begin{figure}
\includegraphics[width=0.999\linewidth]{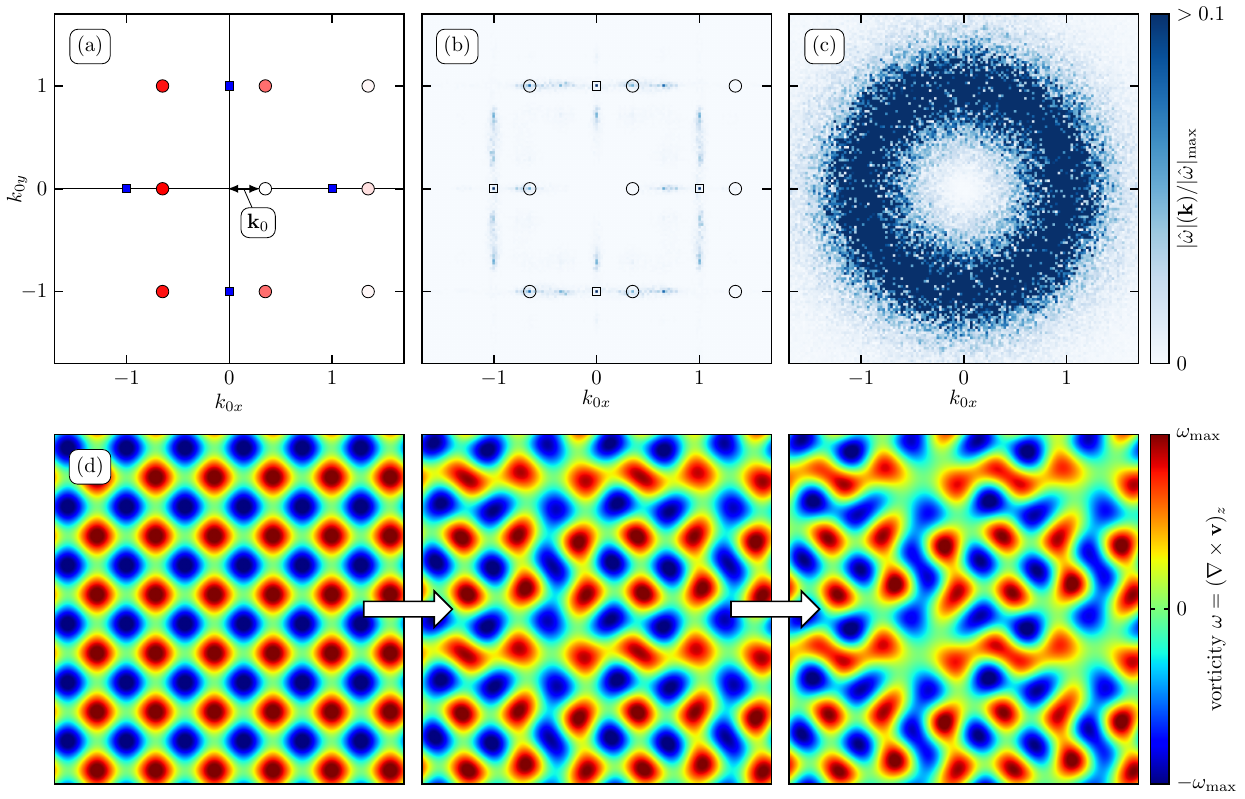}
\caption{\label{fig: Fourier spectrum eigenvectors}Comparison of the eigenvector of the fastest-growing perturbation with the Fourier spectrum observed in the numerical simulations. 
The relative importance of eigenvector components is shown in (a) for the eigenvector corresponding to the eigenvalue with largest real part at $a = 0.5$ and $\lambda = 3.16$ and $\mathbf{k}_0 = (0.35,0)$.
The saturation of the red color is determined by the absolute value of the eigenvalue components that correspond to the growth of the modes located at the respective positions on the grid and, thus, shows which of the modes are expected to grow the fastest when the instability sets in.
The blue squares show the reciprocal lattice corresponding to the square lattice pattern.
For comparison, (b) and (c) show the Fourier spectrum observed when numerically solving Eq.~(\ref{eq: dynamic equation}) at $a = 0.5$ and $\lambda = 3.16$.
Here, (b) is obtained at an intermediate time where the initial square lattice pattern is still present (indicated by the square boxes) but the growth of other modes is already visible in the spectrum (indicated by the circles).
Comparing with (a), we find good agreement with the analysis based on the eigenvector of the fastest-growing perturbation.
The Fourier spectrum of fully developed mesoscale turbulence is shown in (c).
(d) ``Time evolution'' (snapshots) of the vorticity field starting from the square vortex lattice (left) and showing increasingly perturbed patterns to the right.
The form of the perturbation is given by the eigenvector shown in (a).
The size of the snapshots is $12\pi \times 12\pi$.}
\end{figure}

Figure~\ref{fig: state diagram} depicts the state diagram in $a$-$\lambda$-space.
The dashed red line shows the result obtained via the extended linear stability analysis.
The background color indicates results obtained via direct numerical solution of Eq.~(\ref{eq: dynamic equation}) for specific sets of parameters $\lambda$ and $a$.
Here, we start the numerical calculations from an already ordered state and observe if initially small perturbations start to grow.
In this way, we determine whether the system develops a turbulent state (indicated by the light blue color) or the lattice structure remains stable (indicated by the dark blue color).
Remarkably, the extended linear stability analysis predicts the onset of the turbulent state quite accurately.
Figure~\ref{fig: state diagram} demonstrates that the vortex lattice is stable for much higher values of $\lambda$ when $a$ is small.
This observation is explained by the difference in the stationary amplitude $A_\mathrm{s}$ of the modes of the vortex lattice, which is given by $A_\mathrm{s} = \sqrt{a/5}$, see Section~\ref{sec: stationary patterns}.
Increasing $a$ leads to a larger amplitude.
The nonlinear nature of the advection term implies that its impact grows with increasing amplitude of the patterns, thus explaining the earlier onset of turbulence for larger $a$.

Up until this point, we have only looked at the eigenvalues of the matrix $\textbf{M}^\ast$.
However, the corresponding eigenvectors also contain valuable information.
In particular, we can determine which of the modes that constitute the perturbation will start growing the fastest once the instability sets in.
To this end, we take a closer look at the eigenvector $\delta \mathbf{V}_\mathrm{c}$ corresponding to the eigenvalue with largest real part.
This eigenvector contains the growth amplitudes of all $N_\mathrm{p}$ modes of the perturbation in terms of components $\delta \hat{v}_{mn,x}$ and $\delta \hat{v}_{mn,y}$.
Calculating the absolute value of the components of $\delta \mathbf{V}_\mathrm{c}$ corresponding to a particular mode, we determine the growth of that mode once $\lambda$ is increased above $\lambda_\mathrm{c}$.

As an illustration, the absolute eigenvector components are plotted in Fig.~\ref{fig: Fourier spectrum eigenvectors}(a) as the saturation of the red coloring at the respective positions in the grid of perturbative modes in Fourier space.
The coefficients are set to $a = 0.5$ and $\lambda = 3.16$ and we have chosen $\mathbf{k}_0 = (0.35,0)$, which corresponds to the maximum of the growth rate for this parameter set.
As the red color indicates, we find that six modes grow particularly quickly once the instability sets in.
This result is easily verified by comparison with Fourier spectra obtained via direct numerical solution of Eq.~(\ref{eq: dynamic equation}).
To this end, we initialize the numerical calculations with a square lattice pattern according to Eq.~(\ref{eq: lowest-order square lattice}) and add very small random perturbations not favoring any particular wavevector.
Taking a closer look at the evolution of the Fourier spectrum during the onset of the instability, we observe which of the wavevectors start growing first.
The Fourier spectrum at such an intermediate time step is shown in Fig.~\ref{fig: Fourier spectrum eigenvectors}(b).
There, we plot the absolute value of the vorticity transformed to Fourier space, $|\hat{\omega}|(\mathbf{k})$ for the same values of the coefficients used in Fig.~\ref{fig: Fourier spectrum eigenvectors}(a). 
The reciprocal lattice corresponding to the square vortex lattice is clearly visible (indicated by the square boxes).
The plot also shows which other modes are starting to grow at this point.
For comparison, we have overlaid the spectrum with the grid of the perturbative modes shown in Fig.~\ref{fig: Fourier spectrum eigenvectors}(a) as a frame indicated by the circles.
The spectrum agrees very well with the analysis based on the eigenvector $\delta \mathbf{V}_\mathrm{c}$, corroborating the results obtained via the extended linear stability analysis.
Note that the four-fold rotational symmetry in Fourier space is clearly visible in the spectrum, which was already discussed in the context of Fig.~\ref{fig: largest eigenvalue}(b) to (d).
Thus, in addition to the six modes determined in Fig.~\ref{fig: Fourier spectrum eigenvectors}(a), there are 18 more modes that are equally relevant at the onset of the instability, yielding a total of $24$ modes.
To visualize these elaborate perturbations, we transform back to real space in Fig.~\ref{fig: Fourier spectrum eigenvectors}(d).
Here, we show the vorticity field of the square vortex lattice on the left-hand side.
Moving to the right, we add the perturbative modes according to the eigenvector shown in Fig.~\ref{fig: Fourier spectrum eigenvectors}(a) (plus the grid of perturbative modes rotated by $\SI{90}{\degree}$, $\SI{180}{\degree}$ and $\SI{270}{\degree}$).
Increasing the amplitude of the perturbation, we can thus visualize how the square vortex lattice becomes distorted right after the onset of the instability leading to the turbulent state, thereby mimicking the time evolution occurring in the full system.
This is further visualized in the Supplementary Movie, where we show the evolution of both Fourier spectrum and vorticity field obtained via numerical solution of Eq.~(\ref{eq: dynamic equation}) at the onset of the instability.
Fig.~\ref{fig: Fourier spectrum eigenvectors}(c) further includes the Fourier spectrum observed at a time step in the fully developed mesoscale-turbulent state.
Here, the dominant modes form a broad ring with radius $|\mathbf{k}| \approx 1$ and, as expected, the system exhibits complete rotational symmetry.

\section{Hysteretic behavior}
\label{sec: hysteresis}

\begin{figure}
\includegraphics[width=0.95\linewidth]{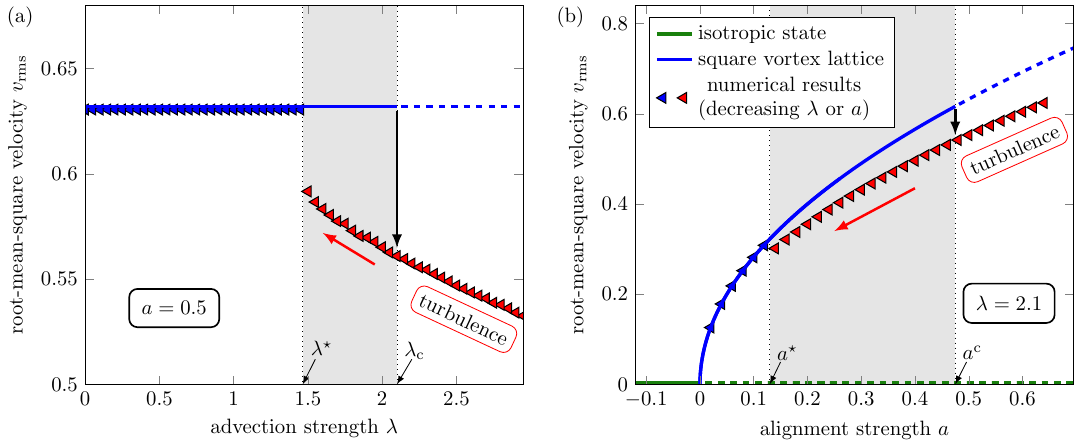}
\caption{\label{fig: hysteresis}Hysteretic behavior.
Solid lines show stable branches, while dashed lines represent unstable branches. 
(a) Root-mean-square velocity $v_\mathrm{rms}$ as a function of $\lambda$  at $a = 0.5$.
The blue line denotes the square vortex lattice branch, which becomes unstable at some critical value of $\lambda$, as discussed in Section~\ref{sec: extended linear stability analysis}.
Here, the system develops turbulence and $v_\mathrm{rms}$ drops significantly.
The red data points are determined via numerical solution of the full Eq.~(\ref{eq: dynamic equation}) by starting in the turbulent regime and slowly decreasing $\lambda$.
(b) Root-mean-square velocity $v_\mathrm{rms}$ as a function of $a$ at $\lambda = 2.1$ for different spatiotemporal states.
The isotropic state is denoted by the green line, the square vortex lattice by the blue line.
Increasing $a$ leads to growing amplitude and thus also growing $v_\mathrm{rms}$, until the lattice becomes unstable.
Again, the turbulence branch of the hysteresis loop is obtained via numerical calculations by starting in the turbulent regime and slowly decreasing $a$. 
In both plots, the shaded area denotes the region where both stable vortex lattice and turbulent state can be encountered.}
\end{figure}

Up until now, the discussion has focused on the stability of the square vortex lattice.
The instability we uncovered indicates the critical advection strength $\lambda_\mathrm{c}(a)$, above which turbulence starts to develop from an initial square vortex lattice.
In particular, this gives the expected behavior when $\lambda$ or $a$ are slowly increased from initially small values.
Here, when numerically solving the full nonlinear equation~(\ref{eq: dynamic equation}), we observe a discontinuous jump from the stationary lattice to dynamic turbulent behavior at $\lambda_\mathrm{c}$ or $a_\mathrm{c}$.
The discontinuous nature of the transition already hints at the coexistence of different solutions and the possibility of hysteresis, i.e., different behavior when $\lambda$ or $a$ are instead slowly decreased from initially large values. 

Investigating this issue within an analytical framework is not feasible due to the complexity of the turbulent state.
Thus, we continue with a numerical study.
Here, we solve the full nonlinear equation~(\ref{eq: dynamic equation}) for a particular value of $\lambda$ and wait until we obtain a statistically stationary state.
This state then acts as the starting point for a subsequent calculation with a slightly smaller value of $\lambda$.
We continue in this way until the flow field becomes stationary and turbulence is superseded by a stationary vortex lattice.
To quantify the results, we here employ the root-mean-square velocity $v_\mathrm{rms} = \sqrt{\langle |\mathbf{v}|^2 \rangle}$, where the mean value $\langle \dots \rangle$ is obtained in the statistically stationary state as both a spatial and temporal average.
Note that the average velocity $\langle \mathbf{v}\rangle$ vanishes for both square vortex lattice and turbulence and, thus, is not a suitable measure to distinguish between different spatiotemporal states.
In the stationary vortex lattice, $v_\mathrm{rms}$ can be calculated analytically, which yields $v_\mathrm{rms} = 2 A_\mathrm{s}$.
In the turbulent state, $v_\mathrm{rms}$ is generally smaller because the nonlinear advection term leads to energy transfer between scales and subsequent dissipation, which comes on top of the amplitude saturation via the cubic term in Eq.~(\ref{eq: dynamic equation})~\cite{reinken2022optimal,james2018turbulence}.

Fig.~\ref{fig: hysteresis}(a) shows the root-mean-square velocity as a function of advection strength at $a = 0.5$.
The root-mean square velocity of the vortex lattice is simply given by a constant value in this diagram because the stationary amplitude solely depends on $a$.
As we determined via the extended stability analysis in Section~\ref{sec: extended linear stability analysis}, at $\lambda = \lambda_\mathrm{c}$ the vortex lattice becomes unstable to the simultaneous growth of multiple modes.
In Fig.~\ref{fig: hysteresis}(a), this is indicated by the transition to the dashed line.
Here, a turbulent state emerges, which manifests itself as discontinuous drop in the value of $v_\mathrm{rms}$.
As discussed above, the turbulent branch has to be determined numerically via solution of Eq.~(\ref{eq: dynamic equation}).
In Fig.~\ref{fig: hysteresis}, the results are shown by the red data points.
Slowly decreasing $\lambda$, we move along the turbulent branch from the right-hand side to the left-hand side until $v_\mathrm{rms}$ jumps to the vortex lattice branch at $\lambda^\star$, at which point the system has settled into a stationary state.
As Fig.~\ref{fig: hysteresis}(a) shows, the corresponding value of $\lambda$ is significantly smaller than the point where the square vortex lattice becomes unstable.
The region in between is characterized by the existence of both stable regular vortex lattice and turbulent state.
As a result, we observe hysteretic behavior.

To further illustrate the hysteresis loop, we plot $v_\mathrm{rms}$ for constant $\lambda$ as a function of $a$ in Fig.~\ref{fig: hysteresis}(b).
This diagram also gives a complete overview of the various points of interest and motivates the representative sketch in Fig.~\ref{fig: bifurcation and snapshots}.
First, the isotropic solution ($\mathbf{v} = \mathbf{0}$), which is indicated by the green line, becomes unstable at $a=0$.
The subsequent unstable branch is shown as the dashed green line.
From this instability emerges the square vortex lattice (blue line), which becomes unstable at $a_\mathrm{c}$, resulting in a turbulent state, compare Fig~\ref{fig: state diagram}.
Again, the red data points show the results obtained via direct numerical solution of Eq.~(\ref{eq: dynamic equation}), starting in the turbulent regime and slowly decreasing $a$.
Performing additional numerical calculations for different values of $a$, we determine the hysteresis region in $a$-$\lambda$-space, which is shown as the area shaded in green in Fig.~\ref{fig: state diagram}.
We obtained similar results for the lower limit of the hysteresis from numerical calculations starting from an initially isotropic state with small random perturbations~\cite{reinken2022optimal}.
Due to the random initial values, the development of a disordered, turbulent state is favored.
However, when the activity parameters are small enough, the system eventually settled into a regular square vortex lattice.
In this way, the hysteresis region shown in Fig.~\ref{fig: hysteresis} is reproduced independently for different initial conditions.
Further, a hysteretic transition was also found in a recent investigation of the dynamics in the Toner--Tu--Swift--Hohenberg model under circular confinement~\cite{shiratani2023route}.
In that case, the transition is observed between regularly oscillating vortices and chaotic dynamics.

Since the transition from turbulence to stationary lattice is not  associated with a crossing of an eigenvalue of the spectra obtained in the respective linear stability analysis above, we hypothesize that there might be a connection to a statistical phase transition analogous to classical turbulence systems like pipe flow~\cite{avila2023transition}. Alternatively,  a long-living spatiotemporally chaotic transient~\cite{tel2008chaotic} is possible, but can be largely ruled out due to the fact that the hysteresis region is reproduced in simulations with random initial conditions as mentioned above. 
This conclusion is further corroborated by long-running numerical simulations, which are presented in Appendix~\ref{app: numerical methods} and show no sign of transient behavior. 
Comparing Fig.~\ref{fig: hysteresis}(a) and (b), we observe that the jump in $v_\mathrm{rms}$ at the lower end of the hysteresis region decreases from $\Delta v_\mathrm{rms} \approx 0.04$ at $a^\star = 0.5$ and $\lambda^\star \approx 1.48$ to $\Delta v_\mathrm{rms} \approx 0.02$ at $a^\star \approx 0.13$ and $\lambda^\star = 2.1$. Additional simulations used to determine the hysteresis region shown in Fig.~\ref{fig: state diagram} confirm the observation that the jump diminishes when moving to the left of the state diagram (decreasing $a$ and increasing $\lambda$) and might even vanish altogether. More detailed numerical studies will be necessary to give a complete characterization of the transition at the end of the hysteretic branch at low activity.

\section{Stability of regular vortex lattices with arbitrary wavenumber}
\label{sec: arbitrary wavenumber}

Up until this point, we only considered vortex lattices comprised of two modes with wavenumber $k = k_\mathrm{c}$, which is the only possibility exactly at the onset of the instability at $a=0$.
However, increasing $a$, vortex patterns consisting of modes with wavenumbers $k \neq k_\mathrm{c}$ might be stable as well.
The extended stability analysis proposed and discussed in Section~\ref{sec: extended linear stability analysis} may also be utilized to investigate these more general structures.
Motivated by the results on stationary pattern selection in Section~\ref{sec: stationary patterns}, we will restrict ourselves to square vortex lattices consisting of two perpendicular modes with wavevectors $|\mathbf{k}_1| = |\mathbf{k}_2| = k_\mathrm{s}$.
Similarly to the procedure employed in Section~\ref{sec: stationary patterns}, evolution equations for the amplitudes of the modes, $A_1$ and $A_2$, can be derived.
Again neglecting spatial modulations and higher harmonics yields
\begin{equation}
\label{eq: amplitude equations square vortex lattice arbitrary wavenumber}
\begin{aligned}
\frac{\partial A_1}{\partial t} &= \alpha(k_\mathrm{s}) A_1 - A_1 \big(3 A_1^2 + 2 A_2^2 \big)\, ,\\
\frac{\partial A_2}{\partial t} &= \alpha(k_\mathrm{s}) A_2 - A_2 \big(3 A_2^2 + 2 A_1^2 \big)\, .
\end{aligned}
\end{equation}
Compared to Eqs.~(\ref{eq: amplitude and angle equations three modes}) and (\ref{eq: amplitude equations WNLA}), the linear coefficient now depends on the wavenumber of the lattice and is given by $\alpha(k_\mathrm{s}) = a - 1 + 2 k_\mathrm{s}^2 - k_\mathrm{s}^4$.
The stationary solution is determined via $A_1 = A_2 = A_\mathrm{s} = \sqrt{\alpha(k)/5}$.
In terms of the velocity field, the square vortex lattice in this two-mode approximation is thus given for arbitrary lattice wavenumber $k_\mathrm{s}$ as
\begin{equation}
\mathbf{v}_\mathrm{s}(k_\mathrm{s}) = \sqrt{\frac{\alpha(k_\mathrm{s})}{5}} \begin{pmatrix}
e^{ik_\mathrm{s}y} \\ e^{ik_\mathrm{s}x} 
\end{pmatrix} + \mathrm{c.c.}\, .
\end{equation}
Setting $k_\mathrm{s} = k_\mathrm{c} = 1$ reproduces the vortex lattice investigated before, see Eq.~(\ref{eq: lowest-order square lattice}).

\begin{figure}
\includegraphics[width=0.999\linewidth]{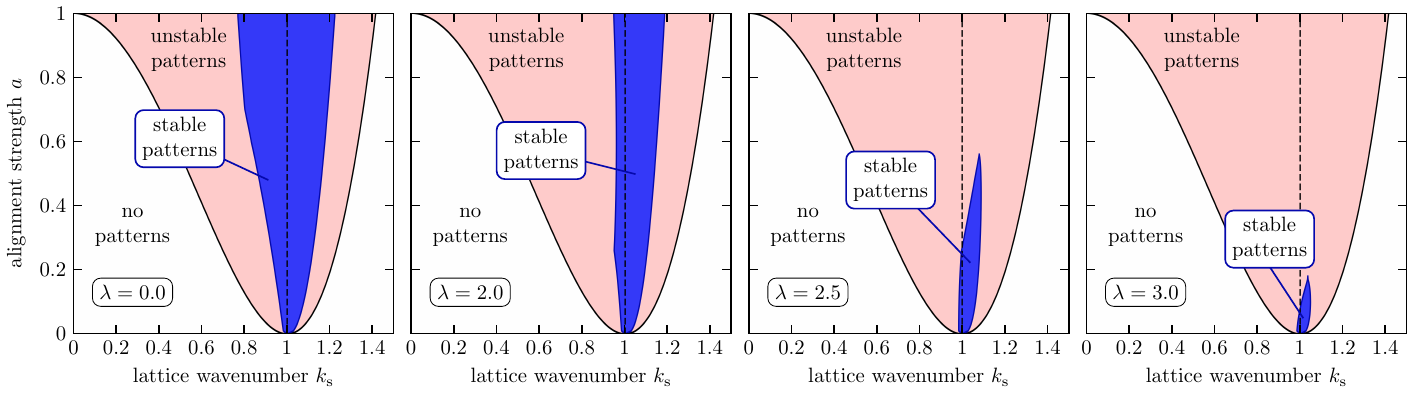}
\caption{\label{fig: Busse}Stability of regular patterns with arbitrary wavenumber $k_\mathrm{s}$.
The plots show the regions (blue) where the square vortex lattice is stable as a function of $k_\mathrm{s}$ and $a$ at different values of $\lambda$.
The light red area represents the band of unstable modes that might grow from the isoptropic state according to Eq.~(\ref{eq: band of unstable modes}).
The dashed line shows the critical wavenumber $k_\mathrm{c} = 1$.
In compliance with the emergence of turbulence, we observe that the stability balloons rapidly shrink when increasing $\lambda$.}
\end{figure}

To determine the stability of the vortex lattice of arbitrary wavenumber, we again add a perturbation consisting of multiple modes.
As before, we take into account $3\times 3$ modes that are located on a grid in Fourier space, albeit with a grid spacing of $k_\mathrm{s}$ instead of $k_\mathrm{c}$ [compare Eq.~(\ref{eq: ansatz wavevectors multiple modes})].
Performing the same procedure as introduced in Section~\ref{sec: extended linear stability analysis}, we then calculate the maximum growth rate now depending on not only $a$ and $\lambda$ but also $k_\mathrm{s}$.
Observing if the growth rate becomes positive, we determine whether the square vortex lattice is unstable with respect to the simultaneous growth of multiple modes.
The results can be illustrated using the concept of stability balloons~\cite{busse1978non,cross2009pattern}.
A few examples are plotted in Fig.~\ref{fig: Busse}, where we show the regions of stable patterns in the space spanned by $k_\mathrm{s}$ and $a$ for various values of $\lambda$.
The light red region in these plots is limited by Eq.~(\ref{eq: band of unstable modes}).
Here, perturbations of the isotropic, uniform state will grow, but no regular patterns are stable.
As expected, we find that the region of stable patterns is largest for vanishing advection strength, $\lambda = 0$.
Increasing $\lambda$, the stability balloon shrinks, which complies with the emergence of the turbulent state.
Note that vortex lattices characterized by the critical wavenumber, $k_\mathrm{s} = k_\mathrm{c}$, are located on the dashed line in Fig.~\ref{fig: Busse}.
Remarkably, these lattice are not necessarily the most stable.
We rather find that lattices with slightly increased $k_\mathrm{s}$ remain stable for larger values of $a$.

The stability of regular structures in other systems, e.g., stripe-like structures in Rayleigh--B\'{e}nard convection~\cite{cross2009pattern}, is limited by different types of instabilities depending on the particular boundary of the stability balloon.
With that in mind, Appendix~\ref{app: boundaries balloons} explores the specific instabilities at the boundaries for smaller and larger square lattice constant $k_\mathrm{s}$ at $\lambda = 0$, i.e., to the left and the right of the stability balloon shown in Fig.~\ref{fig: Busse}(a).
Here, we indeed observe very different instabilities at the left and right boundary.
However, both are dominated by the growth of modes with $k \approx k_\mathrm{c}$, thus leading to structures closer to the ones that emerge when starting from an isotropic state.
These results indicate that the instabilities of square lattices with arbitrary wavenumber are similarly complex and diverse as the situation in other systems, such as Rayleigh--B\'{e}nard convection~\cite{cross2009pattern}.

\section{Conclusions}
\label{sec: conclusions}

In summary, this work investigates the onset of the turbulent state in an established minimal model for active fluids, where the evolution of the effective velocity is determined as a competition between gradient dynamics governed by a functional $\mathcal{F}$ and nonlinear advection.
We first establish that a square lattice of vortices represents a minimum of $\mathcal{F}$ and thus a stable solution in the absence of nonlinear advection.
In analogy to high-Reynolds-number turbulence, nonlinear advection destabilizes the stationary pattern and induces a turbulent state.
Here, we show, by means of an extended stability analysis that the transition follows from a linear instability.
The approach is based on linearization around an approximation of the analytical vortex lattice solution.
Analogous results are found if the polar alignment strength is increased above a critical value. Since both parameters increase with activity, 
we can state that strong enough activity is a necessary ingredient for active turbulence in the model for polar active fluids studied here and in related experiments.
Remarkably, the instability is only revealed when taking into account the mutual excitement and simultaneous growth of multiple perturbative modes.
Moreover, our results show that hysteretic behavior is a characteristic feature tightly connected to the onset and disappearance of mesoscale turbulence.
The phenomenology observed in the hysteretic transition region between stable vortex lattices
and active turbulence is reminiscent of the picture of a chaotic saddle separating laminar and turbulent
regimes~\cite{schneider2007turbulence} and of a saddle-node bifurcation of coherent structures~\cite{faisst2003traveling} to the first
occurence of turbulence from the laminar state in classical pipe flow~\cite{eckhardt2007turbulence,avila2023transition}.

Our work complements earlier studies that include stability analyses of nonuniform states, e.g., a similar approach was recently used to determine the stability of traveling states in active phase field crystals~\cite{menzel2014active}.
Nonuniform states are also sometimes investigated in an equilibrium situation, as in microphase-separated diblock copolymer melts~\cite{laradji1997stability}, which allows for methods not available here due to the non-equilibrium nature of active fluids.

Regarding the experimental relevance of the present study, it is important to note that square and triangular lattices, see Fig.~\ref{fig: patterns}(b) and (c), can be stabilized in bacterial suspensions by placing appropriate arrangements of small obstacles in the flow field~\cite{nishiguchi2018engineering}.
In these experiments~\cite{nishiguchi2018engineering}, it was also shown that the square vortex lattice is more robust against changes of the arrangement than the triangular lattice.
This observation is in line with our analytical results, i.e., that only the square vortex lattice is in fact an intrinsically stable pattern in the model.
The stability analysis suggests that, in principle, square vortex lattices may be achieved in bacterial suspensions without external stabilization if sufficient control of the activity is possible (e.g., by varying the availability of oxygen~\cite{sokolov2012physical}).
Further, analytical studies of a similar model for mesoscale turbulence have shown that a shear-thickening solvent fluid may induce regular vortex lattices as well~\cite{reinken2024vortex}.
Note that structures similar to the earlier predicted hexagonal vortex lattice~\cite{james2018turbulence,james2021emergence} have recently been observed in experiments~\cite{xu2024self}.

In the context of stabilized vortex lattices, the same model that we investigated in this study was employed in Refs.~\cite{reinken2020organizing,reinken2022ising} to explore the dynamics of microswimmer suspensions within arrangements of small obstacles.
These obstacles are able to stabilize regular vortex patterns even for values of the nonlinear advection strength $\lambda$ much higher than the critical value $\lambda_\mathrm{c}$ determined in the present work.
Interestingly, the transition between a stabilized square vortex lattice and the disordered state facilitated by increasing $\lambda$ showed features of a non-equilibrium phase transition in the universality class of the two-dimenstional Ising model~\cite{reinken2022ising}.
This striking contrast to the linear instability observed in the present work in the absence of obstacles demonstrates that the route to turbulence in active polar fluids is strongly affected by the
geometry of the system and by confinement,
as is the transition to inertial turbulence~\cite{eckert2010troublesome,barkley2015rise}.

In active nematics the onset of the turbulent state was recently linked to a directed percolation transition~\cite{doostmohammadi2017onset}, similar to the route to sustained inertial turbulence in channel, Couette and pipe flows~\cite{sano2016universal,lemoult2016directed,avila2023transition}.
In fact, Ref.~\cite{doostmohammadi2017onset} also investigated active nematic turbulence in a channel geometry that emerged from increasing the channel width and hence decreased the influence
of spatial confinement.
This is in contrast to the bulk system without spatial constraints investigated in the present study.
Clearly, further work is necessary to establish whether the different routes to turbulence in these active systems are a consequence of different types of setup or possibly due to the relevance of different order parameters in these systems.

\begin{acknowledgments}
We thank Igor S. Aronson, Martin James, Andreas M. Menzel and Michael Wilczek for stimulating discussions.
This work was funded by the Deutsche Forschungsgemeinschaft (DFG, German Research Foundation) - Projektnummer 163436311 - SFB 910. H.R. acknowledges partial support by the DFG through the research grant ME 3571/4-1.
\end{acknowledgments}

\appendix

\section{Numerical methods}
\label{app: numerical methods}

We employ a pseudo-spectral method to numerically solve Eq.~(\ref{eq: dynamic equation}) in a two-dimensional system with periodic boundary conditions.
For the time integration we use Euler’s method combined with an operator splitting technique treating linear and nonlinear parts of Eq.~(\ref{eq: dynamic equation}) separately.
For the calculations in the context of the state diagram in Fig.~\ref{fig: state diagram}, the spatial resolution is set to $128 \times 128$ grid points and the system size to $12\pi \times 12\pi$. For every set of parameters $a$ and $\lambda$, we initialize the velocity field with a lattice-like pattern according to Eq.~(\ref{eq: lowest-order square lattice}) and add a small random perturbation to the value of $\mathbf{v}$ at every grid point, which is taken from a uniform distribution over $[-10^{-3},10^{-3}]$.
Then, we evolve the system for $10000$ time units and determine whether the initial square vortex lattice state remains or the velocity has significantly changed and a turbulent state starts to develop.
The state diagram in Fig.~\ref{fig: state diagram} is colored in according to the result of this analysis.
In order to determine the Fourier spectra shown in Fig.~\ref{fig: Fourier spectrum eigenvectors}, we use a higher spatial resolution of $400\times 400$ grid points and set the system size to $80\pi \times 80 \pi$.
Further, the snapshots shown in Fig.~\ref{fig: bifurcation and snapshots} are extracted from a system with $256\times 256 $ grid points and a size of $32\pi \times 32\pi$.

\begin{figure}
\includegraphics[width=0.999\linewidth]{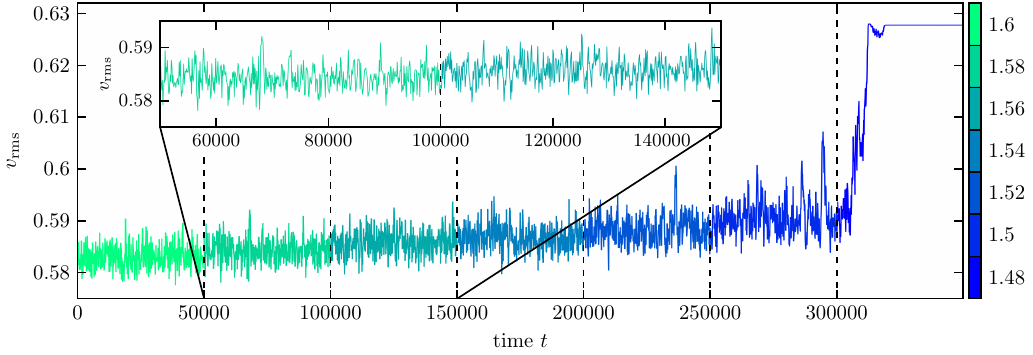}
\caption{\label{fig: vrms hysteresis}Time evolution of the root-mean-square velocity $v_\mathrm{rms}$ in the simulation investigating hysteresis.
In the depicted run, we start from a turbulent state and successively decrease $\lambda$ in steps of $\Delta \lambda = 0.02$.
After every step, we wait for $\Delta t = 50000$ time units (dashed vertical lines) to ensure a statistically stationary state.
Here, we only show the last few steps before the system develops a stationary vortex lattice.
The inset shows a magnified view of two subsequent time intervals, demonstrating fluctuations around a mean value of $v_\mathrm{rms}$, which increases for decreasing $\lambda$.}
\end{figure}

In the context of the calculations to explore hysteretic behavior (see Fig.~\ref{fig: hysteresis}), we set the number of grid points to $256\times256$ and the system size to $48\pi \times 48\pi$. 
To explore the hysteresis region we start from a turbulent state at high activity and then successively decrease one of the activity parameters, $\lambda$ or $a$.
The data points in Fig.~\ref{fig: hysteresis}(a) are determined by decreasing $\lambda$ by $\Delta \lambda = 0.05$ every $\Delta t = 10000$ time units and then measuring $v_\mathrm{rms}$ at the end of the respective time periods.
It is possible that the persistent turbulent states observed in the hysteretic region are long-living spatiotemporally chaotic transients.
In order to investigate this possibility, we further performed simulations with very large running times where we set the time periods after every successive step of decreasing the activity to $\Delta t = 50000$.
Here, we use $a = 0.5$ and decrease $\lambda$ in steps of $\Delta \lambda = 0.02$.
Fig.~\ref{fig: vrms hysteresis} shows the evolution of the root-mean-square velocity for the last few steps before the system develops a stationary state.
Using these very large time periods $\Delta t$ and smaller successive steps $\Delta \lambda$, we find that the system settles into the vortex lattice at same value $\lambda^\ast \approx 1.48$ as observed when using smaller $\Delta t$ and larger $\Delta \lambda$, compare Fig.~\ref{fig: hysteresis}(a).
Thus, there are no signs of a transient phenomenon.
The data points in Fig.~\ref{fig: hysteresis}(b) are determined by decreasing $a$ by $\Delta a = 0.02$ every $\Delta t = 50000$ time units. 
Finally, in order to determine the hysteresis region in the state diagram in Fig.~\ref{fig: state diagram}, we use $\Delta t = 50000$ as well as steps of $\Delta a = 0.02$ or $\Delta \lambda = 0.05$.

Concerning the extended linear stability analysis described in Section~\ref{sec: extended linear stability analysis}, we determine the matrix $\mathbf{M}^\ast$ with the help of the Python library Sympy~\cite{sympy}.
Subsequently, inserting specific values for the coefficients $a$ and $\lambda$, the wavevector of the center of the grid of perturbative modes, $\mathbf{k}_0$, and, if applicable, the wavenumber of the square lattice, $k_\mathrm{s}$, we evaluate the eigenvalues of $\mathbf{M}^\ast$ numerically.
When determining the critical advection strength, $\lambda_\mathrm{c}$, in the context of the state diagram in Fig.~\ref{fig: state diagram}, we regard the square vortex lattice as unstable when the largest eigenvalue is larger than $\sigma^\mathrm{max} > 10^{-3}$.
This is done in order to avoid artifacts that may result from neglecting higher harmonics.
The same approach is applied to calculate the stability balloons in Fig.~\ref{fig: Busse}.

\section{Amplitude equations for different stationary patterns}
\label{app: amplitude equations}

As shown in the main text, in the case of regular patterns that can be represented by three modes [Eq.~(\ref{eq: superposition modes})], we are able to approximate the phase space by only five variables, the amplitudes $A_1$, $A_2$ and $A_3$, as well as the angles $\varphi_2$ and $\varphi_3$.
In the following, we will show how evolution equations can be derived for these quantities by exploiting the fact that we only have to deal with the function $f_0(A_1,A_2,A_3,\varphi_2,\varphi_3)$, which is determined as 
\begin{equation}
\label{eq: f_0}
\begin{aligned}
f_0 = &- a \big(|A_1|^2 + |A_2|^2 + |A_3|^2\big) + \frac{3}{2}\big(|A_1|^4 + |A_2|^4 + |A_3|^4\big) + 2|A_1|^2 |A_2|^2 \big[1 + 2 \cos^2(\varphi_2) \big]\\ 
&+ 2|A_1|^2 |A_3|^2 \big[1 + 2 \cos^2(\varphi_2) \big] + 2 |A_2|^2 |A_3|^2 \big[ 1 + 2 \cos^2(\varphi_2 - \varphi_3) \big]\, ,
\end{aligned}
\end{equation}
instead of the full functional $\mathcal{F}$ [see Eq.~(\ref{eq: functional})].
First, we utilize the dynamic equation governing the velocity field $\mathbf{v}(\mathbf{x},t)$ [Eq.~(\ref{eq: dynamic equation})] and write
\begin{equation}
\label{eq: amplitude and angle equations  step 1}
\frac{\partial \mathbf{v}}{\partial A_i} \frac{\partial A_i}{\partial t}  = - \frac{\partial f_0}{\partial A_i} \frac{\partial A_i}{\partial \mathbf{v}}  \, , \qquad \frac{\partial \mathbf{v}}{\partial \varphi_i} \frac{\partial \varphi_i}{\partial t}  = - \frac{\partial f_0}{\partial \varphi_i} \frac{\partial \varphi_i}{\partial \mathbf{v}}  \, .
\end{equation}
Rearranging yields
\begin{equation}
\label{eq: amplitude and angle equations  step 2}
\bigg|\frac{\partial \mathbf{v}}{\partial A_i}\bigg|^{2} \,\frac{\partial A_i}{\partial t}  = -  \frac{\partial f_0}{\partial A_i} \, , \qquad 
\bigg|\frac{\partial \mathbf{v}}{\partial \varphi_i}\bigg|^{2} \,\frac{\partial \varphi_i}{\partial t} = - \frac{\partial f_0}{\partial \varphi_i}\, ,
\end{equation}
which determines the amplitude equations for amplitudes $A_1$, $A_2$, $A_3$ as well as the evolution equations for the angles $\varphi_2$ and $\varphi_3$.

The prefactors in Eq.~(\ref{eq: amplitude and angle equations  step 2}) are quadratic in the partial derivatives of $\mathbf{v}$ with respect to $A_1$, $A_2$, $A_3$, $\varphi_2$ and $\varphi_3$.
The same argument as for $f$ applies, see Eq.~(\ref{eq: modes in f}), and the prefactors are sums of a homogeneous term and higher order modes $e^{2il\mathbf{k}\cdot\mathbf{x}}$ with integer $l$, i.e.,
\begin{equation}
\label{eq: amplitude equations prefactors}
\begin{aligned}
\bigg|\frac{\partial \mathbf{v}}{\partial A_i}\bigg|^{2} &= 2 + \mathcal{O}\big(e^{2i\mathbf{k}\cdot\mathbf{x}}\big)\, , \\
\bigg|\frac{\partial \mathbf{v}}{\partial \varphi_i}\bigg|^{2} &= 2|A_i|^2 + \mathcal{O}\big(e^{2i\mathbf{k}\cdot\mathbf{x}}\big)\, .
\end{aligned}
\end{equation}
Again, we neglect higher order contributions and insert Eq.~(\ref{eq: amplitude equations prefactors}) into Eq.~(\ref{eq: amplitude and angle equations  step 2}).
As a final step we evaluate the derivatives of $f_0$ with respect to $A_1$, $A_2$, $A_3$, $\varphi_2$ and $\varphi_3$ and arrive at a closed set of ordinary differential equations for the amplitudes and angles, see Eqs.~(\ref{eq: amplitude and angle equations three modes}).

Stationary solutions of Eqs.~(\ref{eq: amplitude and angle equations three modes}) represent the regular patterns introduced in the main text, i.e., the isotropic solution, stripe patterns, as well as square and hexagonal lattice-like structures.
In the following, we determine the linear stability of these solutions.
First, for the sake of a simplified notation, we define the vector $ \mathbf{c} = (A_1, A_2, A_3, \varphi_2, \varphi_3)$, which contains the phase space variables, and write the set of Eqs.~(\ref{eq: amplitude and angle equations three modes}) as
\begin{equation}
\label{eq: amplitude and angle equations vector notation}
\partial_t \mathbf{c} =  \mathbf{G}(\mathbf{c})\, ,
\end{equation}
where $\mathbf{G}= (G_{A_1},G_{A_2},G_{A_3},G_{\varphi_2},G_{\varphi_3})$ contains the right-hand sides.
Stationary solutions are then denoted by $\mathbf{c}_0$.
In order to determine the stability, we add a small perturbation to $\mathbf{c}_0$,
\begin{equation}
\label{eq: A0 plus perturbation}
\mathbf{c} =  \mathbf{c}_0 + \delta\hat{\mathbf{c}} e^{\sigma t}\, ,
\end{equation}
where $\delta\hat{\mathbf{c}} = (\delta \hat{A}_1, \delta \hat{A}_2, \delta \hat{A}_3, \delta \hat{\varphi}_2, \delta \hat{\varphi}_3)$ are the perturbation amplitudes and $\sigma$ denotes the growth rate.
Inserting Eq.~(\ref{eq: A0 plus perturbation}) into Eq.~(\ref{eq: amplitude and angle equations vector notation}) and linearizing yields 
\begin{equation}
\label{eq: amplitude and angle linearized}
\sigma \delta \hat{\mathbf{c}} =  \mathbf{J}\vert_{\mathbf{c}_0} \cdot \delta \hat{\mathbf{c}}\, ,
\end{equation}
where $\mathbf{J}\vert_{\mathbf{c}_0}$ denotes the Jacobian $\mathbf{J}$
evaluated at $\mathbf{c}_0$. 
If $\mathbf{J}\vert_{\mathbf{c}_0}$ possesses eigenvalues with positive real part, the stationary solution $\mathbf{c}_0$ is linearly unstable.
Furthermore, the eigenvectors corresponding to the positive eigenvalues indicate the kind of perturbations expected to grow.

\paragraph*{Isotropic state.}

The trivial solution of Eq.~(\ref{eq: amplitude and angle equations vector notation}), $A_1 = A_2 = A_3 = 0$, represents the isotropic state, i.e., $\mathbf{v} = \mathbf{0}$ everywhere.
Here, the two angles $\varphi_2$ and $\varphi_3$ are kept as free parameters for the linear stability analysis.
We write down the Jacobian $\mathbf{J}\vert_{\mathbf{c}_0}$ for $\mathbf{c}_0 = (0, 0, 0, \varphi_2, \varphi_3)$ and calculate the eigenvalues, which yields
\begin{equation}
\label{eq: mesoscale turbulence: eigenvaluesisotropic state}
\sigma_1 = \sigma_2 = \sigma_3 = a \, ,\qquad 
\sigma_4 = \sigma_5 = 0 \, ,
\end{equation}
with corresponding eigenvectors 
\begin{equation}
\label{eq: mesoscale turbulence: eigenvectors isotropic solution}
\begin{gathered}
\delta \hat{\mathbf{c}}_1 = (1,0,0,0,0) \, , \quad
\delta \hat{\mathbf{c}}_2 = (0,1,0,0,0) \, , \quad
\delta \hat{\mathbf{c}}_3 = (0,0,1,0,0) \, , \quad
\delta \hat{\mathbf{c}}_4 = (0,0,0,1,0) \, , \quad
\delta \hat{\mathbf{c}}_5 = (0,0,0,0,1) \, . \quad
\end{gathered}
\end{equation}
The first three eigenvalues are negative for $a < 0$, which means the isotropic solution is stable and represents a minimum of $\mathcal{F}$.
For $a > 0$, however, all three eigenvalues become positive, the isotropic solution becomes linearly unstable with respect to perturbations characterized by the eigenvectors corresponding to $\sigma_1$, $\sigma_2$ and $\sigma_3$.
These eigenvectors describe growth of one of the amplitudes of the modes $A_1$, $A_2$ and $A_3$, respectively.
The remaining eigenvalues $\sigma_4$, $\sigma_5$ vanish regardless of $a$. 
This means that the orientation of the modes is irrelevant and the amplitudes start to grow for any $\varphi_2$ and $\varphi_3$, provided $a > 0$.
This immediately raises the question which of the possible configurations of the modes represented via Eq.~(\ref{eq: superposition modes}) will be selected in the emergent patterns, i.e., which of the stationary solutions of Eq.~(\ref{eq: amplitude and angle equations vector notation}) is linearly stable.
We will investigate this in the following, starting with a state containing a single mode, i.e., a stripe pattern.

\paragraph*{Stripe pattern.}

\begin{figure}
\centering
\includegraphics[width=0.4\linewidth]{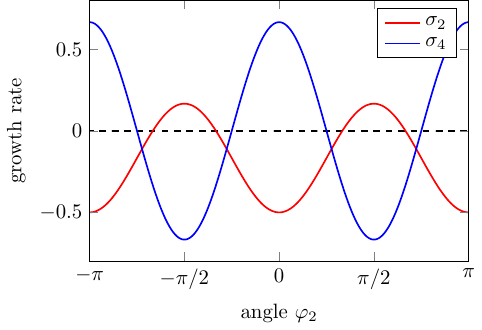}
\caption{\label{fig: growth rate stripe pattern}Stability of the stripe pattern consisting of one mode with angle $\varphi_1 = 0$ and amplitude $A_1 = \sqrt{a/(3b)}$. Shown are the growth rates $\sigma_2$ and $\sigma_4$ connected to mode $i=2$ as a function of the angle $\varphi_2$, where $\sigma_2$ is connected to a growth of the amplitude $A_2$ and $\sigma_4$ is connected to a change of the angle $\varphi_2$. The stripe pattern is unstable with respect to modes that are approximately perpendicular, i.e, $\varphi_2 \approx \pi/2$. The growth rates connected to the third mode ($\sigma_3$, $\sigma_5$) behave analogously with respect to the angle $\varphi_3$.}
\end{figure}

For $a > 0$, one of the stationary solutions of Eqs.~(\ref{eq: amplitude and angle equations three modes}) is characterized by two of the amplitudes being zero.
The other, non-zero mode yields a stripe pattern in the vorticity field, see Fig.~\ref{fig: patterns}(a).
Without loss of generality, we set $A_2 = 0$ and $A_3 = 0$ and find $A_1 = \sqrt{a/3}$.
The angles $\varphi_2$, $\varphi_3$ are undetermined for this solution and will be considered as free parameters.
Writing down the Jacobian $\mathbf{J}\vert_{\mathbf{c}_0}$ for $\mathbf{c}_0 = (\sqrt{a/3}, 0, 0, \varphi_2, \varphi_3)$ and calculating the eigenvalues yields
\begin{equation}
\label{eq: eigenvalues stripe pattern}
\begin{aligned}
\sigma_1 &= -2 a \, , \quad
\sigma_2 = \frac{1}{3} a \big(1 - 4 \cos^2(\varphi_2) \big) \, , \quad
\sigma_2 = \frac{1}{3} a \big(1 - 4 \cos^2(\varphi_3) \big) \, ,\\
\sigma_4 &= \frac{4}{3} a \big(2 \cos^2(\varphi_2) - 1 \big)\, ,\quad
\sigma_5 = \frac{4}{3} a \big(2 \cos^2(\varphi_3) - 1 \big)\, .
\end{aligned}
\end{equation}
Depending on the angles $\varphi_2$ and $\varphi_3$, the eigenvalues $\sigma_2$, $\sigma_3$, $\sigma_4$ and $\sigma_5$ can be positive or negative.
Taking a closer look at the eigenvectors $\delta \hat{\mathbf{c}}_2$, $\delta \hat{\mathbf{c}}_3$, $\delta \hat{\mathbf{c}}_4$ and $\delta \hat{\mathbf{c}}_5$ corresponding to these eigenvalues,
\begin{equation}
\label{eq: eigenvectors stripe pattern}
\begin{aligned}
\delta \hat{\mathbf{c}}_2 = (0,1,0,0,0) \, ,\quad
\delta \hat{\mathbf{c}}_3 = (0,0,1,0,0) \, ,\quad
\delta \hat{\mathbf{c}}_4 = (0,0,0,1,0) \, ,\quad
\delta \hat{\mathbf{c}}_5 = (0,0,0,0,1) \, ,
\end{aligned}
\end{equation}
we find that the perturbations connected to $\sigma_2$ and $\sigma_3$ describe growing amplitudes $A_2$ and $A_3$, respectively, whereas
the perturbations connected to $\sigma_3$ and $\sigma_4$ describe changing the angles $\varphi_2$ and $\varphi_3$.
Looking at a plot of the growth rates $\sigma_2$ and $\sigma_4$ as a function of the angle $\varphi_2$, see Fig.~\ref{fig: growth rate stripe pattern}, the behavior becomes clearer.
For angles close to $\pi/2$, i.e., perpendicular to the first mode, $\sigma_2$ is positive, which means that the amplitude $A_2$ will grow.
In contrast, for angles close to $0$ or $\pi$, $\sigma_2$ is negative, which means that perturbations described by modes that are not approximately perpendicular will not grow.
Simultaneously, $\sigma_4$ is positive for these damped modes, which means that their orientations start to rotate.
The behavior of the third mode is analogous.
To sum up, the stripe pattern is unstable with respect to additional modes growing, with perpendicular modes exhibiting the highest growth rate, already hinting at the next solution we will investigate: the square lattice pattern.

\paragraph*{Square lattice pattern.}

The square lattice pattern is characterized by two modes with the same amplitude oriented perpendicular to each other, see Fig.~\ref{fig: patterns}(b).
Without loss of generality, we set $A_1=A_2$, $A_3=0$, $\varphi_2=\pi/2$ and keep $\varphi_3$ as a free parameter.
Solving Eq.~(\ref{eq: amplitude and angle equations three modes}) for these assumptions yields $A_1 = A_2 = \sqrt{a/5}$ for the stationary value of the amplitudes.
The solution under investigation is then given by $\mathbf{c}_0 = (\sqrt{a/5},\sqrt{a/5},0,\pi/2,\varphi_3)$.
Writing down the Jacobian $\mathbf{J}\vert_{\mathbf{c}_0}$ and calculating the eigenvalues we find
\begin{equation}
\label{eq: eigenvalues square lattice pattern}
\begin{aligned}
\sigma_1 =  0 \, ,\quad  
\sigma_2 = -2a \, ,\quad 
\sigma_3 = -\frac{4}{5}a \, , \quad
\sigma_4 = -\frac{3}{5}a \, ,\quad 
\sigma_5 = -\frac{2}{5}a\, .
\end{aligned}
\end{equation}
All except one eigenvalue are negative, provided $a > 0$.
The remaining eigenvalue $\sigma_1 = 0$ warrants a closer look.
The corresponding eigenvector is $\delta \hat{\mathbf{c}}_1 = (0,0,0,0,1)$, which describes a change of the angle $\varphi_3$ of the third mode.
This mode, however, vanishes for a square lattice pattern and there is also no unstable perturbation that leads to a growth of $A_3$.
Thus, the eigenvalue $\sigma_1 = 0$ can be disregarded when determining the stability.
We conclude that the square lattice pattern is linearly stable and therefore represents a minimum of the functional $\mathcal{F}$.

\paragraph*{Triangular lattice pattern.}

A third solution of Eqs.~(\ref{eq: amplitude and angle equations three modes}) for $a > 0$ is characterized by all amplitudes being equal and non-zero, $A_1=A_2=A_3$, and the modes oriented in a triangular configuration, see Fig.~\ref{fig: patterns}(c).
Without loss of generality, we set $\varphi_2 = 2\pi/3$ and $\varphi_3=4\pi/3$ and find for the stationary value of the amplitudes $A_1 = A_2 = A_3 = \sqrt{a/9}$.
The solution under investigation is then given by $\mathbf{c}_0 = (\sqrt{a/9},\sqrt{a/9},\sqrt{a/9},2\pi/3,4\pi/3)$.
Again, writing down the Jacobian $\mathbf{J}\vert_{\mathbf{c}_0}$ and calculating the eigenvalues we find
\begin{equation}
\label{eq:  eigenvalues triangular lattice pattern}
\begin{aligned}
\sigma_1 &= -2a \, ,\quad
\sigma_2 = -\frac{1}{3} a (1 + \sqrt{5}) \, , \qquad
\sigma_3 = -\frac{1}{3} a (1 - \sqrt{5}) \, ,\\
\sigma_4 &= -\frac{1}{9} a (1 + \sqrt{13}) \, ,\quad \ \,
\sigma_5 = -\frac{1}{9} a (1 - \sqrt{13})\, .
\end{aligned}
\end{equation}
The eigenvalues $\sigma_1$, $\sigma_2$ and $\sigma_4$ are negative, whereas $\sigma_3$ and $\sigma_5$ are positive, making the triangular lattice state linearly unstable.
We take a look at the corresponding eigenvectors $\delta \hat{\mathbf{c}}_3$ and $\delta \hat{\mathbf{c}}_5$ to find out how the growing perturbations can be characterized.
These are
\begin{equation}
\label{eq: eigenvectors trinagular pattern}
\begin{aligned}
\delta \hat{\mathbf{c}}_3 &= \bigg(0,-\sqrt{a}\frac{\sqrt{3}+\sqrt{39}}{18},\sqrt{a}\frac{\sqrt{3}+\sqrt{39}}{18},1,1\bigg) \, ,\\ 
\delta \hat{\mathbf{c}}_5 &= \bigg(\sqrt{a}\frac{\sqrt{3}+\sqrt{15}}{9},-\sqrt{a}\frac{\sqrt{3}+\sqrt{15}}{9},-\sqrt{a}\frac{\sqrt{3}+\sqrt{15}}{9},-1,1\bigg) \, .
\end{aligned}
\end{equation}
Both of these eigenvectors correspond to simultaneous growth of amplitudes and change of angles.
In particular, $\delta \hat{\mathbf{c}}_3$ describes the second or third mode reorienting itself to reduce the angle between it and the first mode while simultaneously growing in amplitude.
Meanwhile, the other mode reorients itself, while decreasing in amplitude. 
This might represent a road to the square lattice state, where the mode with decreasing amplitude vanishes completely and the growing mode rotates until orientated perpendicular to the first mode.
The eigenvector $\delta \hat{\mathbf{c}}_5$ describes a different kind of perturbation: 
Either the amplitude of the first mode grows, while both the second and third mode decrease in amplitude and reorient their directions,
or the amplitude of the first mode decreases, while the amplitudes of the second and third mode grow.
In any case, the hexagonal lattice-like pattern is apparently not a minimum of $\mathcal{F}$.

\section{Weakly nonlinear analysis for the square vortex lattice}
\label{app: weakly nonlinear analysis}

The aim of the weakly nonlinear analysis is to determine equations that govern the spatiotemporal evolution of the amplitudes of the two dominant modes in the square vortex lattice.
To this end, we employ the standard procedure~\cite{cross1993pattern,newell1993order} and perform a multiple scale analysis.
First, using the small expansion parameter $\varepsilon$ characterizing the distance to the linear instability [$\varepsilon^2 = \sigma(|\mathbf{k}|=k_\mathrm{c}) = a$], we write down an expansion of the velocity field $\mathbf{v}$ as given in Eq.~(\ref{eq: expansion velocity WNLA}).
As introduced in the main text, we use $A_1 = A_{y,1,1,0}$ and $A_2 = A_{x,1,0,1}$ for the dominant modes of the square vortex lattice, mirroring the notation in Section~\ref{sec: stationary patterns} and Appendix~\ref{app: amplitude equations}.
Further, the pressure-like quantity $q$ is expanded similarly, i.e.,
\begin{equation}
q(\mathbf{x},t,\mathbf{X},T) = \sum_{l=1}^{\infty}\sum_{m=0}^{l}\sum_{n=0}^{l} \varepsilon^l q_{l,m,n}(\mathbf{X},T) e^{i m x + i n y} + \mathrm{c.c.}\, .
\end{equation}
Based on the definition of slow time and long spatial scale, $T = \varepsilon^2 t$ and $\mathbf{X} = \varepsilon \mathbf{x}$, we replace the derivatives in the full dynamic equation~(\ref{eq: dynamic equation}) with 
\begin{equation}
\label{eq: app: WNLA derivatives replacement}
\begin{aligned}
\partial_t \quad &\rightarrow \quad\partial_t  + \varepsilon^2 \partial_T,\\
\partial_{x} \quad &\rightarrow \quad \partial_{x} + \varepsilon \partial_{X}, \qquad\quad \partial_{y} \rightarrow \quad \partial_{y} + \varepsilon \partial_{Y},\\
\nabla^2 \quad &\rightarrow \quad \partial^2_{x} + \partial^2_{y} + \varepsilon \Big(2\partial_{x}\partial_{X} + 2\partial_{y}\partial_{Y}\Big) + \varepsilon^2\Big(\partial^2_{X} + \partial^2_{Y}\Big),\\
\nabla^4 \quad &\rightarrow \quad \partial^4_{x} + \partial^4_{y} + 2 \partial^2_{x}\partial^2_{y} + \varepsilon \Big( 4\partial^3_{x}\partial_{X} + 4\partial^3_{y}\partial_{Y} + 4\partial^2_{x}\partial_{y}\partial_{Y} + 4\partial^2_{y}\partial_{x}\partial_{X}\Big) \\
&\qquad + \varepsilon^2 \Big( 6\partial^2_{x}\partial^2_{X} + 6\partial^2_{y}\partial^2_{Y} + 2\partial^2_{x}\partial^2_{Y} + 2\partial^2_{y}\partial^2_{X} + 8\partial_{x}\partial_{y}\partial_{X}\partial_{Y} \Big) + \mathcal{O}(\varepsilon^3)\, . 
\end{aligned}
\end{equation}

\begin{figure}
\centering
\includegraphics[width=0.99\linewidth]{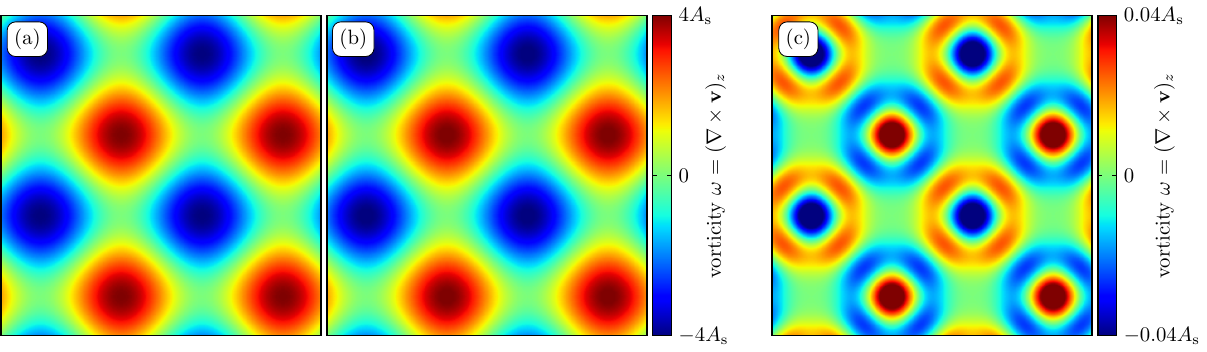}
\caption{\label{fig: comparison two modes and numerics}Comparison of the two-mode approximation of the square vortex lattice with the full numerical solution.
(a) Vorticity field in the square vortex lattice represented by two perpendicular modes according to Eq.~(\ref{eq: lowest-order square lattice}). (b) Vorticity field obtained by numerically solving the full evolution equation (\ref{eq: dynamic equation}).
Note that the value of $\lambda$ does not impact the patterns as long as it is below the threshold to turbulence.
The two vorticity fields are virtually indistinguishable.
This observation is corroborated by (c) where the difference of the two fields shown in (a) and (b) is plotted.
Note that the colorbar is scaled differently for visualization purposes.
For the calculation, we have set $a=0.5$.}
\end{figure}

The subsequent analysis is now based on the fact that the resulting equation has to be fulfilled in every order of $\varepsilon$ as well as every harmonic contribution separately.
In first order, $\mathcal{O}(\varepsilon^1)$, we find that the contributions $A_{x,1,0,0}$, $A_{x,1,1,0}$, $A_{y,1,0,0}$ $A_{y,1,0,1}$, $A_{x,1,1,1}$ and $A_{y,1,1,1}$ (as well as first-order contribution to $q$) must vanish to satisfy simultaneously the evolution equation for $\mathbf{v}$ and the incompressibility condition $ \nabla \cdot \mathbf{v} = 0$. 
In second order, $\mathcal{O}(\varepsilon^2)$, matching terms containing the same harmonics yields
\begin{equation}
\begin{aligned}
q_{2,1,1} &= - \lambda A_1 A_2\, , \\
q_{2,1,-1} &= \lambda A_1 A_2^\ast\, ,
\end{aligned}
\end{equation}
where the asterisked quantities denote the complex conjugates.
Thus, contributions from the nonlinear advection term are compensated by the higher-order terms in $q$.
From the incompressibility condition we obtain the relations
\begin{equation}
\begin{aligned}
A_{x,2,0,1} = i \partial_X A_2 \, , \\
A_{y,2,1,0} = i \partial_Y A_1 \, .
\end{aligned}
\end{equation}
We further find that all other contributions $\propto \varepsilon^2$ in the expansion must vanish.
In third order, $\mathcal{O}(\varepsilon^3)$, we obtain the evolution equations for the lowest order amplitudes of the square vortex lattice, which are given in Eq.~(\ref{eq: amplitude equations WNLA})

Further matching terms containing higher harmonics in third order yields the relations
\begin{equation}
\begin{aligned}
A_{x,3,0,3} &= A_2^3/(a - 64)\, , \qquad
A_{x,3,2,1} = A_1^{2} A_2/(a - 16)\, ,\qquad
A_{x,3,2,-1} = A_1^{2} A_2^\ast/(a - 16)\, ,\\
A_{y,3,3,0} &= A_1^3/(a - 64)\, ,\qquad
A_{y,3,1,2} = A_1 A_2^2/(a - 16)\, ,\qquad
A_{y,3,-1,2} = A_1^\ast A_2^2/(a - 16)\, .
\end{aligned}
\end{equation}
These higher-order amplitudes are thus ``slaved'' to the dominant modes characterized by $A_1$ and $A_2$.
They do not contribute to the derived amplitude equations, but rather represent higher-order corrections to the square vortex lattice, thus going beyond the two-mode approximation [see Eq.~(\ref{eq: lowest-order square lattice})].
Assuming that the amplitudes $A_1$ and $A_2$ are uniform and stationary, we thus obtain for the velocity field 
\begin{equation}
\label{eq: vortex lattice higher harmonics}
\begin{aligned}
v_x = &A_2 e^{iy} + \frac{A_2^3}{a-64} e^{3iy} + \frac{A_1^2 A_2}{a-16}e^{2ix+iy}+\frac{A_1^2 A_2^\ast}{a-16}e^{2ix-iy} + \mathrm{c.c.} + \mathcal{O}(\varepsilon^4) \, ,\\
v_y = &A_1 e^{ix} + \frac{A_1^3}{a-64} e^{3ix} + \frac{A_1 A_2^2}{a-16}e^{2iy+ix}+\frac{A_1^\ast A_2^2}{a-16}e^{2iy-ix} + \mathrm{c.c.} + \mathcal{O}(\varepsilon^4) \, ,
\end{aligned}
\end{equation}
where the stationary amplitudes are given as solution of Eqs.~(\ref{eq: amplitude equations WNLA}), i.e., $A_1 = A_2 = \sqrt{a/5}$.
Close to the instability, i.e., for small values of $a$, Eq.~(\ref{eq: vortex lattice higher harmonics}) clearly demonstrates that the higher harmonics are orders of magnitude smaller than the amplitudes $A_1$ and $A_2$.
This is corroborated by numerical calculations.
Solving the full nonlinear equation~(\ref{eq: dynamic equation}) and comparing with the two-mode vortex lattice approximation [Eq.~(\ref{eq: lowest-order square lattice})], we find only very small differences between the two patterns.
This is shown in Fig.~\ref{fig: comparison two modes and numerics}, where the two-mode approximation and the full solution are plotted and found to be virtually indistinguishable.
We further plot the difference in vorticity in Fig.~\ref{fig: comparison two modes and numerics}(c) and observe only a small discrepancy of up to $\SI{1}{\percent}$ locally.
Note the different scale of the colorbar in (c) compared to (a) and (b).
These observations justify our approach in the main text, where we have neglected higher-order harmonics in the stability analysis.

Remarkably, the derived amplitude equations as well as the higher-order harmonics are independent of the nonlinear advection term.
At least up to the order considered here, all terms were compensated by higher-order contributions in $q$.
Performing numerical calculations with the full nonlinear equation~(\ref{eq: dynamic equation}) corroborates this observation.
The remaining, yet small discrepancies between the two-mode lattice approximation and full numerical solution (as shown in Fig.~\ref{fig: comparison two modes and numerics}) do not change when varying $\lambda$ (up until the point when turbulence sets in).
Determining the stability of the square vortex lattice with respect to nonlinear advection thus necessitates a different approach.
One possibility is the extended stability analysis outlined in the main text, which takes the simultaneous growth of multiple perturbative modes into account.

\section{Components of the matrix $\mathbf{B}$}
\label{app: details linear stability}

As discussed in the main text in the beginning of Section~\ref{sec: extended linear stability analysis}, the naive approach to test the stability of the square lattice pattern is to consider a single perturbative mode similar to the stability analysis of the isotropic state (see Section~\ref{sec: stationary patterns}).
In this context, we introduce the matrix $\mathbf{B}(\mathbf{x})$ in Eq.~(\ref{eq: LSA square lattice one mode step 1}).
The components of $\mathbf{B}(\mathbf{x})$ are given as
\begin{equation}
\label{eq: LSA square lattice one mode Bxx}
\begin{aligned}
B_{xx} = 
&\big(a - 1 - 8A^2  + 2|\mathbf{k}|^2 - |\mathbf{k}|^4\big) - A^2 \big(e^{- 2ix} + e^{2ix}\big) - 3A^2  \big( e^{- 2 iy} + e^{ 2i y}\big) \\
&+ ik_y\lambda A \big( e^{- ix} + e^{ix }\big) - ik_x\lambda A \big( e^{-iy} +  e^{iy}\big)\, ,
\end{aligned}
\end{equation}
\begin{equation}
\label{eq: LSA square lattice one mode Bxy}
\begin{aligned}
B_{xy} = &2A^2 \big(e^{-ix -iy} + e^{ix + iy} + e^{ix - iy} + e^{-ix + iy}\big)+i\lambda A \big( e^{-iy} - e^{iy}\big)\, ,
\end{aligned}
\end{equation}
\begin{equation}
\label{eq: LSA square lattice one mode Byx}
\begin{aligned}
B_{yx} = &2A^2 \big(e^{-ix -iy} + e^{ix + iy} + e^{ix - iy} + e^{-ix + iy}\big) +i\lambda A \big( e^{ix} - e^{-ix}\big)\, ,
\end{aligned}
\end{equation}
\begin{equation}
\label{eq: LSA square lattice one mode Byy}
\begin{aligned}
B_{yy} = 
&\big(a - 1 - 8A^2  + 2|\mathbf{k}|^2 - |\mathbf{k}|^4\big) - 3A^2 \big(e^{- 2ix} + e^{2ix}\big) - A^2  \big( e^{- 2 iy} + e^{ 2i y}\big) \\
&+ ik_y\lambda A \big( e^{- ix} + e^{ix }\big) - ik_x\lambda A \big( e^{-iy} +  e^{iy}\big)\, ,
\end{aligned}
\end{equation}
Note that the factor $e^{i\mathbf{k}\cdot\mathbf{x}}$ is already factorized out.
Thus, as discussed in the main text, $\mathbf{B}(\mathbf{x})$ contains higher-order modes.
As a result, the matrix is still dependent on the spatial variable $\mathbf{x} = (x,y)$.
Neglecting all of these higher-order modes and keeping only terms independent of $\mathbf{x}$ yields the matrix $\mathbf{B}_0$, see Eq.~(\ref{eq: LSA square lattice one mode B lowest order}).

\section{Imaginary part of growth rate}
\label{app: imaginary}

\begin{figure}
\includegraphics[width=0.999\linewidth]{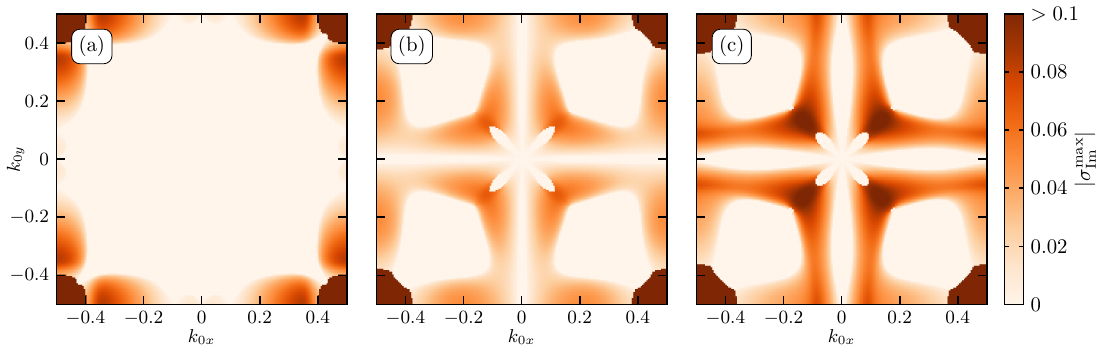}
\caption{\label{fig: imaginary}Absolute value of the imaginary part of the eigenvalue of matrix $\textbf{M}^\ast$ with largest real part, denoted here by $\sigma_\mathrm{Im}^\mathrm{max}$.
In order to compare with Fig.~\ref{fig: largest eigenvalue}(b)/(c)/(d) where the real part is shown, the imaginary part is similarly plotted as a function of $\mathbf{k}_0$ at $a = 0.1$ and (b) $\lambda = 2$, (c) $\lambda = 4$ and (d) $\lambda = 4.8$
On the axes $\mathbf{k}_0 = (k_{0x},0)$ and $\mathbf{k}_0 = (0,k_{0y})$, i.e., where $\sigma_\mathrm{Re}^\mathrm{max}$ obtains the maxima (see Fig.~\ref{fig: largest eigenvalue}), the $\sigma_\mathrm{Im}^\mathrm{max}$ vanishes.}
\end{figure}

In the main text, we focus on the real part of the eigenvalues of the matrix $\textbf{M}^\ast$ in order to determine the stability of the square vortex lattice state.
An instability that has been identified in this way can be further classified via the imaginary part of the corresponding eigenvalue.
To this end, we first determine the eigenvalue $\sigma^\mathrm{max}$ with the largest real part and then take a closer look at the imaginary part $\sigma_\mathrm{Im}^\mathrm{max}$.
We plot the absolute value of the imaginary part in Fig.~\ref{fig: imaginary} as a function of $\mathbf{k}_0$ at the same values of $\lambda$ that are used in Fig.~\ref{fig: largest eigenvalue}(b)/(c)/(d) where the real part is shown.
The imaginary part is nonzero for large regions of $\mathbf{k}_0$-space but vanishes on the axes $\mathbf{k}_0 = (k_{0x},0)$ and $\mathbf{k}_0 = (0,k_{0y})$.
Recall that the maximum growth rate is observed for perturbations with $\mathbf{k}_0$ located on one of these axes, see Fig.~\ref{fig: largest eigenvalue}.
Solely looking at the perturbation associated with the largest growth rate would thus indicate a stationary instability not directly connected to oscillatory or more complex dynamic behavior.
However, above the onset of the instability ($\lambda > \lambda_\mathrm{c}$), there is a set of unstable perturbations close to the fastest-growing perturbation, as is clearly visible in Fig.~\ref{fig: largest eigenvalue}(c).
As Fig.~\ref{fig: imaginary} shows, these perturbations with $\mathbf{k}_0$ not on either axis [$\mathbf{k}_0 = (k_{0x},0)$ or $\mathbf{k}_0 = (0,k_{0y})$] are actually characterized by $\sigma_\mathrm{Im}^\mathrm{max} \neq 0$.
Thus, the unstable perturbations close to the fastest-growing perturbation are instead associated with oscillatory behavior and the emergence of dynamic states.
Further research is necessary to determine whether additional instabilities are involved in the emergence of fully turbulent states.

\section{Instabilities at the borders of the stability balloons}
\label{app: boundaries balloons}

\begin{figure}
\includegraphics[width=0.999\linewidth]{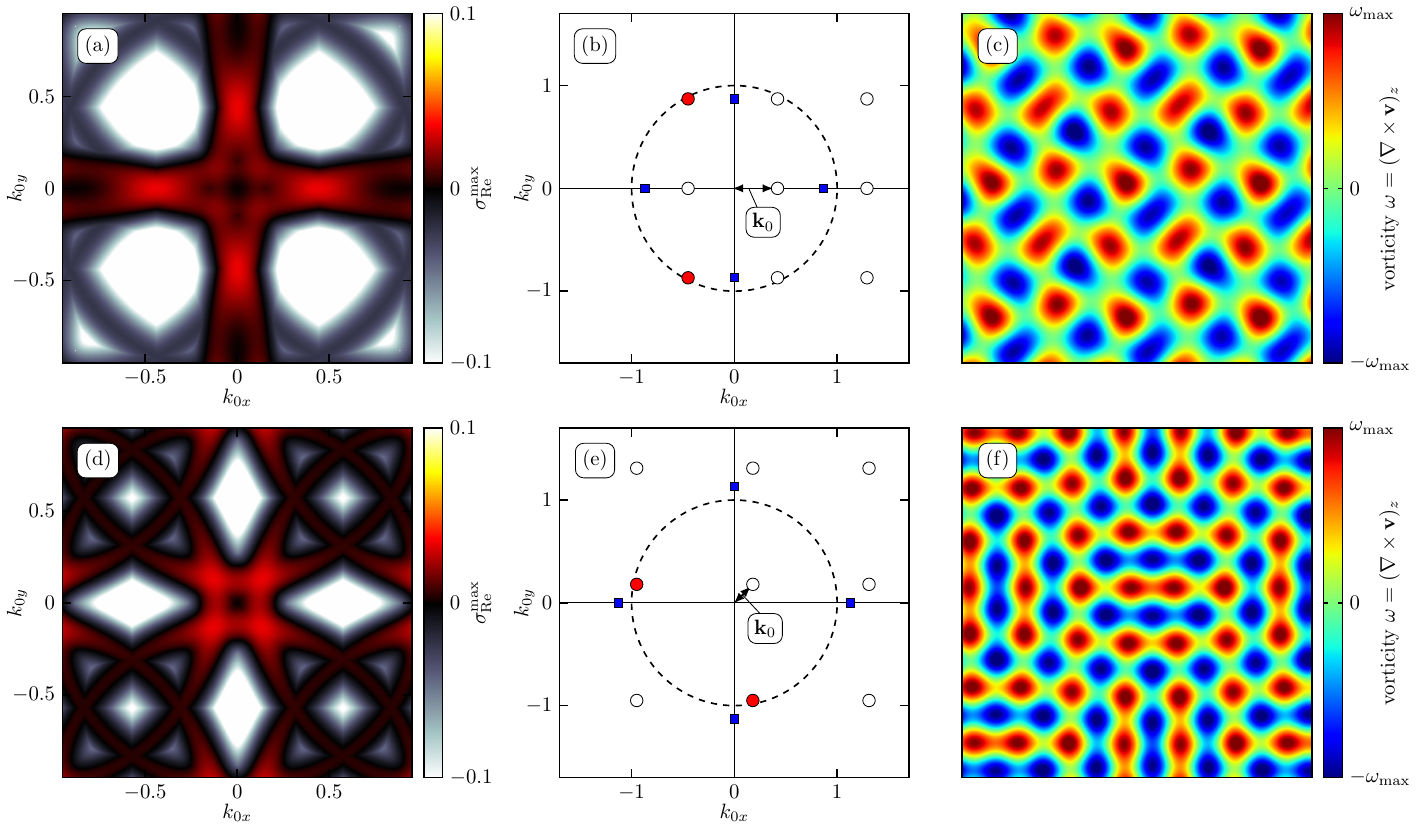}
\caption{\label{fig: borders}Instabilities at the borders of the stability balloon at $a = 0.2$ and $\lambda = 0$.
(a)/(b)/(c) show plots for $k_\mathrm{s}= 0.87$, i.e., close to the left border of the stability balloon shown in Fig.~\ref{fig: Busse}(a).
(d)/(e)/(f) show plots for $k_\mathrm{s}= 1.13$, i.e., close to the right border.
(a)/(d) Maximum growth rate of perturbations to the square lattice with $k_\mathrm{s}=0.87$/$k_\mathrm{s}=1.13$.
(b)/(e) Representation of the eigenvector corresponding to the eigenvalue with largest real part at $\mathbf{k}_0=(0.42,0)$/$\mathbf{k}_0=(0.18,18)$, i.e., for the fastest-growing perturbation.
The relative importance of the perturbative modes is determined by the absolute value of the corresponding eigenvector components and indicated by the saturation of the red color.
The blue squares show the reciprocal lattice corresponding to the square lattice pattern.
The circle indicates $|\mathbf{k}| = k_\mathrm{c} = 1$.
(c)/(f) Vorticity field of the square vortex lattice with $k_\mathrm{s}=0.87$/$k_\mathrm{s}=1.13$ with added perturbations according to the eigenvector shown in (b)/(e).
The size of the snapshots is $12\pi \times 12\pi$.}
\end{figure}

As discussed in Section~\ref{sec: arbitrary wavenumber}, the procedure introduced in Section~\ref{sec: extended linear stability analysis} can also be employed for lattices with arbitrary wavenumber.
Utilizing the approach, we determined the stability balloons shown in Fig.~\ref{fig: Busse}.
In order to explore the instabilities that determine the boundaries of the stability regions, we take a closer look at the eigenvalues and eigenvectors determined by the extended linear stability analysis.
In particular, we compare the left and right boundaries, i.e., square lattice patterns with smaller and larger lattice constants $k_\mathrm{s}$.
To this end, we focus on two exemplary values, $k_\mathrm{s}=0.87$ and $k_\mathrm{s}=1.13$, to explore the instabilities determining the borders of the stability region.
We further set $a = 0.2$ and $\lambda = 0$.
Again, we consider perturbations consisting of $3\times 3$ modes on a grid in Fourier space as discussed in Section~\ref{sec: extended linear stability analysis} and \ref{sec: arbitrary wavenumber}.
We determine the eigenvalues of the matrix $\mathbf{M}^\star$, which are found to be reel-valued for $\lambda=0$.
The growth rate of perturbations is thus directly given by the largest eigenvalue.
Fig.~\ref{fig: borders}(a) and (d) show this growth rate for square lattices with $k_\mathrm{s}=0.87$ and $k_\mathrm{s}=1.13$.
As is clearly visible, it obtains a maximum for perturbations characterized by a grid offset of $\mathbf{k}_0 = (0.42,0)$ and $\mathbf{k}_0 = (0.18,0.18)$ for the two lattice constants, respectively.
The eigenvectors corresponding to these maxima are visualized in Fig.~\ref{fig: borders}(b) and (e), where we show the perturbative modes as a grid in Fourier space.
The red color indicates the relative importance of the particular modes, determined by the absolute value of the corresponding eigenvector components.
The blue squares show the reciprocal lattices corresponding to the respective square lattice structure. 
The dashed circle indicates $|\mathbf{k}| = k_\mathrm{c} = 1$, i.e., the critical wavenumber of instabilities of the isotropic state.
We find that the growing modes of the perturbation are located on the circle, thus leading to emerging structures closer to $k_\mathrm{c}$.
Although this is the case for both lattice constants, $k_\mathrm{s}=0.87$ and $k_\mathrm{s}=1.13$, the instabilities are indeed very different due to their orientation with respect to the square lattice.
This difference becomes much clearer when visualizing the perturbations in real space.
To this end, we plot the perturbed vorticity field in Fig.~\ref{fig: borders}(c) and (f) for the two lattice constants.
Here, a moderate perturbation according to the eigenvectors shown in (b) and (e) is added to the vorticity field in the respective square vortex lattice. 
For symmetry reasons, we also include the perturbative modes rotated in $\SI{90}{\degree}$ steps.
The occurrence of different types of secondary instabilities at the boundaries of the stability region of particular patterns is also encountered in other systems.
For example, the stability of stripe-like states in Rayleigh--B\'{e}nard convection is limited by the zigzag instability for smaller and the Eckhaus instability for larger wavenumbers~\cite{cross2009pattern}.


\begin{thebibliography}{79}%
\makeatletter
\providecommand \@ifxundefined [1]{%
 \@ifx{#1\undefined}
}%
\providecommand \@ifnum [1]{%
 \ifnum #1\expandafter \@firstoftwo
 \else \expandafter \@secondoftwo
 \fi
}%
\providecommand \@ifx [1]{%
 \ifx #1\expandafter \@firstoftwo
 \else \expandafter \@secondoftwo
 \fi
}%
\providecommand \natexlab [1]{#1}%
\providecommand \enquote  [1]{``#1''}%
\providecommand \bibnamefont  [1]{#1}%
\providecommand \bibfnamefont [1]{#1}%
\providecommand \citenamefont [1]{#1}%
\providecommand \href@noop [0]{\@secondoftwo}%
\providecommand \href [0]{\begingroup \@sanitize@url \@href}%
\providecommand \@href[1]{\@@startlink{#1}\@@href}%
\providecommand \@@href[1]{\endgroup#1\@@endlink}%
\providecommand \@sanitize@url [0]{\catcode `\\12\catcode `\$12\catcode
  `\&12\catcode `\#12\catcode `\^12\catcode `\_12\catcode `\%12\relax}%
\providecommand \@@startlink[1]{}%
\providecommand \@@endlink[0]{}%
\providecommand \url  [0]{\begingroup\@sanitize@url \@url }%
\providecommand \@url [1]{\endgroup\@href {#1}{\urlprefix }}%
\providecommand \urlprefix  [0]{URL }%
\providecommand \Eprint [0]{\href }%
\providecommand \doibase [0]{https://doi.org/}%
\providecommand \selectlanguage [0]{\@gobble}%
\providecommand \bibinfo  [0]{\@secondoftwo}%
\providecommand \bibfield  [0]{\@secondoftwo}%
\providecommand \translation [1]{[#1]}%
\providecommand \BibitemOpen [0]{}%
\providecommand \bibitemStop [0]{}%
\providecommand \bibitemNoStop [0]{.\EOS\space}%
\providecommand \EOS [0]{\spacefactor3000\relax}%
\providecommand \BibitemShut  [1]{\csname bibitem#1\endcsname}%
\let\auto@bib@innerbib\@empty
\bibitem [{\citenamefont {Eckert}(2010)}]{eckert2010troublesome}%
  \BibitemOpen
  \bibfield  {author} {\bibinfo {author} {\bibfnamefont {M.}~\bibnamefont
  {Eckert}},\ }\href@noop {} {\bibfield  {journal} {\bibinfo  {journal} {Eur.
  Phys. J. H}\ }\textbf {\bibinfo {volume} {35}},\ \bibinfo {pages} {29}
  (\bibinfo {year} {2010})}\BibitemShut {NoStop}%
\bibitem [{\citenamefont {Barkley}\ \emph {et~al.}(2015)\citenamefont
  {Barkley}, \citenamefont {Song}, \citenamefont {Mukund}, \citenamefont
  {Lemoult}, \citenamefont {Avila},\ and\ \citenamefont
  {Hof}}]{barkley2015rise}%
  \BibitemOpen
  \bibfield  {author} {\bibinfo {author} {\bibfnamefont {D.}~\bibnamefont
  {Barkley}}, \bibinfo {author} {\bibfnamefont {B.}~\bibnamefont {Song}},
  \bibinfo {author} {\bibfnamefont {V.}~\bibnamefont {Mukund}}, \bibinfo
  {author} {\bibfnamefont {G.}~\bibnamefont {Lemoult}}, \bibinfo {author}
  {\bibfnamefont {M.}~\bibnamefont {Avila}},\ and\ \bibinfo {author}
  {\bibfnamefont {B.}~\bibnamefont {Hof}},\ }\href@noop {} {\bibfield
  {journal} {\bibinfo  {journal} {Nature}\ }\textbf {\bibinfo {volume} {526}},\
  \bibinfo {pages} {550} (\bibinfo {year} {2015})}\BibitemShut {NoStop}%
\bibitem [{\citenamefont {Taylor}(1923)}]{taylor1923viii}%
  \BibitemOpen
  \bibfield  {author} {\bibinfo {author} {\bibfnamefont {G.~I.}\ \bibnamefont
  {Taylor}},\ }\href@noop {} {\bibfield  {journal} {\bibinfo  {journal}
  {Philos. Trans. R. Soc. A}\ }\textbf {\bibinfo {volume} {223}},\ \bibinfo
  {pages} {289} (\bibinfo {year} {1923})}\BibitemShut {NoStop}%
\bibitem [{\citenamefont {Barkley}(2016)}]{barkley2016theoretical}%
  \BibitemOpen
  \bibfield  {author} {\bibinfo {author} {\bibfnamefont {D.}~\bibnamefont
  {Barkley}},\ }\href@noop {} {\bibfield  {journal} {\bibinfo  {journal} {J.
  Fluid Mech.}\ }\textbf {\bibinfo {volume} {803}} (\bibinfo {year}
  {2016})}\BibitemShut {NoStop}%
\bibitem [{\citenamefont {Feldmann}\ \emph {et~al.}(2023)\citenamefont
  {Feldmann}, \citenamefont {Borrero-Echeverry}, \citenamefont {Burin},
  \citenamefont {Avila},\ and\ \citenamefont {Avila}}]{feldmann2023routes}%
  \BibitemOpen
  \bibfield  {author} {\bibinfo {author} {\bibfnamefont {D.}~\bibnamefont
  {Feldmann}}, \bibinfo {author} {\bibfnamefont {D.}~\bibnamefont
  {Borrero-Echeverry}}, \bibinfo {author} {\bibfnamefont {M.~J.}\ \bibnamefont
  {Burin}}, \bibinfo {author} {\bibfnamefont {K.}~\bibnamefont {Avila}},\ and\
  \bibinfo {author} {\bibfnamefont {M.}~\bibnamefont {Avila}},\ }\href@noop {}
  {\bibfield  {journal} {\bibinfo  {journal} {Philos. Trans. R. Soc. A}\
  }\textbf {\bibinfo {volume} {381}},\ \bibinfo {pages} {20220114} (\bibinfo
  {year} {2023})}\BibitemShut {NoStop}%
\bibitem [{\citenamefont {Sano}\ and\ \citenamefont
  {Tamai}(2016)}]{sano2016universal}%
  \BibitemOpen
  \bibfield  {author} {\bibinfo {author} {\bibfnamefont {M.}~\bibnamefont
  {Sano}}\ and\ \bibinfo {author} {\bibfnamefont {K.}~\bibnamefont {Tamai}},\
  }\href@noop {} {\bibfield  {journal} {\bibinfo  {journal} {Nat. Phys.}\
  }\textbf {\bibinfo {volume} {12}},\ \bibinfo {pages} {249} (\bibinfo {year}
  {2016})}\BibitemShut {NoStop}%
\bibitem [{\citenamefont {Grossmann}(2000)}]{grossmann2000onset}%
  \BibitemOpen
  \bibfield  {author} {\bibinfo {author} {\bibfnamefont {S.}~\bibnamefont
  {Grossmann}},\ }\href@noop {} {\bibfield  {journal} {\bibinfo  {journal}
  {Rev. Mod. Phys.}\ }\textbf {\bibinfo {volume} {72}},\ \bibinfo {pages} {603}
  (\bibinfo {year} {2000})}\BibitemShut {NoStop}%
\bibitem [{\citenamefont {Lemoult}\ \emph {et~al.}(2016)\citenamefont
  {Lemoult}, \citenamefont {Shi}, \citenamefont {Avila}, \citenamefont
  {Jalikop}, \citenamefont {Avila},\ and\ \citenamefont
  {Hof}}]{lemoult2016directed}%
  \BibitemOpen
  \bibfield  {author} {\bibinfo {author} {\bibfnamefont {G.}~\bibnamefont
  {Lemoult}}, \bibinfo {author} {\bibfnamefont {L.}~\bibnamefont {Shi}},
  \bibinfo {author} {\bibfnamefont {K.}~\bibnamefont {Avila}}, \bibinfo
  {author} {\bibfnamefont {S.~V.}\ \bibnamefont {Jalikop}}, \bibinfo {author}
  {\bibfnamefont {M.}~\bibnamefont {Avila}},\ and\ \bibinfo {author}
  {\bibfnamefont {B.}~\bibnamefont {Hof}},\ }\href@noop {} {\bibfield
  {journal} {\bibinfo  {journal} {Nat. Phys.}\ }\textbf {\bibinfo {volume}
  {12}},\ \bibinfo {pages} {254} (\bibinfo {year} {2016})}\BibitemShut
  {NoStop}%
\bibitem [{\citenamefont {Manneville}(2015)}]{manneville2015transition}%
  \BibitemOpen
  \bibfield  {author} {\bibinfo {author} {\bibfnamefont {P.}~\bibnamefont
  {Manneville}},\ }\href@noop {} {\bibfield  {journal} {\bibinfo  {journal}
  {Eur. J. Mech. B Fluids}\ }\textbf {\bibinfo {volume} {49}},\ \bibinfo
  {pages} {345} (\bibinfo {year} {2015})}\BibitemShut {NoStop}%
\bibitem [{\citenamefont {Drazin}\ and\ \citenamefont
  {Reid}(2004)}]{drazin2004hydrodynamic}%
  \BibitemOpen
  \bibfield  {author} {\bibinfo {author} {\bibfnamefont {P.~G.}\ \bibnamefont
  {Drazin}}\ and\ \bibinfo {author} {\bibfnamefont {W.~H.}\ \bibnamefont
  {Reid}},\ }\href@noop {} {\emph {\bibinfo {title} {Hydrodynamic stability}}}\
  (\bibinfo  {publisher} {Cambridge University Press},\ \bibinfo {year}
  {2004})\BibitemShut {NoStop}%
\bibitem [{\citenamefont {Nishi}\ \emph {et~al.}(2008)\citenamefont {Nishi},
  \citenamefont {{\"U}nsal}, \citenamefont {Durst},\ and\ \citenamefont
  {Biswas}}]{nishi2008laminar}%
  \BibitemOpen
  \bibfield  {author} {\bibinfo {author} {\bibfnamefont {M.}~\bibnamefont
  {Nishi}}, \bibinfo {author} {\bibfnamefont {B.}~\bibnamefont {{\"U}nsal}},
  \bibinfo {author} {\bibfnamefont {F.}~\bibnamefont {Durst}},\ and\ \bibinfo
  {author} {\bibfnamefont {G.}~\bibnamefont {Biswas}},\ }\href@noop {}
  {\bibfield  {journal} {\bibinfo  {journal} {J. Fluid Mech.}\ }\textbf
  {\bibinfo {volume} {614}},\ \bibinfo {pages} {425} (\bibinfo {year}
  {2008})}\BibitemShut {NoStop}%
\bibitem [{\citenamefont {Avila}\ \emph {et~al.}(2011)\citenamefont {Avila},
  \citenamefont {Moxey}, \citenamefont {de~Lozar}, \citenamefont {Avila},
  \citenamefont {Barkley},\ and\ \citenamefont {Hof}}]{avila2011onset}%
  \BibitemOpen
  \bibfield  {author} {\bibinfo {author} {\bibfnamefont {K.}~\bibnamefont
  {Avila}}, \bibinfo {author} {\bibfnamefont {D.}~\bibnamefont {Moxey}},
  \bibinfo {author} {\bibfnamefont {A.}~\bibnamefont {de~Lozar}}, \bibinfo
  {author} {\bibfnamefont {M.}~\bibnamefont {Avila}}, \bibinfo {author}
  {\bibfnamefont {D.}~\bibnamefont {Barkley}},\ and\ \bibinfo {author}
  {\bibfnamefont {B.}~\bibnamefont {Hof}},\ }\href@noop {} {\bibfield
  {journal} {\bibinfo  {journal} {Science}\ }\textbf {\bibinfo {volume}
  {333}},\ \bibinfo {pages} {192} (\bibinfo {year} {2011})}\BibitemShut
  {NoStop}%
\bibitem [{\citenamefont {Avila}\ \emph {et~al.}(2023)\citenamefont {Avila},
  \citenamefont {Barkley},\ and\ \citenamefont {Hof}}]{avila2023transition}%
  \BibitemOpen
  \bibfield  {author} {\bibinfo {author} {\bibfnamefont {M.}~\bibnamefont
  {Avila}}, \bibinfo {author} {\bibfnamefont {D.}~\bibnamefont {Barkley}},\
  and\ \bibinfo {author} {\bibfnamefont {B.}~\bibnamefont {Hof}},\ }\href@noop
  {} {\bibfield  {journal} {\bibinfo  {journal} {Annu. Rev. Fluid Mech.}\
  }\textbf {\bibinfo {volume} {55}},\ \bibinfo {pages} {575} (\bibinfo {year}
  {2023})}\BibitemShut {NoStop}%
\bibitem [{\citenamefont {Hinrichsen}(2000)}]{hinrichsen2000non}%
  \BibitemOpen
  \bibfield  {author} {\bibinfo {author} {\bibfnamefont {H.}~\bibnamefont
  {Hinrichsen}},\ }\href@noop {} {\bibfield  {journal} {\bibinfo  {journal}
  {Adv. Phys.}\ }\textbf {\bibinfo {volume} {49}},\ \bibinfo {pages} {815}
  (\bibinfo {year} {2000})}\BibitemShut {NoStop}%
\bibitem [{\citenamefont {Marchetti}\ \emph {et~al.}(2013)\citenamefont
  {Marchetti}, \citenamefont {Joanny}, \citenamefont {Ramaswamy}, \citenamefont
  {Liverpool}, \citenamefont {Prost}, \citenamefont {Rao},\ and\ \citenamefont
  {Simha}}]{marchetti2013hydrodynamics}%
  \BibitemOpen
  \bibfield  {author} {\bibinfo {author} {\bibfnamefont {M.~C.}\ \bibnamefont
  {Marchetti}}, \bibinfo {author} {\bibfnamefont {J.-F.}\ \bibnamefont
  {Joanny}}, \bibinfo {author} {\bibfnamefont {S.}~\bibnamefont {Ramaswamy}},
  \bibinfo {author} {\bibfnamefont {T.~B.}\ \bibnamefont {Liverpool}}, \bibinfo
  {author} {\bibfnamefont {J.}~\bibnamefont {Prost}}, \bibinfo {author}
  {\bibfnamefont {M.}~\bibnamefont {Rao}},\ and\ \bibinfo {author}
  {\bibfnamefont {R.~A.}\ \bibnamefont {Simha}},\ }\href@noop {} {\bibfield
  {journal} {\bibinfo  {journal} {Rev. Mod. Phys.}\ }\textbf {\bibinfo {volume}
  {85}},\ \bibinfo {pages} {1143} (\bibinfo {year} {2013})}\BibitemShut
  {NoStop}%
\bibitem [{\citenamefont {Cates}\ and\ \citenamefont
  {Tailleur}(2015)}]{cates2015motility}%
  \BibitemOpen
  \bibfield  {author} {\bibinfo {author} {\bibfnamefont {M.~E.}\ \bibnamefont
  {Cates}}\ and\ \bibinfo {author} {\bibfnamefont {J.}~\bibnamefont
  {Tailleur}},\ }\href@noop {} {\bibfield  {journal} {\bibinfo  {journal}
  {Annu. Rev. Condens. Matter Phys.}\ }\textbf {\bibinfo {volume} {6}},\
  \bibinfo {pages} {219} (\bibinfo {year} {2015})}\BibitemShut {NoStop}%
\bibitem [{\citenamefont {Bechinger}\ \emph {et~al.}(2016)\citenamefont
  {Bechinger}, \citenamefont {Di~Leonardo}, \citenamefont {L\"owen},
  \citenamefont {Reichhardt}, \citenamefont {Volpe},\ and\ \citenamefont
  {Volpe}}]{bechinger2016active}%
  \BibitemOpen
  \bibfield  {author} {\bibinfo {author} {\bibfnamefont {C.}~\bibnamefont
  {Bechinger}}, \bibinfo {author} {\bibfnamefont {R.}~\bibnamefont
  {Di~Leonardo}}, \bibinfo {author} {\bibfnamefont {H.}~\bibnamefont
  {L\"owen}}, \bibinfo {author} {\bibfnamefont {C.}~\bibnamefont {Reichhardt}},
  \bibinfo {author} {\bibfnamefont {G.}~\bibnamefont {Volpe}},\ and\ \bibinfo
  {author} {\bibfnamefont {G.}~\bibnamefont {Volpe}},\ }\href@noop {}
  {\bibfield  {journal} {\bibinfo  {journal} {Rev. Mod. Phys.}\ }\textbf
  {\bibinfo {volume} {88}},\ \bibinfo {pages} {045006} (\bibinfo {year}
  {2016})}\BibitemShut {NoStop}%
\bibitem [{\citenamefont {B{\"a}r}\ \emph {et~al.}(2020)\citenamefont
  {B{\"a}r}, \citenamefont {Gro{\ss}mann}, \citenamefont {Heidenreich},\ and\
  \citenamefont {Peruani}}]{bar2020self}%
  \BibitemOpen
  \bibfield  {author} {\bibinfo {author} {\bibfnamefont {M.}~\bibnamefont
  {B{\"a}r}}, \bibinfo {author} {\bibfnamefont {R.}~\bibnamefont
  {Gro{\ss}mann}}, \bibinfo {author} {\bibfnamefont {S.}~\bibnamefont
  {Heidenreich}},\ and\ \bibinfo {author} {\bibfnamefont {F.}~\bibnamefont
  {Peruani}},\ }\href@noop {} {\bibfield  {journal} {\bibinfo  {journal} {Annu.
  Rev. Condens. Matter Phys.}\ }\textbf {\bibinfo {volume} {11}},\ \bibinfo
  {pages} {441} (\bibinfo {year} {2020})}\BibitemShut {NoStop}%
\bibitem [{\citenamefont {Gompper}\ \emph {et~al.}(2020)\citenamefont
  {Gompper}, \citenamefont {Winkler}, \citenamefont {Speck}, \citenamefont
  {Solon}, \citenamefont {Nardini}, \citenamefont {Peruani}, \citenamefont
  {L{\"o}wen}, \citenamefont {Golestanian}, \citenamefont {Kaupp},
  \citenamefont {Alvarez} \emph {et~al.}}]{gompper20202020}%
  \BibitemOpen
  \bibfield  {author} {\bibinfo {author} {\bibfnamefont {G.}~\bibnamefont
  {Gompper}}, \bibinfo {author} {\bibfnamefont {R.~G.}\ \bibnamefont
  {Winkler}}, \bibinfo {author} {\bibfnamefont {T.}~\bibnamefont {Speck}},
  \bibinfo {author} {\bibfnamefont {A.}~\bibnamefont {Solon}}, \bibinfo
  {author} {\bibfnamefont {C.}~\bibnamefont {Nardini}}, \bibinfo {author}
  {\bibfnamefont {F.}~\bibnamefont {Peruani}}, \bibinfo {author} {\bibfnamefont
  {H.}~\bibnamefont {L{\"o}wen}}, \bibinfo {author} {\bibfnamefont
  {R.}~\bibnamefont {Golestanian}}, \bibinfo {author} {\bibfnamefont {U.~B.}\
  \bibnamefont {Kaupp}}, \bibinfo {author} {\bibfnamefont {L.}~\bibnamefont
  {Alvarez}}, \emph {et~al.},\ }\href@noop {} {\bibfield  {journal} {\bibinfo
  {journal} {J. Phys. Condens. Matter}\ }\textbf {\bibinfo {volume} {32}},\
  \bibinfo {pages} {193001} (\bibinfo {year} {2020})}\BibitemShut {NoStop}%
\bibitem [{\citenamefont {Chat{\'e}}(2020)}]{chate2020dry}%
  \BibitemOpen
  \bibfield  {author} {\bibinfo {author} {\bibfnamefont {H.}~\bibnamefont
  {Chat{\'e}}},\ }\href@noop {} {\bibfield  {journal} {\bibinfo  {journal}
  {Annu. Rev. Condens. Matter Phys.}\ }\textbf {\bibinfo {volume} {11}},\
  \bibinfo {pages} {189} (\bibinfo {year} {2020})}\BibitemShut {NoStop}%
\bibitem [{\citenamefont {Alert}\ \emph {et~al.}(2022)\citenamefont {Alert},
  \citenamefont {Casademunt},\ and\ \citenamefont {Joanny}}]{alert2021active}%
  \BibitemOpen
  \bibfield  {author} {\bibinfo {author} {\bibfnamefont {R.}~\bibnamefont
  {Alert}}, \bibinfo {author} {\bibfnamefont {J.}~\bibnamefont {Casademunt}},\
  and\ \bibinfo {author} {\bibfnamefont {J.-F.}\ \bibnamefont {Joanny}},\
  }\href@noop {} {\bibfield  {journal} {\bibinfo  {journal} {Annu. Rev.
  Condens. Matter Phys.}\ }\textbf {\bibinfo {volume} {13}} (\bibinfo {year}
  {2022})}\BibitemShut {NoStop}%
\bibitem [{\citenamefont {Dombrowski}\ \emph {et~al.}(2004)\citenamefont
  {Dombrowski}, \citenamefont {Cisneros}, \citenamefont {Chatkaew},
  \citenamefont {Goldstein},\ and\ \citenamefont
  {Kessler}}]{dombrowski2004self}%
  \BibitemOpen
  \bibfield  {author} {\bibinfo {author} {\bibfnamefont {C.}~\bibnamefont
  {Dombrowski}}, \bibinfo {author} {\bibfnamefont {L.}~\bibnamefont
  {Cisneros}}, \bibinfo {author} {\bibfnamefont {S.}~\bibnamefont {Chatkaew}},
  \bibinfo {author} {\bibfnamefont {R.~E.}\ \bibnamefont {Goldstein}},\ and\
  \bibinfo {author} {\bibfnamefont {J.~O.}\ \bibnamefont {Kessler}},\
  }\href@noop {} {\bibfield  {journal} {\bibinfo  {journal} {Phys. Rev. Lett.}\
  }\textbf {\bibinfo {volume} {93}},\ \bibinfo {pages} {098103} (\bibinfo
  {year} {2004})}\BibitemShut {NoStop}%
\bibitem [{\citenamefont {Wensink}\ \emph {et~al.}(2012)\citenamefont
  {Wensink}, \citenamefont {Dunkel}, \citenamefont {Heidenreich}, \citenamefont
  {Drescher}, \citenamefont {Goldstein}, \citenamefont {L{\"o}wen},\ and\
  \citenamefont {Yeomans}}]{wensink2012meso}%
  \BibitemOpen
  \bibfield  {author} {\bibinfo {author} {\bibfnamefont {H.~H.}\ \bibnamefont
  {Wensink}}, \bibinfo {author} {\bibfnamefont {J.}~\bibnamefont {Dunkel}},
  \bibinfo {author} {\bibfnamefont {S.}~\bibnamefont {Heidenreich}}, \bibinfo
  {author} {\bibfnamefont {K.}~\bibnamefont {Drescher}}, \bibinfo {author}
  {\bibfnamefont {R.~E.}\ \bibnamefont {Goldstein}}, \bibinfo {author}
  {\bibfnamefont {H.}~\bibnamefont {L{\"o}wen}},\ and\ \bibinfo {author}
  {\bibfnamefont {J.~M.}\ \bibnamefont {Yeomans}},\ }\href@noop {} {\bibfield
  {journal} {\bibinfo  {journal} {Proc. Natl. Acad. Sci. U.S.A.}\ }\textbf
  {\bibinfo {volume} {109}},\ \bibinfo {pages} {14308} (\bibinfo {year}
  {2012})}\BibitemShut {NoStop}%
\bibitem [{\citenamefont {Doostmohammadi}\ \emph {et~al.}(2018)\citenamefont
  {Doostmohammadi}, \citenamefont {Ign{\'e}s-Mullol}, \citenamefont {Yeomans},\
  and\ \citenamefont {Sagu{\'e}s}}]{doostmohammadi2018active}%
  \BibitemOpen
  \bibfield  {author} {\bibinfo {author} {\bibfnamefont {A.}~\bibnamefont
  {Doostmohammadi}}, \bibinfo {author} {\bibfnamefont {J.}~\bibnamefont
  {Ign{\'e}s-Mullol}}, \bibinfo {author} {\bibfnamefont {J.~M.}\ \bibnamefont
  {Yeomans}},\ and\ \bibinfo {author} {\bibfnamefont {F.}~\bibnamefont
  {Sagu{\'e}s}},\ }\href@noop {} {\bibfield  {journal} {\bibinfo  {journal}
  {Nat. Commun.}\ }\textbf {\bibinfo {volume} {9}},\ \bibinfo {pages} {3246}
  (\bibinfo {year} {2018})}\BibitemShut {NoStop}%
\bibitem [{\citenamefont {Doostmohammadi}\ \emph {et~al.}(2017)\citenamefont
  {Doostmohammadi}, \citenamefont {Shendruk}, \citenamefont {Thijssen},\ and\
  \citenamefont {Yeomans}}]{doostmohammadi2017onset}%
  \BibitemOpen
  \bibfield  {author} {\bibinfo {author} {\bibfnamefont {A.}~\bibnamefont
  {Doostmohammadi}}, \bibinfo {author} {\bibfnamefont {T.~N.}\ \bibnamefont
  {Shendruk}}, \bibinfo {author} {\bibfnamefont {K.}~\bibnamefont {Thijssen}},\
  and\ \bibinfo {author} {\bibfnamefont {J.~M.}\ \bibnamefont {Yeomans}},\
  }\href@noop {} {\bibfield  {journal} {\bibinfo  {journal} {Nat. Commun.}\
  }\textbf {\bibinfo {volume} {8}},\ \bibinfo {pages} {15326} (\bibinfo {year}
  {2017})}\BibitemShut {NoStop}%
\bibitem [{\citenamefont {Qi}\ \emph {et~al.}(2022)\citenamefont {Qi},
  \citenamefont {Westphal}, \citenamefont {Gompper},\ and\ \citenamefont
  {Winkler}}]{qi2022emergence}%
  \BibitemOpen
  \bibfield  {author} {\bibinfo {author} {\bibfnamefont {K.}~\bibnamefont
  {Qi}}, \bibinfo {author} {\bibfnamefont {E.}~\bibnamefont {Westphal}},
  \bibinfo {author} {\bibfnamefont {G.}~\bibnamefont {Gompper}},\ and\ \bibinfo
  {author} {\bibfnamefont {R.~G.}\ \bibnamefont {Winkler}},\ }\href@noop {}
  {\bibfield  {journal} {\bibinfo  {journal} {Commun. Phys.}\ }\textbf
  {\bibinfo {volume} {5}},\ \bibinfo {pages} {49} (\bibinfo {year}
  {2022})}\BibitemShut {NoStop}%
\bibitem [{\citenamefont {Zantop}\ and\ \citenamefont
  {Stark}(2022)}]{zantop2022emergent}%
  \BibitemOpen
  \bibfield  {author} {\bibinfo {author} {\bibfnamefont {A.~W.}\ \bibnamefont
  {Zantop}}\ and\ \bibinfo {author} {\bibfnamefont {H.}~\bibnamefont {Stark}},\
  }\href@noop {} {\bibfield  {journal} {\bibinfo  {journal} {Soft Matter}\
  }\textbf {\bibinfo {volume} {18}},\ \bibinfo {pages} {6179} (\bibinfo {year}
  {2022})}\BibitemShut {NoStop}%
\bibitem [{\citenamefont {Gro{\ss}mann}\ \emph {et~al.}(2014)\citenamefont
  {Gro{\ss}mann}, \citenamefont {Romanczuk}, \citenamefont {B{\"a}r},\ and\
  \citenamefont {Schimansky-Geier}}]{grossmann2014vortex}%
  \BibitemOpen
  \bibfield  {author} {\bibinfo {author} {\bibfnamefont {R.}~\bibnamefont
  {Gro{\ss}mann}}, \bibinfo {author} {\bibfnamefont {P.}~\bibnamefont
  {Romanczuk}}, \bibinfo {author} {\bibfnamefont {M.}~\bibnamefont {B{\"a}r}},\
  and\ \bibinfo {author} {\bibfnamefont {L.}~\bibnamefont {Schimansky-Geier}},\
  }\href@noop {} {\bibfield  {journal} {\bibinfo  {journal} {Phys. Rev. Lett.}\
  }\textbf {\bibinfo {volume} {113}},\ \bibinfo {pages} {258104} (\bibinfo
  {year} {2014})}\BibitemShut {NoStop}%
\bibitem [{\citenamefont {Gro{\ss}mann}\ \emph {et~al.}(2015)\citenamefont
  {Gro{\ss}mann}, \citenamefont {Romanczuk}, \citenamefont {B{\"a}r},\ and\
  \citenamefont {Schimansky-Geier}}]{grossmann2015pattern}%
  \BibitemOpen
  \bibfield  {author} {\bibinfo {author} {\bibfnamefont {R.}~\bibnamefont
  {Gro{\ss}mann}}, \bibinfo {author} {\bibfnamefont {P.}~\bibnamefont
  {Romanczuk}}, \bibinfo {author} {\bibfnamefont {M.}~\bibnamefont {B{\"a}r}},\
  and\ \bibinfo {author} {\bibfnamefont {L.}~\bibnamefont {Schimansky-Geier}},\
  }\href@noop {} {\bibfield  {journal} {\bibinfo  {journal} {Eur. Phys. J.
  Special Topics}\ }\textbf {\bibinfo {volume} {224}},\ \bibinfo {pages} {1325}
  (\bibinfo {year} {2015})}\BibitemShut {NoStop}%
\bibitem [{\citenamefont {Dunkel}\ \emph
  {et~al.}(2013{\natexlab{a}})\citenamefont {Dunkel}, \citenamefont
  {Heidenreich}, \citenamefont {B{\"a}r},\ and\ \citenamefont
  {Goldstein}}]{dunkel2013minimal}%
  \BibitemOpen
  \bibfield  {author} {\bibinfo {author} {\bibfnamefont {J.}~\bibnamefont
  {Dunkel}}, \bibinfo {author} {\bibfnamefont {S.}~\bibnamefont {Heidenreich}},
  \bibinfo {author} {\bibfnamefont {M.}~\bibnamefont {B{\"a}r}},\ and\ \bibinfo
  {author} {\bibfnamefont {R.~E.}\ \bibnamefont {Goldstein}},\ }\href@noop {}
  {\bibfield  {journal} {\bibinfo  {journal} {New J. Phys.}\ }\textbf {\bibinfo
  {volume} {15}},\ \bibinfo {pages} {045016} (\bibinfo {year}
  {2013}{\natexlab{a}})}\BibitemShut {NoStop}%
\bibitem [{\citenamefont {Dunkel}\ \emph
  {et~al.}(2013{\natexlab{b}})\citenamefont {Dunkel}, \citenamefont
  {Heidenreich}, \citenamefont {Drescher}, \citenamefont {Wensink},
  \citenamefont {B{\"a}r},\ and\ \citenamefont {Goldstein}}]{dunkel2013fluid}%
  \BibitemOpen
  \bibfield  {author} {\bibinfo {author} {\bibfnamefont {J.}~\bibnamefont
  {Dunkel}}, \bibinfo {author} {\bibfnamefont {S.}~\bibnamefont {Heidenreich}},
  \bibinfo {author} {\bibfnamefont {K.}~\bibnamefont {Drescher}}, \bibinfo
  {author} {\bibfnamefont {H.~H.}\ \bibnamefont {Wensink}}, \bibinfo {author}
  {\bibfnamefont {M.}~\bibnamefont {B{\"a}r}},\ and\ \bibinfo {author}
  {\bibfnamefont {R.~E.}\ \bibnamefont {Goldstein}},\ }\href@noop {} {\bibfield
   {journal} {\bibinfo  {journal} {Phys. Rev. Lett.}\ }\textbf {\bibinfo
  {volume} {110}},\ \bibinfo {pages} {228102} (\bibinfo {year}
  {2013}{\natexlab{b}})}\BibitemShut {NoStop}%
\bibitem [{\citenamefont {Riedel}\ \emph {et~al.}(2005)\citenamefont {Riedel},
  \citenamefont {Kruse},\ and\ \citenamefont {Howard}}]{riedel2005self}%
  \BibitemOpen
  \bibfield  {author} {\bibinfo {author} {\bibfnamefont {I.~H.}\ \bibnamefont
  {Riedel}}, \bibinfo {author} {\bibfnamefont {K.}~\bibnamefont {Kruse}},\ and\
  \bibinfo {author} {\bibfnamefont {J.}~\bibnamefont {Howard}},\ }\href@noop {}
  {\bibfield  {journal} {\bibinfo  {journal} {Science}\ }\textbf {\bibinfo
  {volume} {309}},\ \bibinfo {pages} {300} (\bibinfo {year}
  {2005})}\BibitemShut {NoStop}%
\bibitem [{\citenamefont {Sumino}\ \emph {et~al.}(2012)\citenamefont {Sumino},
  \citenamefont {Nagai}, \citenamefont {Shitaka}, \citenamefont {Tanaka},
  \citenamefont {Yoshikawa}, \citenamefont {Chat{\'e}},\ and\ \citenamefont
  {Oiwa}}]{sumino2012large}%
  \BibitemOpen
  \bibfield  {author} {\bibinfo {author} {\bibfnamefont {Y.}~\bibnamefont
  {Sumino}}, \bibinfo {author} {\bibfnamefont {K.~H.}\ \bibnamefont {Nagai}},
  \bibinfo {author} {\bibfnamefont {Y.}~\bibnamefont {Shitaka}}, \bibinfo
  {author} {\bibfnamefont {D.}~\bibnamefont {Tanaka}}, \bibinfo {author}
  {\bibfnamefont {K.}~\bibnamefont {Yoshikawa}}, \bibinfo {author}
  {\bibfnamefont {H.}~\bibnamefont {Chat{\'e}}},\ and\ \bibinfo {author}
  {\bibfnamefont {K.}~\bibnamefont {Oiwa}},\ }\href@noop {} {\bibfield
  {journal} {\bibinfo  {journal} {Nature}\ }\textbf {\bibinfo {volume} {483}},\
  \bibinfo {pages} {448} (\bibinfo {year} {2012})}\BibitemShut {NoStop}%
\bibitem [{\citenamefont {Doostmohammadi}\ \emph {et~al.}(2016)\citenamefont
  {Doostmohammadi}, \citenamefont {Adamer}, \citenamefont {Thampi},\ and\
  \citenamefont {Yeomans}}]{doostmohammadi2016stabilization}%
  \BibitemOpen
  \bibfield  {author} {\bibinfo {author} {\bibfnamefont {A.}~\bibnamefont
  {Doostmohammadi}}, \bibinfo {author} {\bibfnamefont {M.~F.}\ \bibnamefont
  {Adamer}}, \bibinfo {author} {\bibfnamefont {S.~P.}\ \bibnamefont {Thampi}},\
  and\ \bibinfo {author} {\bibfnamefont {J.~M.}\ \bibnamefont {Yeomans}},\
  }\href@noop {} {\bibfield  {journal} {\bibinfo  {journal} {Nat. Commun.}\
  }\textbf {\bibinfo {volume} {7}},\ \bibinfo {pages} {10557} (\bibinfo {year}
  {2016})}\BibitemShut {NoStop}%
\bibitem [{\citenamefont {Thijssen}\ \emph {et~al.}(2020)\citenamefont
  {Thijssen}, \citenamefont {Nejad},\ and\ \citenamefont
  {Yeomans}}]{thijssen2020role}%
  \BibitemOpen
  \bibfield  {author} {\bibinfo {author} {\bibfnamefont {K.}~\bibnamefont
  {Thijssen}}, \bibinfo {author} {\bibfnamefont {M.~R.}\ \bibnamefont
  {Nejad}},\ and\ \bibinfo {author} {\bibfnamefont {J.~M.}\ \bibnamefont
  {Yeomans}},\ }\href@noop {} {\bibfield  {journal} {\bibinfo  {journal} {Phys.
  Rev. Lett.}\ }\textbf {\bibinfo {volume} {125}},\ \bibinfo {pages} {218004}
  (\bibinfo {year} {2020})}\BibitemShut {NoStop}%
\bibitem [{\citenamefont {Caballero}\ \emph {et~al.}(2023)\citenamefont
  {Caballero}, \citenamefont {You},\ and\ \citenamefont
  {Marchetti}}]{caballero2023vorticity}%
  \BibitemOpen
  \bibfield  {author} {\bibinfo {author} {\bibfnamefont {F.}~\bibnamefont
  {Caballero}}, \bibinfo {author} {\bibfnamefont {Z.}~\bibnamefont {You}},\
  and\ \bibinfo {author} {\bibfnamefont {M.~C.}\ \bibnamefont {Marchetti}},\
  }\href@noop {} {\bibfield  {journal} {\bibinfo  {journal} {Soft Matter}\
  }\textbf {\bibinfo {volume} {19}},\ \bibinfo {pages} {7828} (\bibinfo {year}
  {2023})}\BibitemShut {NoStop}%
\bibitem [{\citenamefont {James}\ \emph {et~al.}(2021)\citenamefont {James},
  \citenamefont {Suchla}, \citenamefont {Dunkel},\ and\ \citenamefont
  {Wilczek}}]{james2021emergence}%
  \BibitemOpen
  \bibfield  {author} {\bibinfo {author} {\bibfnamefont {M.}~\bibnamefont
  {James}}, \bibinfo {author} {\bibfnamefont {D.~A.}\ \bibnamefont {Suchla}},
  \bibinfo {author} {\bibfnamefont {J.}~\bibnamefont {Dunkel}},\ and\ \bibinfo
  {author} {\bibfnamefont {M.}~\bibnamefont {Wilczek}},\ }\href@noop {}
  {\bibfield  {journal} {\bibinfo  {journal} {Nat. Commun.}\ }\textbf {\bibinfo
  {volume} {12}},\ \bibinfo {pages} {5630} (\bibinfo {year}
  {2021})}\BibitemShut {NoStop}%
\bibitem [{\citenamefont {Worlitzer}\ \emph {et~al.}(2021)\citenamefont
  {Worlitzer}, \citenamefont {Ariel}, \citenamefont {Be'er}, \citenamefont
  {Stark}, \citenamefont {B{\"a}r},\ and\ \citenamefont
  {Heidenreich}}]{worlitzer2021turbulence}%
  \BibitemOpen
  \bibfield  {author} {\bibinfo {author} {\bibfnamefont {V.~M.}\ \bibnamefont
  {Worlitzer}}, \bibinfo {author} {\bibfnamefont {G.}~\bibnamefont {Ariel}},
  \bibinfo {author} {\bibfnamefont {A.}~\bibnamefont {Be'er}}, \bibinfo
  {author} {\bibfnamefont {H.}~\bibnamefont {Stark}}, \bibinfo {author}
  {\bibfnamefont {M.}~\bibnamefont {B{\"a}r}},\ and\ \bibinfo {author}
  {\bibfnamefont {S.}~\bibnamefont {Heidenreich}},\ }\href@noop {} {\bibfield
  {journal} {\bibinfo  {journal} {Soft Matter}\ }\textbf {\bibinfo {volume}
  {17}},\ \bibinfo {pages} {10447} (\bibinfo {year} {2021})}\BibitemShut
  {NoStop}%
\bibitem [{\citenamefont {Nishiguchi}\ \emph {et~al.}(2018)\citenamefont
  {Nishiguchi}, \citenamefont {Aranson}, \citenamefont {Snezhko},\ and\
  \citenamefont {Sokolov}}]{nishiguchi2018engineering}%
  \BibitemOpen
  \bibfield  {author} {\bibinfo {author} {\bibfnamefont {D.}~\bibnamefont
  {Nishiguchi}}, \bibinfo {author} {\bibfnamefont {I.~S.}\ \bibnamefont
  {Aranson}}, \bibinfo {author} {\bibfnamefont {A.}~\bibnamefont {Snezhko}},\
  and\ \bibinfo {author} {\bibfnamefont {A.}~\bibnamefont {Sokolov}},\
  }\href@noop {} {\bibfield  {journal} {\bibinfo  {journal} {Nat. Commun.}\
  }\textbf {\bibinfo {volume} {9}},\ \bibinfo {pages} {4486} (\bibinfo {year}
  {2018})}\BibitemShut {NoStop}%
\bibitem [{\citenamefont {Reinken}\ \emph {et~al.}(2020)\citenamefont
  {Reinken}, \citenamefont {Nishiguchi}, \citenamefont {Heidenreich},
  \citenamefont {Sokolov}, \citenamefont {B{\"a}r}, \citenamefont {Klapp},\
  and\ \citenamefont {Aranson}}]{reinken2020organizing}%
  \BibitemOpen
  \bibfield  {author} {\bibinfo {author} {\bibfnamefont {H.}~\bibnamefont
  {Reinken}}, \bibinfo {author} {\bibfnamefont {D.}~\bibnamefont {Nishiguchi}},
  \bibinfo {author} {\bibfnamefont {S.}~\bibnamefont {Heidenreich}}, \bibinfo
  {author} {\bibfnamefont {A.}~\bibnamefont {Sokolov}}, \bibinfo {author}
  {\bibfnamefont {M.}~\bibnamefont {B{\"a}r}}, \bibinfo {author} {\bibfnamefont
  {S.~H.~L.}\ \bibnamefont {Klapp}},\ and\ \bibinfo {author} {\bibfnamefont
  {I.~S.}\ \bibnamefont {Aranson}},\ }\href@noop {} {\bibfield  {journal}
  {\bibinfo  {journal} {Commun. Phys.}\ }\textbf {\bibinfo {volume} {3}},\
  \bibinfo {pages} {1} (\bibinfo {year} {2020})}\BibitemShut {NoStop}%
\bibitem [{\citenamefont {Reinken}\ \emph
  {et~al.}(2022{\natexlab{a}})\citenamefont {Reinken}, \citenamefont
  {Heidenreich}, \citenamefont {B{\"a}r},\ and\ \citenamefont
  {Klapp}}]{reinken2022ising}%
  \BibitemOpen
  \bibfield  {author} {\bibinfo {author} {\bibfnamefont {H.}~\bibnamefont
  {Reinken}}, \bibinfo {author} {\bibfnamefont {S.}~\bibnamefont
  {Heidenreich}}, \bibinfo {author} {\bibfnamefont {M.}~\bibnamefont
  {B{\"a}r}},\ and\ \bibinfo {author} {\bibfnamefont {S.~H.~L.}\ \bibnamefont
  {Klapp}},\ }\href@noop {} {\bibfield  {journal} {\bibinfo  {journal} {Phys.
  Rev. Lett.}\ }\textbf {\bibinfo {volume} {128}},\ \bibinfo {pages} {048004}
  (\bibinfo {year} {2022}{\natexlab{a}})}\BibitemShut {NoStop}%
\bibitem [{\citenamefont {Partovifard}\ \emph {et~al.}(2024)\citenamefont
  {Partovifard}, \citenamefont {Grawitter},\ and\ \citenamefont
  {Stark}}]{partovifard2024controlling}%
  \BibitemOpen
  \bibfield  {author} {\bibinfo {author} {\bibfnamefont {A.}~\bibnamefont
  {Partovifard}}, \bibinfo {author} {\bibfnamefont {J.}~\bibnamefont
  {Grawitter}},\ and\ \bibinfo {author} {\bibfnamefont {H.}~\bibnamefont
  {Stark}},\ }\href@noop {} {\bibfield  {journal} {\bibinfo  {journal} {Soft
  Matter}\ } (\bibinfo {year} {2024})}\BibitemShut {NoStop}%
\bibitem [{\citenamefont {Schimming}\ \emph {et~al.}(2024)\citenamefont
  {Schimming}, \citenamefont {Reichhardt},\ and\ \citenamefont
  {Reichhardt}}]{schimming2024vortex}%
  \BibitemOpen
  \bibfield  {author} {\bibinfo {author} {\bibfnamefont {C.~D.}\ \bibnamefont
  {Schimming}}, \bibinfo {author} {\bibfnamefont {C.}~\bibnamefont
  {Reichhardt}},\ and\ \bibinfo {author} {\bibfnamefont {C.}~\bibnamefont
  {Reichhardt}},\ }\href@noop {} {\bibfield  {journal} {\bibinfo  {journal}
  {Phys. Rev. Lett.}\ }\textbf {\bibinfo {volume} {132}},\ \bibinfo {pages}
  {018301} (\bibinfo {year} {2024})}\BibitemShut {NoStop}%
\bibitem [{\citenamefont {Wioland}\ \emph {et~al.}(2013)\citenamefont
  {Wioland}, \citenamefont {Woodhouse}, \citenamefont {Dunkel}, \citenamefont
  {Kessler},\ and\ \citenamefont {Goldstein}}]{wioland2013confinement}%
  \BibitemOpen
  \bibfield  {author} {\bibinfo {author} {\bibfnamefont {H.}~\bibnamefont
  {Wioland}}, \bibinfo {author} {\bibfnamefont {F.~G.}\ \bibnamefont
  {Woodhouse}}, \bibinfo {author} {\bibfnamefont {J.}~\bibnamefont {Dunkel}},
  \bibinfo {author} {\bibfnamefont {J.~O.}\ \bibnamefont {Kessler}},\ and\
  \bibinfo {author} {\bibfnamefont {R.~E.}\ \bibnamefont {Goldstein}},\
  }\href@noop {} {\bibfield  {journal} {\bibinfo  {journal} {Phys. Rev. Lett.}\
  }\textbf {\bibinfo {volume} {110}},\ \bibinfo {pages} {268102} (\bibinfo
  {year} {2013})}\BibitemShut {NoStop}%
\bibitem [{\citenamefont {Beppu}\ \emph {et~al.}(2021)\citenamefont {Beppu},
  \citenamefont {Izri}, \citenamefont {Sato}, \citenamefont {Yamanishi},
  \citenamefont {Sumino},\ and\ \citenamefont {Maeda}}]{beppu2021edge}%
  \BibitemOpen
  \bibfield  {author} {\bibinfo {author} {\bibfnamefont {K.}~\bibnamefont
  {Beppu}}, \bibinfo {author} {\bibfnamefont {Z.}~\bibnamefont {Izri}},
  \bibinfo {author} {\bibfnamefont {T.}~\bibnamefont {Sato}}, \bibinfo {author}
  {\bibfnamefont {Y.}~\bibnamefont {Yamanishi}}, \bibinfo {author}
  {\bibfnamefont {Y.}~\bibnamefont {Sumino}},\ and\ \bibinfo {author}
  {\bibfnamefont {Y.~T.}\ \bibnamefont {Maeda}},\ }\href@noop {} {\bibfield
  {journal} {\bibinfo  {journal} {Proc. Natl. Acad. Sci. U.S.A.}\ }\textbf
  {\bibinfo {volume} {118}},\ \bibinfo {pages} {e2107461118} (\bibinfo {year}
  {2021})}\BibitemShut {NoStop}%
\bibitem [{\citenamefont {Opathalage}\ \emph {et~al.}(2019)\citenamefont
  {Opathalage}, \citenamefont {Norton}, \citenamefont {Juniper}, \citenamefont
  {Langeslay}, \citenamefont {Aghvami}, \citenamefont {Fraden},\ and\
  \citenamefont {Dogic}}]{opathalage2019self}%
  \BibitemOpen
  \bibfield  {author} {\bibinfo {author} {\bibfnamefont {A.}~\bibnamefont
  {Opathalage}}, \bibinfo {author} {\bibfnamefont {M.~M.}\ \bibnamefont
  {Norton}}, \bibinfo {author} {\bibfnamefont {M.~P.}\ \bibnamefont {Juniper}},
  \bibinfo {author} {\bibfnamefont {B.}~\bibnamefont {Langeslay}}, \bibinfo
  {author} {\bibfnamefont {S.~A.}\ \bibnamefont {Aghvami}}, \bibinfo {author}
  {\bibfnamefont {S.}~\bibnamefont {Fraden}},\ and\ \bibinfo {author}
  {\bibfnamefont {Z.}~\bibnamefont {Dogic}},\ }\href@noop {} {\bibfield
  {journal} {\bibinfo  {journal} {Proc. Natl. Acad. Sci. U.S.A.}\ }\textbf
  {\bibinfo {volume} {116}},\ \bibinfo {pages} {4788} (\bibinfo {year}
  {2019})}\BibitemShut {NoStop}%
\bibitem [{\citenamefont {Beppu}\ and\ \citenamefont
  {Maeda}(2022)}]{beppu2022exploring}%
  \BibitemOpen
  \bibfield  {author} {\bibinfo {author} {\bibfnamefont {K.}~\bibnamefont
  {Beppu}}\ and\ \bibinfo {author} {\bibfnamefont {Y.~T.}\ \bibnamefont
  {Maeda}},\ }\href@noop {} {\bibfield  {journal} {\bibinfo  {journal}
  {Biophys. physicobiology}\ }\textbf {\bibinfo {volume} {19}},\ \bibinfo
  {pages} {e190020} (\bibinfo {year} {2022})}\BibitemShut {NoStop}%
\bibitem [{\citenamefont {Beppu}\ \emph {et~al.}(2023)\citenamefont {Beppu},
  \citenamefont {Matsuura},\ and\ \citenamefont {Maeda}}]{beppu2023geometric}%
  \BibitemOpen
  \bibfield  {author} {\bibinfo {author} {\bibfnamefont {K.}~\bibnamefont
  {Beppu}}, \bibinfo {author} {\bibfnamefont {K.}~\bibnamefont {Matsuura}},\
  and\ \bibinfo {author} {\bibfnamefont {Y.~T.}\ \bibnamefont {Maeda}},\
  }\href@noop {} {\bibfield  {journal} {\bibinfo  {journal} {arXiv preprint
  arXiv:2312.15257}\ } (\bibinfo {year} {2023})}\BibitemShut {NoStop}%
\bibitem [{\citenamefont {Shiratani}\ \emph {et~al.}(2023)\citenamefont
  {Shiratani}, \citenamefont {Takeuchi},\ and\ \citenamefont
  {Nishiguchi}}]{shiratani2023route}%
  \BibitemOpen
  \bibfield  {author} {\bibinfo {author} {\bibfnamefont {S.}~\bibnamefont
  {Shiratani}}, \bibinfo {author} {\bibfnamefont {K.~A.}\ \bibnamefont
  {Takeuchi}},\ and\ \bibinfo {author} {\bibfnamefont {D.}~\bibnamefont
  {Nishiguchi}},\ }\href@noop {} {\bibfield  {journal} {\bibinfo  {journal}
  {arXiv preprint arXiv:2304.03306}\ } (\bibinfo {year} {2023})}\BibitemShut
  {NoStop}%
\bibitem [{\citenamefont {Wu}\ \emph {et~al.}(2017)\citenamefont {Wu},
  \citenamefont {Hishamunda}, \citenamefont {Chen}, \citenamefont {DeCamp},
  \citenamefont {Chang}, \citenamefont {Fern{\'a}ndez-Nieves}, \citenamefont
  {Fraden},\ and\ \citenamefont {Dogic}}]{wu2017transition}%
  \BibitemOpen
  \bibfield  {author} {\bibinfo {author} {\bibfnamefont {K.-T.}\ \bibnamefont
  {Wu}}, \bibinfo {author} {\bibfnamefont {J.~B.}\ \bibnamefont {Hishamunda}},
  \bibinfo {author} {\bibfnamefont {D.~T.}\ \bibnamefont {Chen}}, \bibinfo
  {author} {\bibfnamefont {S.~J.}\ \bibnamefont {DeCamp}}, \bibinfo {author}
  {\bibfnamefont {Y.-W.}\ \bibnamefont {Chang}}, \bibinfo {author}
  {\bibfnamefont {A.}~\bibnamefont {Fern{\'a}ndez-Nieves}}, \bibinfo {author}
  {\bibfnamefont {S.}~\bibnamefont {Fraden}},\ and\ \bibinfo {author}
  {\bibfnamefont {Z.}~\bibnamefont {Dogic}},\ }\href@noop {} {\bibfield
  {journal} {\bibinfo  {journal} {Science}\ }\textbf {\bibinfo {volume}
  {355}},\ \bibinfo {pages} {eaal1979} (\bibinfo {year} {2017})}\BibitemShut
  {NoStop}%
\bibitem [{\citenamefont {Chandragiri}\ \emph {et~al.}(2020)\citenamefont
  {Chandragiri}, \citenamefont {Doostmohammadi}, \citenamefont {Yeomans},\ and\
  \citenamefont {Thampi}}]{chandragiri2020flow}%
  \BibitemOpen
  \bibfield  {author} {\bibinfo {author} {\bibfnamefont {S.}~\bibnamefont
  {Chandragiri}}, \bibinfo {author} {\bibfnamefont {A.}~\bibnamefont
  {Doostmohammadi}}, \bibinfo {author} {\bibfnamefont {J.~M.}\ \bibnamefont
  {Yeomans}},\ and\ \bibinfo {author} {\bibfnamefont {S.~P.}\ \bibnamefont
  {Thampi}},\ }\href@noop {} {\bibfield  {journal} {\bibinfo  {journal} {Phys.
  Rev. Lett.}\ }\textbf {\bibinfo {volume} {125}},\ \bibinfo {pages} {148002}
  (\bibinfo {year} {2020})}\BibitemShut {NoStop}%
\bibitem [{\citenamefont {Chandrakar}\ \emph {et~al.}(2020)\citenamefont
  {Chandrakar}, \citenamefont {Varghese}, \citenamefont {Aghvami},
  \citenamefont {Baskaran}, \citenamefont {Dogic},\ and\ \citenamefont
  {Duclos}}]{chandrakar2020confinement}%
  \BibitemOpen
  \bibfield  {author} {\bibinfo {author} {\bibfnamefont {P.}~\bibnamefont
  {Chandrakar}}, \bibinfo {author} {\bibfnamefont {M.}~\bibnamefont
  {Varghese}}, \bibinfo {author} {\bibfnamefont {S.~A.}\ \bibnamefont
  {Aghvami}}, \bibinfo {author} {\bibfnamefont {A.}~\bibnamefont {Baskaran}},
  \bibinfo {author} {\bibfnamefont {Z.}~\bibnamefont {Dogic}},\ and\ \bibinfo
  {author} {\bibfnamefont {G.}~\bibnamefont {Duclos}},\ }\href@noop {}
  {\bibfield  {journal} {\bibinfo  {journal} {Phys. Rev. Lett.}\ }\textbf
  {\bibinfo {volume} {125}},\ \bibinfo {pages} {257801} (\bibinfo {year}
  {2020})}\BibitemShut {NoStop}%
\bibitem [{\citenamefont {Sokolov}\ and\ \citenamefont
  {Aranson}(2012)}]{sokolov2012physical}%
  \BibitemOpen
  \bibfield  {author} {\bibinfo {author} {\bibfnamefont {A.}~\bibnamefont
  {Sokolov}}\ and\ \bibinfo {author} {\bibfnamefont {I.~S.}\ \bibnamefont
  {Aranson}},\ }\href@noop {} {\bibfield  {journal} {\bibinfo  {journal} {Phys.
  Rev. Lett.}\ }\textbf {\bibinfo {volume} {109}},\ \bibinfo {pages} {248109}
  (\bibinfo {year} {2012})}\BibitemShut {NoStop}%
\bibitem [{\citenamefont {Nishiguchi}\ and\ \citenamefont
  {Sano}(2015)}]{nishiguchi2015mesoscopic}%
  \BibitemOpen
  \bibfield  {author} {\bibinfo {author} {\bibfnamefont {D.}~\bibnamefont
  {Nishiguchi}}\ and\ \bibinfo {author} {\bibfnamefont {M.}~\bibnamefont
  {Sano}},\ }\href@noop {} {\bibfield  {journal} {\bibinfo  {journal} {Phys.
  Rev. E}\ }\textbf {\bibinfo {volume} {92}},\ \bibinfo {pages} {052309}
  (\bibinfo {year} {2015})}\BibitemShut {NoStop}%
\bibitem [{\citenamefont {Heidenreich}\ \emph {et~al.}(2016)\citenamefont
  {Heidenreich}, \citenamefont {Dunkel}, \citenamefont {Klapp},\ and\
  \citenamefont {B{\"a}r}}]{heidenreich2016hydrodynamic}%
  \BibitemOpen
  \bibfield  {author} {\bibinfo {author} {\bibfnamefont {S.}~\bibnamefont
  {Heidenreich}}, \bibinfo {author} {\bibfnamefont {J.}~\bibnamefont {Dunkel}},
  \bibinfo {author} {\bibfnamefont {S.~H.~L.}\ \bibnamefont {Klapp}},\ and\
  \bibinfo {author} {\bibfnamefont {M.}~\bibnamefont {B{\"a}r}},\ }\href@noop
  {} {\bibfield  {journal} {\bibinfo  {journal} {Phys. Rev. E}\ }\textbf
  {\bibinfo {volume} {94}},\ \bibinfo {pages} {020601(R)} (\bibinfo {year}
  {2016})}\BibitemShut {NoStop}%
\bibitem [{\citenamefont {Reinken}\ \emph {et~al.}(2018)\citenamefont
  {Reinken}, \citenamefont {Klapp}, \citenamefont {B{\"a}r},\ and\
  \citenamefont {Heidenreich}}]{reinken2018derivation}%
  \BibitemOpen
  \bibfield  {author} {\bibinfo {author} {\bibfnamefont {H.}~\bibnamefont
  {Reinken}}, \bibinfo {author} {\bibfnamefont {S.~H.~L.}\ \bibnamefont
  {Klapp}}, \bibinfo {author} {\bibfnamefont {M.}~\bibnamefont {B{\"a}r}},\
  and\ \bibinfo {author} {\bibfnamefont {S.}~\bibnamefont {Heidenreich}},\
  }\href@noop {} {\bibfield  {journal} {\bibinfo  {journal} {Phys. Rev. E}\
  }\textbf {\bibinfo {volume} {97}},\ \bibinfo {pages} {022613} (\bibinfo
  {year} {2018})}\BibitemShut {NoStop}%
\bibitem [{\citenamefont {Bratanov}\ \emph {et~al.}(2015)\citenamefont
  {Bratanov}, \citenamefont {Jenko},\ and\ \citenamefont
  {Frey}}]{bratanov2015new}%
  \BibitemOpen
  \bibfield  {author} {\bibinfo {author} {\bibfnamefont {V.}~\bibnamefont
  {Bratanov}}, \bibinfo {author} {\bibfnamefont {F.}~\bibnamefont {Jenko}},\
  and\ \bibinfo {author} {\bibfnamefont {E.}~\bibnamefont {Frey}},\ }\href@noop
  {} {\bibfield  {journal} {\bibinfo  {journal} {Proc. Natl. Acad. Sci.
  U.S.A.}\ }\textbf {\bibinfo {volume} {112}},\ \bibinfo {pages} {15048}
  (\bibinfo {year} {2015})}\BibitemShut {NoStop}%
\bibitem [{\citenamefont {James}\ \emph {et~al.}(2018)\citenamefont {James},
  \citenamefont {Bos},\ and\ \citenamefont {Wilczek}}]{james2018turbulence}%
  \BibitemOpen
  \bibfield  {author} {\bibinfo {author} {\bibfnamefont {M.}~\bibnamefont
  {James}}, \bibinfo {author} {\bibfnamefont {W.~J.~T.}\ \bibnamefont {Bos}},\
  and\ \bibinfo {author} {\bibfnamefont {M.}~\bibnamefont {Wilczek}},\
  }\href@noop {} {\bibfield  {journal} {\bibinfo  {journal} {Phys. Rev.
  Fluids}\ }\textbf {\bibinfo {volume} {3}},\ \bibinfo {pages} {061101(R)}
  (\bibinfo {year} {2018})}\BibitemShut {NoStop}%
\bibitem [{\citenamefont {James}\ and\ \citenamefont
  {Wilczek}(2018)}]{james2018vortex}%
  \BibitemOpen
  \bibfield  {author} {\bibinfo {author} {\bibfnamefont {M.}~\bibnamefont
  {James}}\ and\ \bibinfo {author} {\bibfnamefont {M.}~\bibnamefont
  {Wilczek}},\ }\href@noop {} {\bibfield  {journal} {\bibinfo  {journal} {Eur.
  Phys. J. E}\ }\textbf {\bibinfo {volume} {41}},\ \bibinfo {pages} {21}
  (\bibinfo {year} {2018})}\BibitemShut {NoStop}%
\bibitem [{\citenamefont {Reinken}\ \emph {et~al.}(2019)\citenamefont
  {Reinken}, \citenamefont {Heidenreich}, \citenamefont {Baer},\ and\
  \citenamefont {Klapp}}]{reinken2019anisotropic}%
  \BibitemOpen
  \bibfield  {author} {\bibinfo {author} {\bibfnamefont {H.}~\bibnamefont
  {Reinken}}, \bibinfo {author} {\bibfnamefont {S.}~\bibnamefont
  {Heidenreich}}, \bibinfo {author} {\bibfnamefont {M.}~\bibnamefont {Baer}},\
  and\ \bibinfo {author} {\bibfnamefont {S.}~\bibnamefont {Klapp}},\
  }\href@noop {} {\bibfield  {journal} {\bibinfo  {journal} {New J. Phys.}\
  }\textbf {\bibinfo {volume} {21}},\ \bibinfo {pages} {013037} (\bibinfo
  {year} {2019})}\BibitemShut {NoStop}%
\bibitem [{\citenamefont {Mukherjee}\ \emph {et~al.}(2021)\citenamefont
  {Mukherjee}, \citenamefont {Singh}, \citenamefont {James},\ and\
  \citenamefont {Ray}}]{mukherjee2021anomalous}%
  \BibitemOpen
  \bibfield  {author} {\bibinfo {author} {\bibfnamefont {S.}~\bibnamefont
  {Mukherjee}}, \bibinfo {author} {\bibfnamefont {R.~K.}\ \bibnamefont
  {Singh}}, \bibinfo {author} {\bibfnamefont {M.}~\bibnamefont {James}},\ and\
  \bibinfo {author} {\bibfnamefont {S.~S.}\ \bibnamefont {Ray}},\ }\href@noop
  {} {\bibfield  {journal} {\bibinfo  {journal} {Phys. Rev. Lett.}\ }\textbf
  {\bibinfo {volume} {127}},\ \bibinfo {pages} {118001} (\bibinfo {year}
  {2021})}\BibitemShut {NoStop}%
\bibitem [{\citenamefont {Reinken}\ \emph
  {et~al.}(2022{\natexlab{b}})\citenamefont {Reinken}, \citenamefont {Klapp},\
  and\ \citenamefont {Wilczek}}]{reinken2022optimal}%
  \BibitemOpen
  \bibfield  {author} {\bibinfo {author} {\bibfnamefont {H.}~\bibnamefont
  {Reinken}}, \bibinfo {author} {\bibfnamefont {S.~H.}\ \bibnamefont {Klapp}},\
  and\ \bibinfo {author} {\bibfnamefont {M.}~\bibnamefont {Wilczek}},\
  }\href@noop {} {\bibfield  {journal} {\bibinfo  {journal} {Phys. Rev.
  Fluids}\ }\textbf {\bibinfo {volume} {7}},\ \bibinfo {pages} {084501}
  (\bibinfo {year} {2022}{\natexlab{b}})}\BibitemShut {NoStop}%
\bibitem [{\citenamefont {Toner}\ and\ \citenamefont
  {Tu}(1998)}]{toner1998flocks}%
  \BibitemOpen
  \bibfield  {author} {\bibinfo {author} {\bibfnamefont {J.}~\bibnamefont
  {Toner}}\ and\ \bibinfo {author} {\bibfnamefont {Y.}~\bibnamefont {Tu}},\
  }\href@noop {} {\bibfield  {journal} {\bibinfo  {journal} {Phys. Rev. E}\
  }\textbf {\bibinfo {volume} {58}},\ \bibinfo {pages} {4828} (\bibinfo {year}
  {1998})}\BibitemShut {NoStop}%
\bibitem [{\citenamefont {Toner}\ \emph {et~al.}(2005)\citenamefont {Toner},
  \citenamefont {Tu},\ and\ \citenamefont
  {Ramaswamy}}]{toner2005hydrodynamics}%
  \BibitemOpen
  \bibfield  {author} {\bibinfo {author} {\bibfnamefont {J.}~\bibnamefont
  {Toner}}, \bibinfo {author} {\bibfnamefont {Y.}~\bibnamefont {Tu}},\ and\
  \bibinfo {author} {\bibfnamefont {S.}~\bibnamefont {Ramaswamy}},\ }\href@noop
  {} {\bibfield  {journal} {\bibinfo  {journal} {Ann. Phys.}\ }\textbf
  {\bibinfo {volume} {318}},\ \bibinfo {pages} {170} (\bibinfo {year}
  {2005})}\BibitemShut {NoStop}%
\bibitem [{\citenamefont {Cross}\ and\ \citenamefont
  {Hohenberg}(1993)}]{cross1993pattern}%
  \BibitemOpen
  \bibfield  {author} {\bibinfo {author} {\bibfnamefont {M.~C.}\ \bibnamefont
  {Cross}}\ and\ \bibinfo {author} {\bibfnamefont {P.~C.}\ \bibnamefont
  {Hohenberg}},\ }\href@noop {} {\bibfield  {journal} {\bibinfo  {journal}
  {Rev. Mod. Phys.}\ }\textbf {\bibinfo {volume} {65}},\ \bibinfo {pages} {851}
  (\bibinfo {year} {1993})}\BibitemShut {NoStop}%
\bibitem [{\citenamefont {Newell}\ \emph {et~al.}(1993)\citenamefont {Newell},
  \citenamefont {Passot},\ and\ \citenamefont {Lega}}]{newell1993order}%
  \BibitemOpen
  \bibfield  {author} {\bibinfo {author} {\bibfnamefont {A.~C.}\ \bibnamefont
  {Newell}}, \bibinfo {author} {\bibfnamefont {T.}~\bibnamefont {Passot}},\
  and\ \bibinfo {author} {\bibfnamefont {J.}~\bibnamefont {Lega}},\ }\href@noop
  {} {\bibfield  {journal} {\bibinfo  {journal} {Ann. Rev. Fluid Mech.}\
  }\textbf {\bibinfo {volume} {25}},\ \bibinfo {pages} {399} (\bibinfo {year}
  {1993})}\BibitemShut {NoStop}%
\bibitem [{\citenamefont {Aranson}\ and\ \citenamefont
  {Kramer}(2002)}]{aranson2002world}%
  \BibitemOpen
  \bibfield  {author} {\bibinfo {author} {\bibfnamefont {I.~S.}\ \bibnamefont
  {Aranson}}\ and\ \bibinfo {author} {\bibfnamefont {L.}~\bibnamefont
  {Kramer}},\ }\href@noop {} {\bibfield  {journal} {\bibinfo  {journal} {Rev.
  Mod. Phys.}\ }\textbf {\bibinfo {volume} {74}},\ \bibinfo {pages} {99}
  (\bibinfo {year} {2002})}\BibitemShut {NoStop}%
\bibitem [{\citenamefont {James}(2020)}]{james2020phdthesis}%
  \BibitemOpen
  \bibfield  {author} {\bibinfo {author} {\bibfnamefont {M.}~\bibnamefont
  {James}},\ }\emph {\bibinfo {title} {Turbulence and pattern formation in
  continuum models for active matter}},\ \href@noop {} {\bibinfo {type} {Phd
  thesis}},\ \bibinfo  {school} {Georg-August-Universit\"at G\"ottingen},
  \bibinfo {address} {G\"ottingen, Germany} (\bibinfo {year}
  {2020})\BibitemShut {NoStop}%
\bibitem [{\citenamefont {Meurer}\ \emph {et~al.}(2017)\citenamefont {Meurer},
  \citenamefont {Smith}, \citenamefont {Paprocki}, \citenamefont
  {\v{C}ert\'{i}k}, \citenamefont {Kirpichev}, \citenamefont {Rocklin},
  \citenamefont {Kumar}, \citenamefont {Ivanov}, \citenamefont {Moore},
  \citenamefont {Singh}, \citenamefont {Rathnayake}, \citenamefont {Vig},
  \citenamefont {Granger}, \citenamefont {Muller}, \citenamefont {Bonazzi},
  \citenamefont {Gupta}, \citenamefont {Vats}, \citenamefont {Johansson},
  \citenamefont {Pedregosa}, \citenamefont {Curry}, \citenamefont {Terrel},
  \citenamefont {Rou\v{c}ka}, \citenamefont {Saboo}, \citenamefont {Fernando},
  \citenamefont {Kulal}, \citenamefont {Cimrman},\ and\ \citenamefont
  {Scopatz}}]{sympy}%
  \BibitemOpen
  \bibfield  {author} {\bibinfo {author} {\bibfnamefont {A.}~\bibnamefont
  {Meurer}}, \bibinfo {author} {\bibfnamefont {C.~P.}\ \bibnamefont {Smith}},
  \bibinfo {author} {\bibfnamefont {M.}~\bibnamefont {Paprocki}}, \bibinfo
  {author} {\bibfnamefont {O.}~\bibnamefont {\v{C}ert\'{i}k}}, \bibinfo
  {author} {\bibfnamefont {S.~B.}\ \bibnamefont {Kirpichev}}, \bibinfo {author}
  {\bibfnamefont {M.}~\bibnamefont {Rocklin}}, \bibinfo {author} {\bibfnamefont
  {A.}~\bibnamefont {Kumar}}, \bibinfo {author} {\bibfnamefont
  {S.}~\bibnamefont {Ivanov}}, \bibinfo {author} {\bibfnamefont {J.~K.}\
  \bibnamefont {Moore}}, \bibinfo {author} {\bibfnamefont {S.}~\bibnamefont
  {Singh}}, \bibinfo {author} {\bibfnamefont {T.}~\bibnamefont {Rathnayake}},
  \bibinfo {author} {\bibfnamefont {S.}~\bibnamefont {Vig}}, \bibinfo {author}
  {\bibfnamefont {B.~E.}\ \bibnamefont {Granger}}, \bibinfo {author}
  {\bibfnamefont {R.~P.}\ \bibnamefont {Muller}}, \bibinfo {author}
  {\bibfnamefont {F.}~\bibnamefont {Bonazzi}}, \bibinfo {author} {\bibfnamefont
  {H.}~\bibnamefont {Gupta}}, \bibinfo {author} {\bibfnamefont
  {S.}~\bibnamefont {Vats}}, \bibinfo {author} {\bibfnamefont {F.}~\bibnamefont
  {Johansson}}, \bibinfo {author} {\bibfnamefont {F.}~\bibnamefont
  {Pedregosa}}, \bibinfo {author} {\bibfnamefont {M.~J.}\ \bibnamefont
  {Curry}}, \bibinfo {author} {\bibfnamefont {A.~R.}\ \bibnamefont {Terrel}},
  \bibinfo {author} {\bibfnamefont {v.}~\bibnamefont {Rou\v{c}ka}}, \bibinfo
  {author} {\bibfnamefont {A.}~\bibnamefont {Saboo}}, \bibinfo {author}
  {\bibfnamefont {I.}~\bibnamefont {Fernando}}, \bibinfo {author}
  {\bibfnamefont {S.}~\bibnamefont {Kulal}}, \bibinfo {author} {\bibfnamefont
  {R.}~\bibnamefont {Cimrman}},\ and\ \bibinfo {author} {\bibfnamefont
  {A.}~\bibnamefont {Scopatz}},\ }\href {https://doi.org/10.7717/peerj-cs.103}
  {\bibfield  {journal} {\bibinfo  {journal} {PeerJ Comput. Sci.}\ }\textbf
  {\bibinfo {volume} {3}},\ \bibinfo {pages} {e103} (\bibinfo {year}
  {2017})}\BibitemShut {NoStop}%
\bibitem [{\citenamefont {T{\'e}l}\ and\ \citenamefont
  {Lai}(2008)}]{tel2008chaotic}%
  \BibitemOpen
  \bibfield  {author} {\bibinfo {author} {\bibfnamefont {T.}~\bibnamefont
  {T{\'e}l}}\ and\ \bibinfo {author} {\bibfnamefont {Y.-C.}\ \bibnamefont
  {Lai}},\ }\href@noop {} {\bibfield  {journal} {\bibinfo  {journal} {Physics
  Reports}\ }\textbf {\bibinfo {volume} {460}},\ \bibinfo {pages} {245}
  (\bibinfo {year} {2008})}\BibitemShut {NoStop}%
\bibitem [{\citenamefont {Busse}(1978)}]{busse1978non}%
  \BibitemOpen
  \bibfield  {author} {\bibinfo {author} {\bibfnamefont {F.}~\bibnamefont
  {Busse}},\ }\href@noop {} {\bibfield  {journal} {\bibinfo  {journal} {Rep.
  Prog. Phys.}\ }\textbf {\bibinfo {volume} {41}},\ \bibinfo {pages} {1929}
  (\bibinfo {year} {1978})}\BibitemShut {NoStop}%
\bibitem [{\citenamefont {Cross}\ and\ \citenamefont
  {Greenside}(2009)}]{cross2009pattern}%
  \BibitemOpen
  \bibfield  {author} {\bibinfo {author} {\bibfnamefont {M.}~\bibnamefont
  {Cross}}\ and\ \bibinfo {author} {\bibfnamefont {H.}~\bibnamefont
  {Greenside}},\ }\href@noop {} {\emph {\bibinfo {title} {Pattern formation and
  dynamics in nonequilibrium systems}}}\ (\bibinfo  {publisher} {Cambridge
  University Press},\ \bibinfo {year} {2009})\BibitemShut {NoStop}%
\bibitem [{\citenamefont {Schneider}\ \emph {et~al.}(2007)\citenamefont
  {Schneider}, \citenamefont {Eckhardt},\ and\ \citenamefont
  {Yorke}}]{schneider2007turbulence}%
  \BibitemOpen
  \bibfield  {author} {\bibinfo {author} {\bibfnamefont {T.~M.}\ \bibnamefont
  {Schneider}}, \bibinfo {author} {\bibfnamefont {B.}~\bibnamefont
  {Eckhardt}},\ and\ \bibinfo {author} {\bibfnamefont {J.~A.}\ \bibnamefont
  {Yorke}},\ }\href@noop {} {\bibfield  {journal} {\bibinfo  {journal} {Phys.
  Rev. Lett.}\ }\textbf {\bibinfo {volume} {99}},\ \bibinfo {pages} {034502}
  (\bibinfo {year} {2007})}\BibitemShut {NoStop}%
\bibitem [{\citenamefont {Faisst}\ and\ \citenamefont
  {Eckhardt}(2003)}]{faisst2003traveling}%
  \BibitemOpen
  \bibfield  {author} {\bibinfo {author} {\bibfnamefont {H.}~\bibnamefont
  {Faisst}}\ and\ \bibinfo {author} {\bibfnamefont {B.}~\bibnamefont
  {Eckhardt}},\ }\href@noop {} {\bibfield  {journal} {\bibinfo  {journal}
  {Phys. Rev. Lett.}\ }\textbf {\bibinfo {volume} {91}},\ \bibinfo {pages}
  {224502} (\bibinfo {year} {2003})}\BibitemShut {NoStop}%
\bibitem [{\citenamefont {Eckhardt}\ \emph {et~al.}(2007)\citenamefont
  {Eckhardt}, \citenamefont {Schneider}, \citenamefont {Hof},\ and\
  \citenamefont {Westerweel}}]{eckhardt2007turbulence}%
  \BibitemOpen
  \bibfield  {author} {\bibinfo {author} {\bibfnamefont {B.}~\bibnamefont
  {Eckhardt}}, \bibinfo {author} {\bibfnamefont {T.~M.}\ \bibnamefont
  {Schneider}}, \bibinfo {author} {\bibfnamefont {B.}~\bibnamefont {Hof}},\
  and\ \bibinfo {author} {\bibfnamefont {J.}~\bibnamefont {Westerweel}},\
  }\href@noop {} {\bibfield  {journal} {\bibinfo  {journal} {Annu. Rev. Fluid
  Mech.}\ }\textbf {\bibinfo {volume} {39}},\ \bibinfo {pages} {447} (\bibinfo
  {year} {2007})}\BibitemShut {NoStop}%
\bibitem [{\citenamefont {Menzel}\ \emph {et~al.}(2014)\citenamefont {Menzel},
  \citenamefont {Ohta},\ and\ \citenamefont {L{\"o}wen}}]{menzel2014active}%
  \BibitemOpen
  \bibfield  {author} {\bibinfo {author} {\bibfnamefont {A.~M.}\ \bibnamefont
  {Menzel}}, \bibinfo {author} {\bibfnamefont {T.}~\bibnamefont {Ohta}},\ and\
  \bibinfo {author} {\bibfnamefont {H.}~\bibnamefont {L{\"o}wen}},\ }\href@noop
  {} {\bibfield  {journal} {\bibinfo  {journal} {Phys. Rev. E}\ }\textbf
  {\bibinfo {volume} {89}},\ \bibinfo {pages} {022301} (\bibinfo {year}
  {2014})}\BibitemShut {NoStop}%
\bibitem [{\citenamefont {Laradji}\ \emph {et~al.}(1997)\citenamefont
  {Laradji}, \citenamefont {Shi}, \citenamefont {Noolandi},\ and\ \citenamefont
  {Desai}}]{laradji1997stability}%
  \BibitemOpen
  \bibfield  {author} {\bibinfo {author} {\bibfnamefont {M.}~\bibnamefont
  {Laradji}}, \bibinfo {author} {\bibfnamefont {A.-C.}\ \bibnamefont {Shi}},
  \bibinfo {author} {\bibfnamefont {J.}~\bibnamefont {Noolandi}},\ and\
  \bibinfo {author} {\bibfnamefont {R.~C.}\ \bibnamefont {Desai}},\ }\href@noop
  {} {\bibfield  {journal} {\bibinfo  {journal} {Macromolecules}\ }\textbf
  {\bibinfo {volume} {30}},\ \bibinfo {pages} {3242} (\bibinfo {year}
  {1997})}\BibitemShut {NoStop}%
\bibitem [{\citenamefont {Reinken}\ and\ \citenamefont
  {Menzel}(2024)}]{reinken2024vortex}%
  \BibitemOpen
  \bibfield  {author} {\bibinfo {author} {\bibfnamefont {H.}~\bibnamefont
  {Reinken}}\ and\ \bibinfo {author} {\bibfnamefont {A.~M.}\ \bibnamefont
  {Menzel}},\ }\href@noop {} {\bibfield  {journal} {\bibinfo  {journal} {Phys.
  Rev. Lett.}\ }\textbf {\bibinfo {volume} {132}},\ \bibinfo {pages} {138301}
  (\bibinfo {year} {2024})}\BibitemShut {NoStop}%
\bibitem [{\citenamefont {Xu}\ and\ \citenamefont {Wu}(2024)}]{xu2024self}%
  \BibitemOpen
  \bibfield  {author} {\bibinfo {author} {\bibfnamefont {H.}~\bibnamefont
  {Xu}}\ and\ \bibinfo {author} {\bibfnamefont {Y.}~\bibnamefont {Wu}},\
  }\href@noop {} {\bibfield  {journal} {\bibinfo  {journal} {Nature}\ }\textbf
  {\bibinfo {volume} {627}},\ \bibinfo {pages} {553} (\bibinfo {year}
  {2024})}\BibitemShut {NoStop}%
\end{thebibliography}

%

\end{document}